\documentclass[]{jfm}

\usepackage{graphicx}
\usepackage{newtxtext}
\usepackage{newtxmath}
\usepackage{natbib}
\usepackage{hyperref}
\hypersetup{
    colorlinks = true,
    urlcolor   = blue,
    citecolor  = black,
}

\newcommand{\RomanNumeralCaps}[1]
\linenumbers

\usepackage{amsmath,amssymb}
\usepackage{graphicx}
\usepackage{bm,bbm}
\usepackage[]{subfigure}
\usepackage[font=footnotesize,format=plain,justification=justified,width=\textwidth]{caption}
\usepackage{tikz}
\usepackage{pgfplots}
\usepackage{pgfplotstable}
\usepgfplotslibrary{patchplots}
\usepgfplotslibrary{groupplots}
\usepgfplotslibrary{fillbetween}
\usepackage{pdfpages}
\usepackage{scalefnt}
\usepackage{microtype}
\usepackage{xfrac}
\usepackage{relsize}

\pgfkeys{/pgf/number format/.cd,1000 sep={\,}}
\pgfplotsset{
  every tick label/.append style={scale=0.8}
}


\newcount\ndots
\def\drawline#1#2{\raise 2.5pt\vbox{\hrule width #1pt height #2pt}}

\def\trian{\raise 1.25pt\hbox{$\scriptscriptstyle\triangle$}\nobreak\ }
\def\solidtrian{\raise 1.25pt
\hbox to 3bp{
\hfill}\nobreak\ }
\def\dsolidtrian{\raise 1.25pt
\hbox to 3bp{
\hfill}\nobreak\ }
\def\soliddiamond{\raise 1.25pt
\hbox to 4bp{
\hfill}\nobreak\ }

\def\square{${\vcenter{\hrule height .4pt
              \hbox{\vrule width .4pt height 3pt \kern 3pt \vrule width .4pt}
          \hrule height .4pt}}$\nobreak\ }

\def\plus{\raise 1.25pt \hbox{$\scriptscriptstyle +$}\nobreak\ }
\def\x{\raise 1.25pt \hbox{$\scriptscriptstyle \times$}\nobreak\ }
\def\legendtable#1{\vbox{\baselineskip=10pt\tabskip=0pt\let\\=\cr\halign{\hfil##\hskip 3pt&##\hfil\cr#1\crcr}}}
\def\lllegend#1 #2 #3{\figlab {#1} {#2} {\legendtable{#3}}}
\def\lrlegend#1 #2 #3{\figlab {#1} {#2} {\llap{\legendtable{#3}}}}
\def\ullegend#1 #2 #3{\figlab {#1} {#2} {\vtop{\hrule height 0pt\legendtable{#3}}}}
\def\urlegend#1 #2 #3{\figlab {#1} {#2} {\llap{\vtop{\hrule height
0pt\legendtable{#3}}}}}

\def\Dpartial#1#2{ \frac{\partial #1}{\partial #2} }

\def\Dnorm#1#2{ \frac{d #1 }{ d #2} }

\def\onedot{$\mathsurround0pt\ldotp$}
\def\cddot{
  \mathbin{\vcenter{\baselineskip.67ex
    \hbox{\onedot}\hbox{\onedot}}%
  }}%

\newcommand{\bB}{{\mathbf{B}}}
\newcommand{\bD}{{\mathbf{D}}}
\newcommand{\bF}{{\mathbf{F}}}
\newcommand{\bI}{{\mathbf{I}}}
\newcommand{\bS}{{\mathbf{S}}}
\newcommand{\bU}{{\mathbf{U}}}

\newcommand{\bOmega}{{\boldsymbol{\Omega}}}

\newcommand{\bn}{{\mathbf{n}}}
\newcommand{\bp}{{\mathbf{p}}}
\newcommand{\br}{{\mathbf{r}}}
\newcommand{\bu}{{\mathbf{u}}}
\newcommand{\bv}{{\mathbf{v}}}
\newcommand{\bx}{{\mathbf{x}}}
\newcommand{\bX}{{\mathbf{X}}}

\definecolor{RYB1}{RGB}{207, 37, 37}
\definecolor{RYB2}{RGB}{37, 91, 207}
\definecolor{RYB3}{RGB}{37, 207, 91}
\definecolor{RYB4}{RGB}{163,26,145}
\definecolor{RYB5}{RGB}{253, 180, 98}
\definecolor{RYB6}{RGB}{179, 222, 105}
\definecolor{RYB7}{RGB}{128, 177, 211}

\pgfplotscreateplotcyclelist{newcolors}{
{RYB1,every mark/.append style={fill=RYB1,mark size={2.5}},mark=*},
{RYB2,every mark/.append style={fill=RYB2},mark=square*},
{RYB3,every mark/.append style={fill=RYB3,mark size={3}},mark=triangle*},
{RYB4,every mark/.append style={fill=RYB4,mark size={4}},mark=x},
{RYB5,every mark/.append style={fill=RYB5,mark size={3}},mark=oplus},
{RYB7,every mark/.append style={fill=RYB7},mark=*},
}
\pgfplotsset{
    standard/.style={
    thick,
    compat=1.8,
            scale only axis,
        width=0.45\textwidth,
        enlarge x limits=0.05,
        enlarge y limits=0.05,
        max space between ticks=40,
cycle list name=newcolors
    }
}

\usepackage[colorinlistoftodos, color=blue!20!white, bordercolor=gray,
  textsize=tiny,textwidth=0.8in]{todonotes}

\newcommand{\axislabledfigure}[4]{
\begin{tikzpicture}
\node[anchor=south west, inner sep=0] (image) at (0,0)
    {\includegraphics[width=#4\textwidth]{Figures/#1}};
\begin{scope}[x={(image.south east)},y={(image.north west)}]    
   \node at (0.5,-0.07) {#2}; 
   \node[rotate=90] at (-0.025,0.5) {#3};
\end{scope}
\end{tikzpicture}
}

\newcommand{\axislabledfigureC}[5]{
\begin{tikzpicture}
\node[anchor=south west, inner sep=0] (image) at (0,0)
    {\includegraphics[width=#4\textwidth]{Figures/#1}};
\begin{scope}[x={(image.south east)},y={(image.north west)}]    
   \node at (0.5,-0.07) {#2}; 
   \node[rotate=90] at (-0.025,0.5) {#3};
\draw[blue!40!white, thick] (0.85,0.85) circle [radius=0.105];
\draw[blue,fill=blue!40!white,draw=none] (0.85,0.85) circle [radius=#5];
\end{scope}
\end{tikzpicture}
}

\title{Object Transport by a Confined Active Suspension}
\author{Jonathan B.~Freund
  \corresp{\email{jbfreund@illinois.edu}}}

\affiliation{Department of Aerospace Engineering, University of Illinois Urbana--Champaign, 104 South Wright Street, Urbana, IL 61802, USA}

\begin{document}
\maketitle





















\centerline{\rule{\textwidth}{0.4pt}}
\medskip

\begin{abstract}
    Numerical simulations in two space dimensions are used to examine the dynamics, transport, and equilibrium behaviors of a neutrally buoyant circular object immersed in an active suspension within a larger closed circular container.  The continuum model of Gao et al.\ ({\it Phys.\ Rev.\ Fluids}, 2017) represents the suspension of non-interacting, immotile, extensor-type microscopic agents that have a direction and strength and align in response to strain rate.  Such a suspension is well known to be unstable above an activity strength threshold, which depends upon the length scale of the confinement.  Introducing the object leads to additional phenomenology.  It can confine fluid between it and the container wall, which suppresses local suspension activity.  However, its motion also correlates strain rates near its surface with a concomitant correlated active-stress response.  Depending on the suspension activity strength, these mechanisms lead to either an attraction toward or repulsion away from the container wall.  In addition, a persistent propagating behavior is found for modest activity strength, which provides a mechanism for long-range transport.  When activity is so weak that the mobility of the object is essential to support suspension instability and sustain flow, the object essentially parks and all flow terminates when its mobility is diminished as it nears the container wall.  Together these mechanisms illustrate potential for performing relatively complex tasks with simple active agents, especially if activity strength is scheduled in time.
\end{abstract}

\centerline{\rule{\textwidth}{0.4pt}}

\section{Background and Introduction}

Non-equilibrium active suspensions of microscopic agents that generate internal stress have been studied extensively, typically as a model of living material, such as a cell interior or a suspension of bacteria~\citep{Marchetti:2013,Saintillan:2015,Saintillan:2018,Needleman:2017}.  The root mechanism in simple models is a strain induced alignment of the  suspended agents such that they exert a net internal stress.  The most interesting circumstances are when the overall suspension is unstable~\citep{Simha:2002}, for which perturbations overcome viscous resistance to form large-scale flow patterns~\citep{Dombrowski:2004}.   Unconstrained, this leads to a chaotic swirling flow pattern, whereas confinement in a container organizes the flow.  For a sufficiently small round container, the chaotic swirls are replaced by a lone circulating vortex, or all flow ceases if it is sufficiently small~\citep{Lushi:2014,Wioland:2013}. Similar low-dissipation global flow patterns arise in other geometries~\citep{Opathalage:2019}, and become increasingly complex in more complex containers~\citep{Hardouin:2020}.  Confinement within an immiscible droplet leads to additional behaviors, coupled with the deformability of the drop shape and its motion within the fluid beyond~\citep{Young:2021}.

The suspension we consider is perhaps the simplest in this class, constructed of uniformly distributed advected agents that orient only in response to the local strain-rate histories they experience.  Phenomenologically, this is taken to be an alignment of their otherwise uniformly distributed active axes, yielding a net internal fluid stress.  They also diffuse in space and in time relax their orientation distribution to uniform.  The basic instabilities of such an active fluid are understood~\citep{Marchetti:2013}.  Unconfined, they are long-wavelength, which loosely explains how they are mediated by the length scales of geometric confinement.  In general, stronger activity can overcome viscous resistance at smaller scales to the point where a seemingly chaotic, turbulence-like flow with a range of length scales develops~\citep{Gao:2017}.

Active suspensions can at least loosely be linked to biological tasks, such as  intracellular flow, cell mobility, or embryonic development.  Here we consider how this simplest suspension interacts with an inactive rigid object, freely suspended and much larger than the active agents.   Investigation focuses on how the object is transported by the suspension within a larger container.
The object is anticipated to couple with the suspension in several ways.  For a close spacing between it and the container wall, viscous resistance will suppress local instabilities, as would any narrow confinement.  However, this aspect of confinement is potentially countered by motion of the object, which will apply a correlated and potentially strong strain rate to the fluid between it and the container wall, which will in turn instigate a coordinated response in the suspension.  Several questions are considered regarding the resulting phenomenology. Does it have stable positions or distances from the container wall?  Does the active suspension transport it to the container wall or keep it free floating?  These questions intersect with past observations regarding how a container of decreasing size first organizes and then suppresses suspension instabilities~\citep{Lushi:2014,Wioland:2013}.  A simple circular object suspended in a circular container is used to identify the basic phenomenology.  This configuration builds on multiple studies in fixed annular geometries \citep{Gao:2017,Hardouin:2020}, now with the inner rigid boundary replaced with a zero-inertia free-floating object.

After introducing the simulation model and methods in section~\ref{s:simulationdetails}, results are consider in three stages.  First, weak activity is considered in section~\ref{s:lowactivity}, for which the basic circulating flow is a relatively straightforward generalization of cases considered previously.  However, there is now also a slow geometric instability that eventually leads to cessation of all suspension activity in the container, parking the object near the container wall.  In contrast, for strong activity, which is considered in section~\ref{s:highactivity}, the flow is chaotic and complex.  Still, the gross features of the phenomenology of the suspended object and its transport are identifiable and explain why in this case the free-floating object is effectively repelled from the container wall.  Section~\ref{s:transition} bridges these weak and strong limits.  It is only for these intermediate strengths that persistent transport along the container wall or contact with the container might occur.  Animated visualization of representative cases are provided as supplemental material (movies 1--10).

\section{Simulation Details}
\label{s:simulationdetails} 

\subsection{Active Suspension Model}
\label{ss:activesus}

The model for the active suspension is exactly that of \citet{Gao:2017}, which itself builds on several earlier developments~\citep{Simha:2002,Saintillan:2008}.  The active components are uniformly distributed immotile extensor (or contractor) particles.  Specifically, we use their coarse-grained model with the Bingham closure they introduce.  For simplicity, we only considered the limit of a diffuse suspension, which represents the basic phenomenology without the added complexity of steric alignment of the active agents and their volume fraction (in their notation, $\zeta = 0$ and $\beta = 0$).  This is sufficient to provide a representative phenomenology for an immersed object.   The original discussion of the model is complete~\citep{Gao:2017}, so we only provide a summary of it.  The same model has been used subsequently to study flow in additional geometries~\citep{Chen:2018} and in liquid drop~\citep{Young:2021}.

The viscous flow equations are augmented with an internal stress $\alpha \bD$,
\begin{align}
  -\nabla\cdot\left(\nabla \bu + \nabla \bu^T\right) + \nabla p &= \nabla \cdot (\alpha \bD) \label{e:mom}\\
  \nabla \cdot \bu &= 0 \label{e:incomp},                    
\end{align}
where $\bu$ is the velocity, and $p$ is the pressure that enforces its incompressibility. The strength $\alpha$ of the active stress component is negative for the case of extensors we consider.  More specifically, $\alpha$ is the dipolar strength $\sigma_a$ of the extensors, normalized by a velocity  $U_o$, length $\ell$, and the Newtonian viscosity of the suspension $\mu$:  $\alpha\equiv\sigma_a/\mu U_o \ell^2$.  For most cases, the immersed object has unit radius so $\ell = 1$ is appropriate.  
Alignment of the suspension agents is represented by tensor order parameter $\bD$, which is governed by an advection--diffusion equation,
\begin{equation}
  \Dpartial{\bD}{t} + \bu \cdot \nabla \bD - (\nabla\bu \cdot \bD + \bD \cdot \nabla \bu^T) = - \left(\nabla \bu + \nabla \bu^T\right)\cddot \bS + d_T \nabla^2
  \bD - 4 d_R(\bD - \bI/2).
  \label{e:D}
\end{equation}
Its left-hand side is the usual upper convective Maxwell advection of a tensor.   On its right-hand side, $d_T\equiv vD_T/b U_o$ is a non-dimensional coefficient of translational diffusion, and $d_R \equiv b D_R/v U_o$ is a coefficient of rotational diffusion toward isotropy ($\bD = \bI/2$), here written explicitly for two space dimensions.  The physical parameters that constitute $d_T$ and $d_R$ are the particle rotational diffusivity $D_R$, the particle translational diffusivity $D_T$, the effective particle volume fraction $v$, and particle dimension $b$.  Except when noted, $d_T = d_R = 0.025$ in the current simulations.  The term involving the rank-four tensor $\bS$ represents how strain rates orient the active agents, leading to their net active stress.  For $\bS$, \citet{Gao:2017} employed a closure based on the fourth moment of the microscopic distribution function $\Psi(\bx,\bp,t)$, which underlies the coarse-grained average of the active agent distribution: 
\begin{equation}
  1 = \int_0^{2\pi} \Psi(\bx,t) \;d\theta, \quad \bD(\bx,t) = \int_0^{2\pi} \bp\bp \,\Psi(\bx,t)  \;d\theta,
  \quad \bS(\bx,t) = \int_0^{2\pi} \bp\bp\bp\bp\, \Psi(\bx,t)  \;d\theta,
\end{equation}
where in two space dimensions $\bp = (\cos\theta,\sin\theta)$.  In solving the overall combined system, $\bD$ is evolved per (\ref{e:D}), and $\bS$ is closed by assuming a Bingham~\citep{Bingham:1974} distribution:  $\Psi(\bx,\bp,t) \approx \Psi_B(\bx,\bp,t) = A(\bx,t) \exp [\bB(\bx,t)\cddot\bp\bp]$.  Following the basic approach of \citet{Chaubal:1998}, also used by \citet{Gao:2017}, this is solved through taking $\bB$ to also be trace free and in the principal coordinates of $\bD$, indicated here with a $p$ superscript.  In two space dimensions, this yields explicit expressions in terms of modified Bessel functions,
\begin{align}
  D_{11}^p &= A \pi \left[I_0(B_{11}^p) + I_1(B_{11}^p) \right]\\
  D_{22}^p &= A \pi \left[I_0(B_{11}^p) - I_1(B_{11}^p) \right].
\end{align}
Since $\bD$ is also trace free, $B_{11}^p = -B_{22}^p$ simply solves 
\begin{equation}
  \frac{I_1(B_{11}^p)}{I_0(B_{11}^p)} = - \sqrt{1 - 4 (D_{11} D_{22} - D_{12}^2)},
\end{equation}
which provides values for the non-zero components of $\bS^p$:
\begin{align}
  S_{1111}^p &= A\pi \left[I_0(B_{11}^p) + \frac{2 B_{11}^p-1}{2 B_{11}^p}I_1(B_{11}^p) \right]\\
    S_{1122}^p = S_{1212}^p &= S_{1221}^p = S_{2112}^p = S_{2121}^p = S_{2211}^p = A \pi \frac{I_1(B_{11}^p)}{2 B_{11}^p} \\
    S_{2222}^p &= A \pi \left[I_0(B_{11}^p) - \frac{2 B_{11}^p+1}{2 B_{11}^p}I_1(B_{11}^p) \right],
\end{align}
with
\begin{equation}
  A = \frac{D_{11}^p/\pi}{I_0(B_{11}^p) + I_1(B_{11}^p)}.
\end{equation}
Rotation of $\bS^p$ for use in (\ref{e:D}) is straightforward.

The immersed object is assumed to have a no slip surface, which provides velocity boundary condition
\begin{equation}
  \bu(\bx,t) = \bU(t) + \bOmega(t) \times \br,
  \label{e:ubc}
\end{equation}
with $\bU$ its translational velocity and $\bOmega = [0,0,\Omega]^T$ its rotation rate, which is crossed with the vector from its centroid $\br = [x-x_o,y-y_o,0]^T$.  For most cases, $\bU$ and $\Omega$ are solved in conjunction with the flow equations (\ref{e:mom}) and (\ref{e:incomp}) to enforce zero net drag and torque on the surface $S_o$ of the object:
\begin{align}
  \bF &= \int_{S_o} \big[-p \bI + (\nabla \bu + \nabla \bu^T) + \alpha \bD\big]\cdot \bn\,d\bx = \mathbf{0} \label{e:F}\\
  T &=\int_{S_o} \br \times\big( \big[-p \bI + (\nabla \bu + \nabla \bu^T) + \alpha \bD\big]\cdot \bn\big)\,d\bx =0 \label{e:T}.
\end{align}
To illuminate mechanisms in specific cases, a kinematic constraint $\bv\cdot\bU=0$ replaces part of the $\bF = 0$ constraint.  In this case, the the combined system is instead solved for $\bU_\perp$, such that $\bv\cdot\bU_\perp = 0$, and $F$, such that $\bF = F\bv$ represents an external force resisting motion in the direction of unit vector $\bv$.  Some cases also simply fix the object in place with $\bU = \bOmega = \mathbf{0}$.  The fixed boundary $S_c$ of the container is also no slip, so $\bu=\mathbf{0}$ on $\bx \in S_c$.  It is assumed that boundaries do not orient the active agents, so  $\bn\cdot\nabla\bD =  \mathbf{0}$ on $\bx \in (S_o \cup S_c)$.  
    
\subsection{Numerical methods}

The momentum balance (\ref{e:mom}) and incompressiblity constraint (\ref{e:incomp}) are discretized with Lagrange polynomial quadrilateral finite elements of degree-$n$ for the velocity $\bu$ and degree $n-1$ for the pressure $p$.  Similarly, $\bD$ in (\ref{e:D}) is discretized with degree-$m$ polynomials.  The translational advection term $\bu\cdot \nabla \bD$ in (\ref{e:D}) is incorporated into the time derivative by moving the mesh at the fluid velocity.  Specifically, this is done by advecting the degree-$m$ polynomial function that maps the finite element mesh $\bx(\bX,t)$ to a fixed reference mesh $\bX$ with the local velocity starting from time $t_r$ when they coincide:
\begin{equation}
  \Dnorm{\bx(\bX,t)}{t} = \bu(\bx,t) \quad\text{for}\quad t \ge t_r \quad\text{and}\quad \bx(\bX,t_{r}) = \bX.
  \label{e:meshadvection}
\end{equation}
The distortional tensor advection terms ($\nabla \bu\cdot\bD + \bD \cdot \nabla\bu^T$), the $\bS$ alignment term, and the $d_R$ rotational relaxation terms are evaluated on the distorted $\bx$ mesh and time-integrated with a second-order backward differencing scheme.  The $d_T$ spatial diffusion term is time integrated with a first-order forward difference, which yields an implicit system that is solved in conjunction with the mass-matrix inversion.  

Overall, this moving-mesh approach is selected primarily to accomodate the motion of the immersed object.  Since the mesh distorts significantly in time, the reference $\bX$ mesh is periodically reconstructed for the current location of the object.  The solution at the current and recent time steps is then projected to the new mesh using the same degree-$m$ weak-form discretization.  This is done every $N_r=10$ time steps, which also sets a new $t_r$ in (\ref{e:meshadvection}).  To impose dynamic constraints, such as (\ref{e:F}) and (\ref{e:T}), or the similar $\bv\cdot\bU=0$ kinematic constraint, the linearity of (\ref{e:mom}) and (\ref{e:incomp}) are used to solve for the $\bU$ and $\Omega$ that enforce the boundary condition (\ref{e:ubc}).   The  scheme was implemented with the \textit{deal.ii} finite-element libraries~\citep{dealII94}. 

For most simulations $m = n = 3$, and total number of degrees of freedom for $\bu$, $p$, and $\bD$ ranged from 1460 for the weakest activity case ($\alpha = -0.6$) to 13\,840 for the most active case ($\alpha = -80$).  Time steps and simulation times varied significantly based on the stability restrictions, resolution, and phenomenology of the different cases; some simulations were run for over $10^6$ time step to accumulate statistics.  Most cases were run with multiple resolutions and time steps to confirm discretization independence.


\subsection{Flow configuration}

\begin{figure}
  \begin{center}
    \subfigure[]{\includegraphics[width=0.4\textwidth]{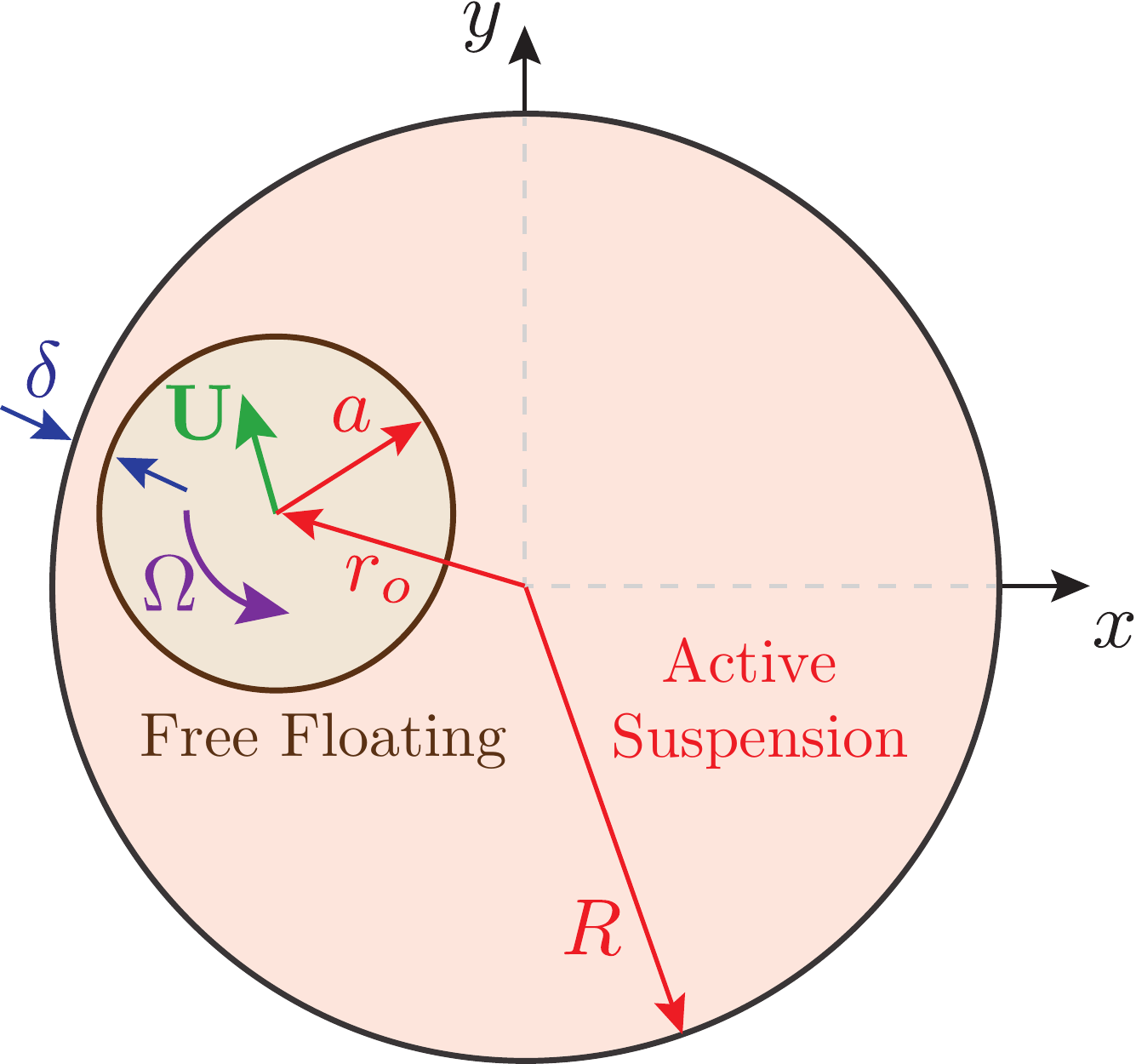}}
    \subfigure[]{\includegraphics[width=0.4\textwidth]{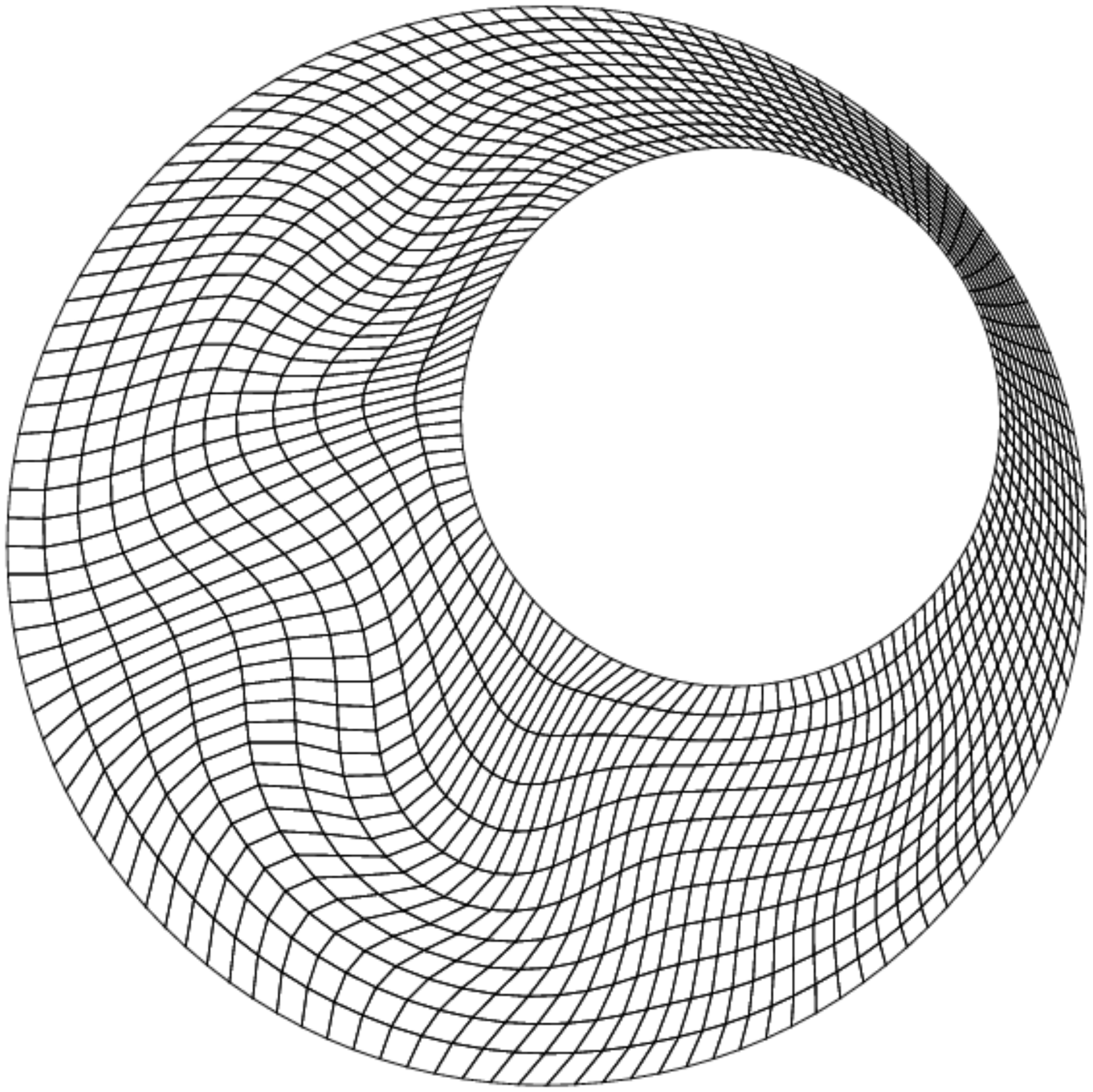}}
    \caption{(a) Configuration schematic; and (b) an example mapped mesh prior to remeshing for an $a  = 1$, $R = 2$, $\alpha = -5.0$ case with 160 quadrilateral cells and 5400 total degrees of freedom for a $n=m=3$ discretization.  For this and all visualizations, solutions are sampled uniformly within each element from the basis functions.  }\label{fig:config}
  \end{center}
\end{figure}

Figure~\ref{fig:config} (a) shows the configuration.  The immersed object is a circle of radius $a=1$ (in most cases) in a larger circular container of radius $R=2$.  Except when noted, the circle is initialized concentric with the container, with $\bx_o(0) = (0,0)$.  Figure~\ref{fig:config} (b) shows an example distorted mesh just prior to projection onto a new one.  New meshes are constructed in a straightforward way from circles and straight lines between the two boundaries.  The suspension is assumed to be initially isotropic:  $\bD(\bx,0) = \bI/2$.  Animated visualizations of the flows are shown in supplemental movies~1--10 for $\alpha = -0.625$, $-1$, $-2.5$, $-5.0$, $-10$, $-20$, $-40$ (for $a = 0.5$, $a=1.0$ and $a = 1.5$), and $-80$, respectively.

When $\alpha$ is such that the fluid is unstable for the given geometry, the flow rapidly evolves from any small perturbation.  None of the results were found to be sensitive to the perturbation details, aside from setting the rotation direction in cases for which the flow is so regular that the rotation direction does not change.  The $\alpha$ stability limit for $R=2$ with $d_T = d_R = 0.05$ without the free-floating object was found to be $\alpha_c \approx -0.63$, matching previous simulations and stability analysis~\citep{Gao:2017,Woodhouse:2012}.  In a corresponding fixed-boundary annulus container [$\bx_o = (0,0)$, $a=1$, and $R=2$, with $\bU = 0$ and $\Omega = 0$], $\alpha = -1.23$ was unstable to a circulating flow while $\alpha = -1.22$ was stable, matching the simulations and analysis of \citet{Chen:2018}.  The propagating wave solutions and transitions to a chaotic flow for the annular case also matched.

\section{Weak activity ($-1\le \alpha \le 0$)}
\label{s:lowactivity}

\subsection{Suspension instability}
\label{s:susstab}

\begin{figure}
  \begin{center}
    \includegraphics[width=0.35\textwidth]{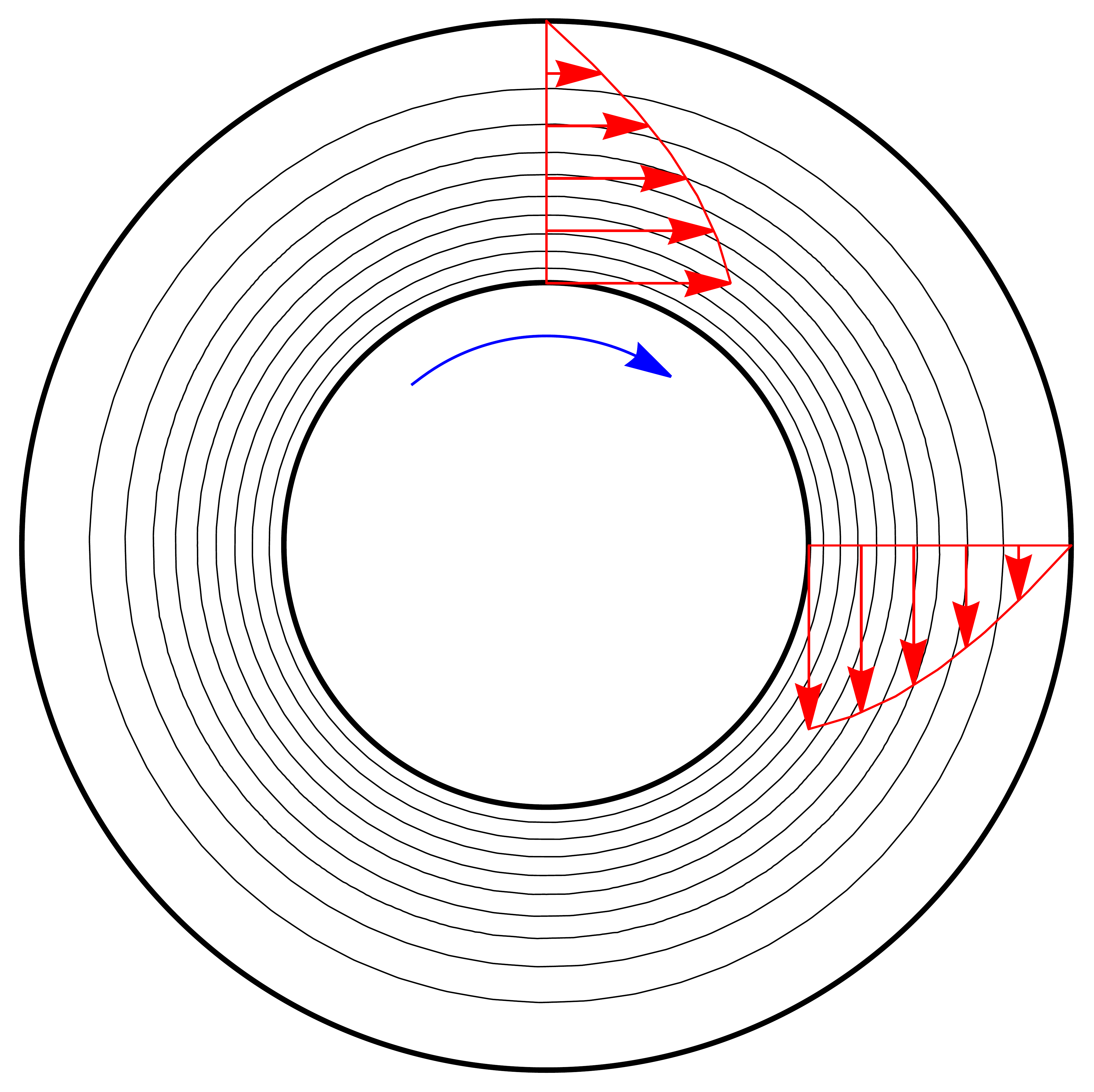}
  \end{center}
  \caption{Fully developed velocity profiles and streamfunction $\psi$ with $\Delta \psi = 0.0011$ contour spacing for $\alpha = -0.6$ for the force- and torque-free immersed circle.}\label{fig:lowalphainitial}
\end{figure}

Although the initial geometry with $|\bx_o| = r_o=0$ matches the annulus of by \citet{Chen:2018}, the mobility of the inner circle is expected to facilitate instability for weaker activity than their fixed-wall limit $\alpha_c \approx -1.23$.   In the fixed-wall case, an axisymmetric circulating flow is observed for small unstable $|\alpha|$, so the circular Couette-like flow in   figure~\ref{fig:lowalphainitial} is as expected, and is observed for  $\alpha \lesssim -0.6$.  The $t = 1500$ time shown is after the activity of the suspension instability has plateaued but before the geometric instability considered in the following section becomes pronounced.  At this time the circular object and container are still nearly concentric with $r_o = 0.001$, having started at $r_o=0$.  The circle rotates at $|\Omega| = 0.017$, with the sign of $\Omega$ depending on the specifics of the initial perturbation.  Of course, this rate is faster for larger $|\alpha|$:  $|\Omega| = 0.026$ for $\alpha = -0.625$, $|\Omega| = 0.052$ for $\alpha = -0.75$, and $|\Omega| = 0.087$ for $\alpha = -1$.  Perturbations decay for $\alpha \ge -0.575$.  Decreasing the rotational mobility of the circle by adding a resistance torque $T_r = - c_\tau \Omega$ increases the $|\alpha|$ suspension stability threshold.  For $\alpha = -1.0$, $c_\tau = 8.5$ is stable whereas $c_\tau = 8.0$ only slows rotation to $|\Omega| = 0.043$.

\subsection{Geometric instability}
\label{ss:geominstab}

\begin{figure}
\begin{center}
  \subfigure[]{
    \begin{tikzpicture}
      \begin{semilogyaxis}
        [ 
        ymin = 5.e-6,
        ymax = 1.1e0,
        xmin = 0,
        xmax = 5000,
        ylabel={$r_o(t)$},
        xlabel={$t$},
        tick scale binop=\times,
        ytick pos=left,
        width=0.45\textwidth,
        height=0.35\textwidth,        
        xminorticks=true,
        ]
        \addplot+[no marks, very thick, color=red] table[x
        expr=\thisrowno{0}, y expr=\thisrowno{1}, col sep=space] {Figures/r-alpha=1.0-smallpert.dat};
      \end{semilogyaxis}
      \begin{axis}
        [ 
        ymin = -0.05,
        ymax = 1.05,
        xmin = 0,
        xmax = 5000,
        ylabel={$r_o(t)$},
        xlabel={$t$},
        axis y line*=right,
        axis x line = none,
        tick scale binop=\times,
        width=0.45\textwidth,
        height=0.35\textwidth,
        grid=major,
        max space between ticks=25,
        ]
        \addplot+[no marks, thick, dashed, color=green!60!black] table[x
        expr=\thisrowno{0}, y expr=\thisrowno{1}, col sep=space] {Figures/r-alpha=1.0-smallpert-delta.dat};
        \addplot+[no marks, very thick, color=blue] table[x
        expr=\thisrowno{0}, y expr=\thisrowno{1}, col sep=space] {Figures/r-alpha=1.0-smallpert.dat};

      \end{axis}
    \end{tikzpicture}
  }
  \subfigure[]{
    \begin{tikzpicture}
      \begin{axis}
        [ 
        ymin = -1.1,
        ymax = 1.1,
        xmin = -1.1,
        xmax = 1.1,
        ylabel={$x_o(t)$},
        xlabel={$y_o(t)$},
        tick scale binop=\times,
        width=0.35\textwidth,
        height=0.35\textwidth,
        ]
        \addplot+[no marks, thick, color=green!60!black,dashed] table[x
        expr=\thisrowno{0}, y expr=\thisrowno{1}, col sep=space] {Figures/xx-alpha=1.0-smallpert-delta.dat};
        \addplot+[no marks, very thick, color=red] table[x
        expr=\thisrowno{0}, y expr=\thisrowno{1}, col sep=space] {Figures/xx-alpha=1.0-smallpert-interp.dat};
        \addplot+[mark=*, black, mark options={fill=black,draw=none}, draw=none, only marks, mark size = 0.3pt] table[x expr=\thisrowno{0}, y expr=\thisrowno{1}, col sep=space] {Figures/xx-alpha=1.0-smallpert.dat};

        \addplot [samples=100, domain=0:2*pi,dotted] ( {cos(deg(x))}, {sin(deg(x))} );
      \end{axis}
    \end{tikzpicture}
  }
\caption{Migration from center point $\bx_o(t) = (0,0)$, $r_o(0) = 0$ outward for $\alpha = -1.0$:  (a) radial distance in log (red) and linear (blue) scales, and (b) the precession over this same time period.  The dotted circle in (b) indicates the radius of contact.  The dashed green line show a case that is constrained with standoff at $\delta \ge 0.05$.  The black dots along the curve in (b) are equally spaced in time, with $\Delta t = 25$.  See also an animated visualization in movie~2.}\label{fig:migrate}
\end{center}

\end{figure}
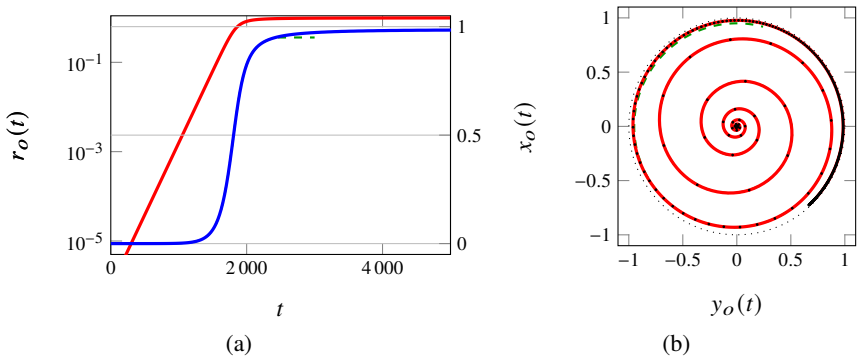

The rotating axisymmetric low-$|\alpha|$ case visualized in figure~\ref{fig:lowalphainitial} is not itself stable, although the subsequent geometric instability develops much slower than that of the suspension itself.  Figure~\ref{fig:migrate} shows the long-time evolution for the $\alpha = -1$ case.  The nearly axisymmetric flow is established with its peek $|\Omega|$ by $t \approx 300$, but the circle then migrates from $r_o \approx 0$ to $0.986$ ($\delta = 0.014$ from contact) for $t \gtrsim 5000$, as seen in figure~\ref{fig:migrate} (a).  
Any small perturbation leads to the same expanding spiral seen in figure~\ref{fig:migrate} (b).  The rate of migration and precession is first slow, then more rapid before it slows again near the container wall.  The migration, and indeed all flow, stops for $t > 5000$, when the wall separation distance is $\delta = 0.014$.   Details of the approach to the container wall and the cessation of flow are discussed in section~\ref{ss:approach}, after the mechanism of the migration is considered here.

\begin{figure}
  \begin{center}  \subfigure[$t=1000$]{\includegraphics[width=0.32\textwidth]{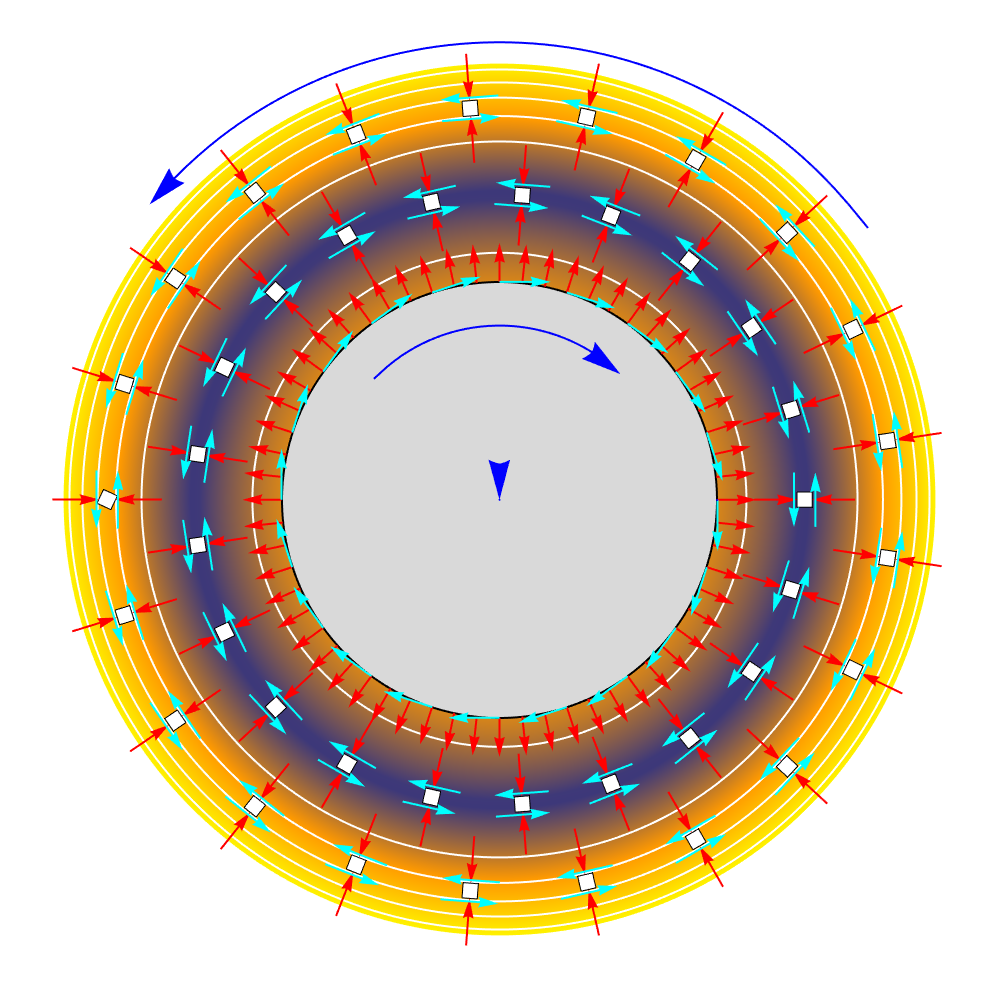}}
\subfigure[$t=1500$]{\includegraphics[width=0.32\textwidth]{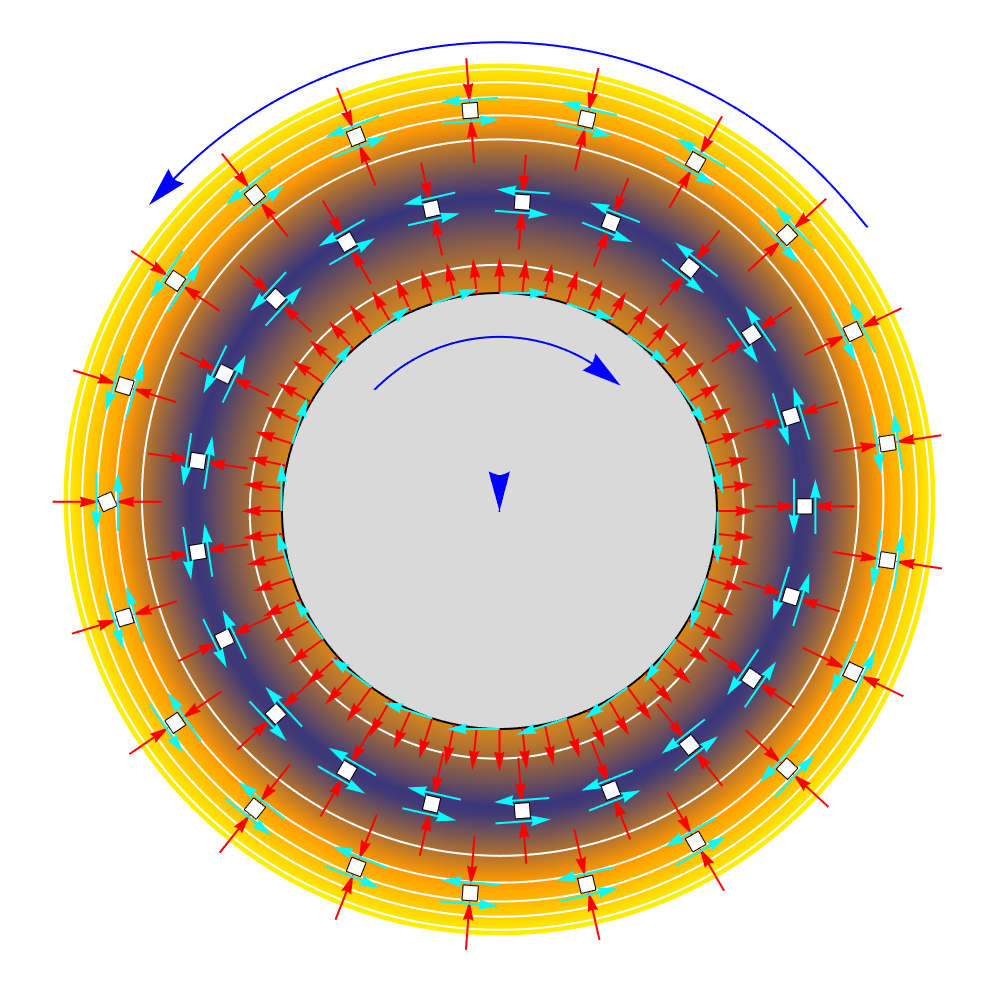}}    \subfigure[$t=1700$]{\includegraphics[width=0.32\textwidth]{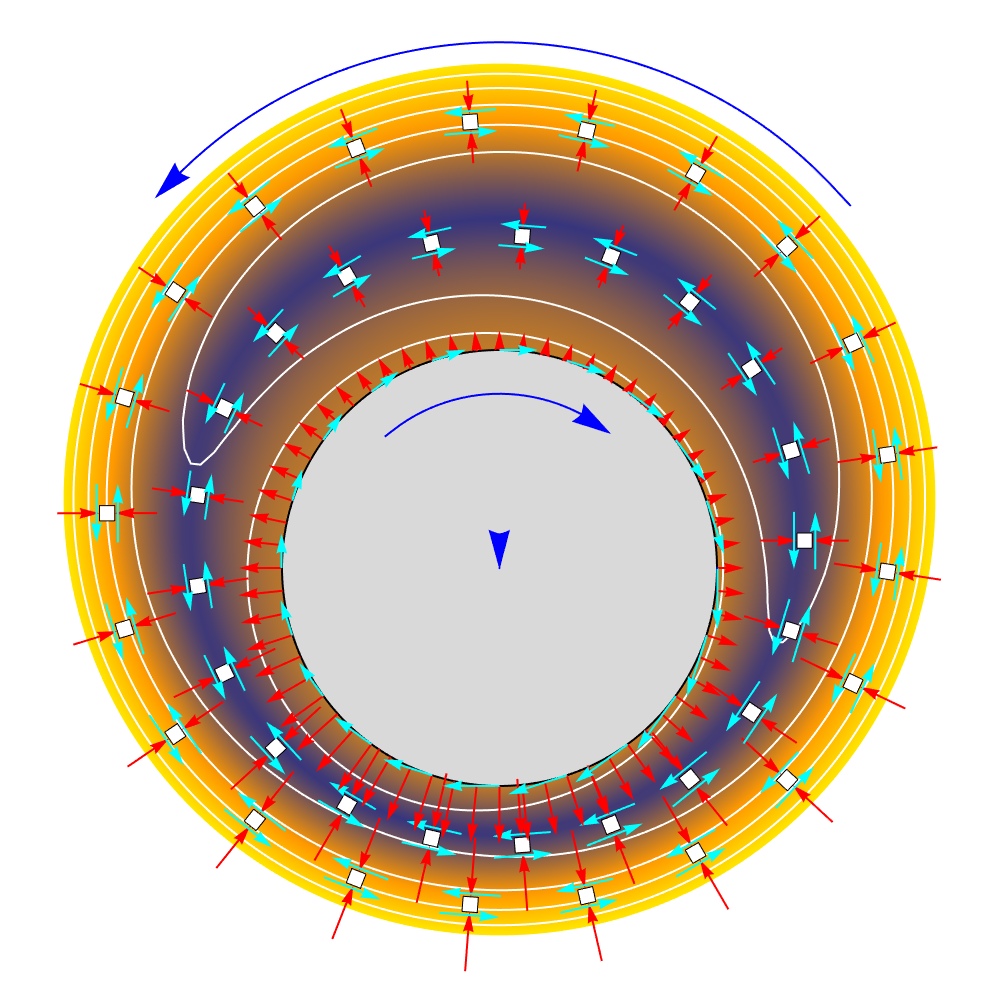}} \subfigure[$t=2000$]{\includegraphics[width=0.32\textwidth]{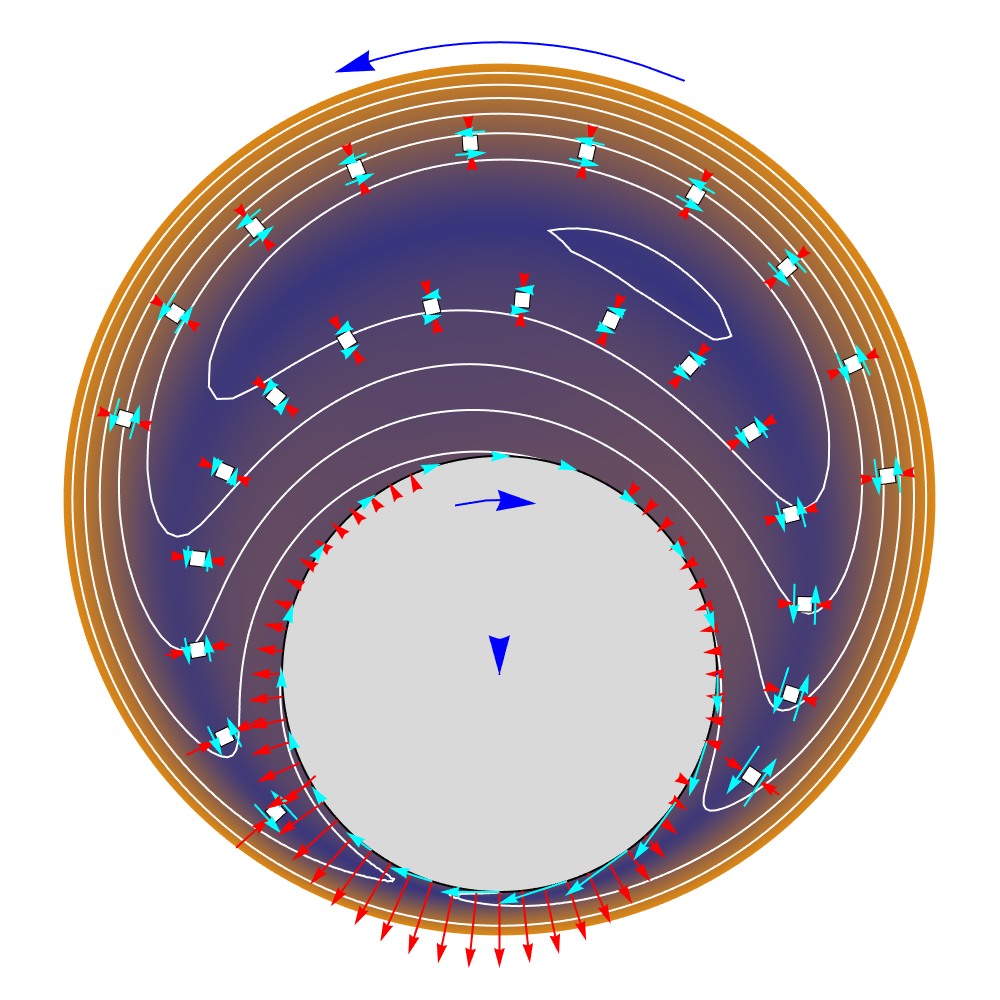}} \subfigure[$t=2200$]{\includegraphics[width=0.32\textwidth]{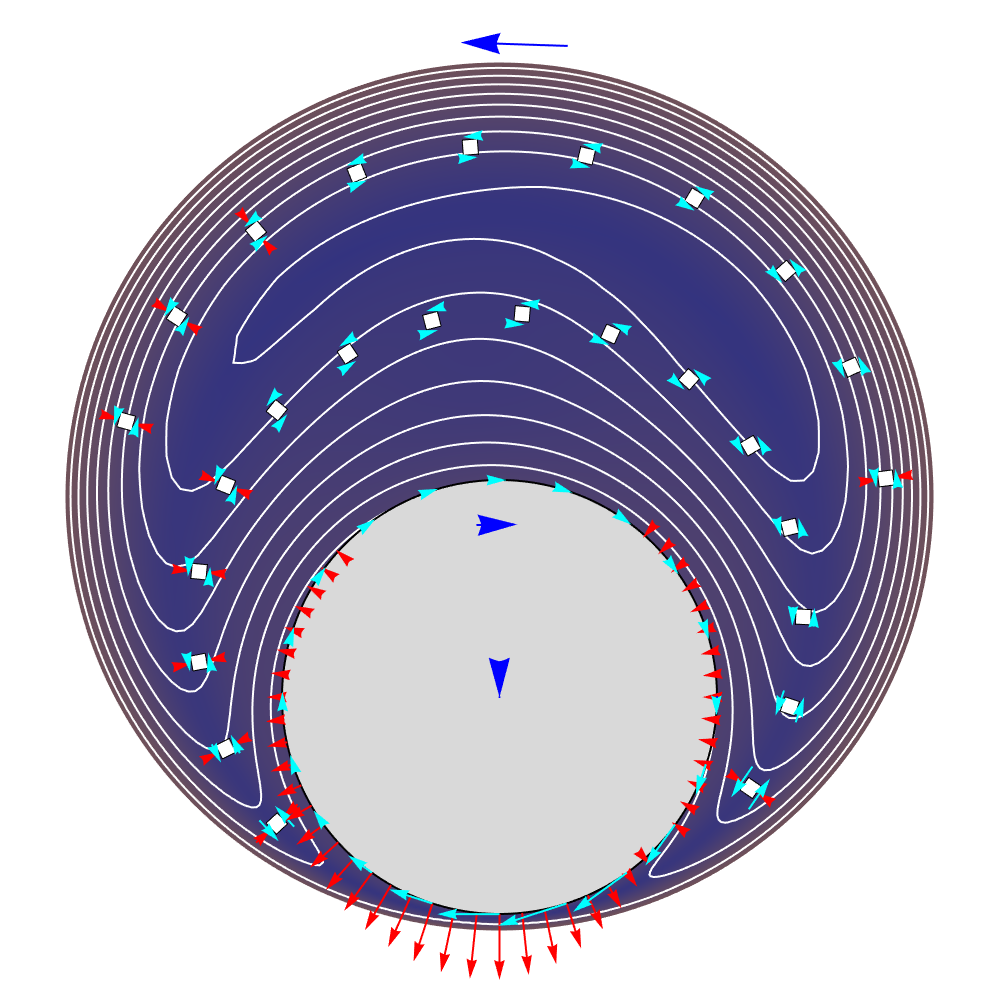}}
\subfigure[$t=7000$]{\includegraphics[width=0.32\textwidth]{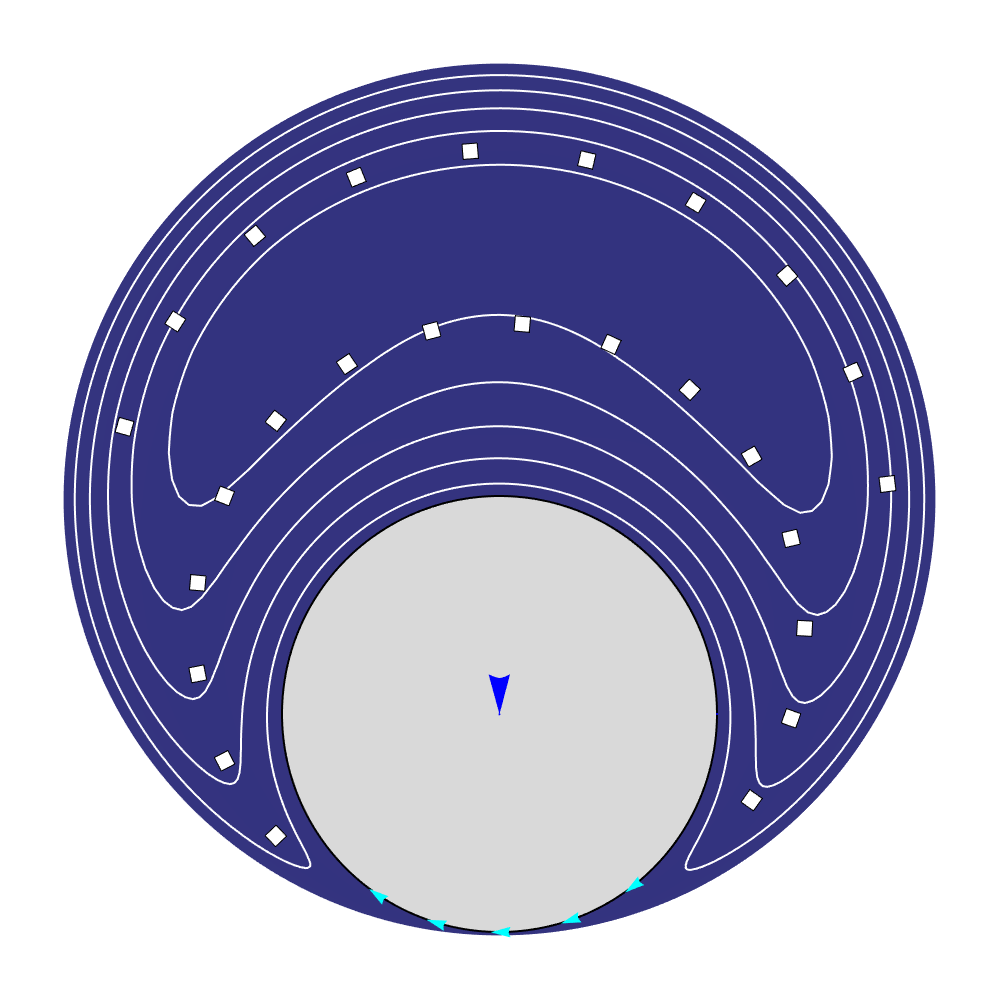}}

\caption{(a--f) Migration toward the wall for $\alpha = -1$ for the times as labeled.  The reference frame is rotating about $\bx = (0,0)$ with the precession rate, with blue arrows visualizing the velocity of the immersed circle and the container in this frame.  The red arrows show the normal component $D_{nn}$ of $\bD$ directed toward the immersed object and outer wall, with its specific angle linearly interpolated between the closest points on each circle.   The cyan arrows visualize the shear component $D_{nt}$ in this same orientation.  The white contours show streamfunction in this rotating frame and the same color levels visualize the time-decreasing velocity magnitude $|\bu|$ in this same frame.  }\label{fig:migrateviz}
\end{center}
\end{figure}

Figures~\ref{fig:migrateviz} (a)--(f) show how a small displacement from symmetric leads to higher strain rates in the fluid in the now narrower region between the object and the closer container wall, which in turn strengthens $\bD$, both its wall normal $D_{nn}$ and tangential $D_{nt}$ components.  As in a planar Couette configuration, there is an active shear stress component, which is sympathetic with the shear strain rate, and a tensile normal stress~\citep{Saintillan:2018}.  The stronger $D_{nt}$ stress in the narrow side versus the wide side drives the precession, while the corresponding imbalanced $D_{nn}$ across the object pulls it toward the nearer container wall.  Together, these yield the spiral pattern of figure~\ref{fig:migrate} (b).  The mismatched $D_{nt}$ also drives a rotation of the circle that is counter to a rolling motion, which in conjunction with the precession also increases the strain rate in the narrow side relative to the wider side.   Exponential growth of the displacement persists until $t\approx 1700$ when $r_o \approx 0.25R$ (figure~\ref{fig:migrate} a), which corresponds approximately to the point at which the streamlines in the frame of the pressessing circle show a distinct recirculation in the larger space (figure~\ref{fig:migrateviz} c).  In the final frame, figure~\ref{fig:migrateviz} (f), $r_o=0.985$ ($\delta = 0.015$) the flow is nearly stopped.  All flow stops at $\delta = 0.014$.

\subsection{Near-wall behavior}
\label{ss:approach}

\begin{figure}
  \begin{center}
    \begin{tikzpicture}
      \begin{axis}
        [ 
        ymin = -0.2,
        ymax = 0.25,
        xmin = -3.2,
        xmax = 3.2,
        ylabel={$D_{nn}'$, $D_{nt}'$, $D_{tt}'$},
        xlabel={$\frac{R}{a}\beta$, $\gamma$},
        tick scale binop=\times,
        width=0.65\textwidth,
        height=0.4\textwidth,
        grid=major,
        legend style={	at={(axis cs:-3.15,.19)},anchor=west,legend
        columns=4,draw=none,legend cell align=left},
         xtick={-3.14159,-2.35619,-1.5708,-0.785398,0.,0.785398,1.5
   708,2.35619,3.14159},
    xticklabels={$-\pi$ ,$-\frac{3 \pi }{4}$,$-\frac{\pi
   }{2}$,$-\frac{\pi }{4}$,$0$,$\frac{\pi }{4}$,$\frac{\pi
   }{2}$,$\frac{3 \pi }{4}$,$\pi$ }
      ]
      \addlegendimage{empty legend}
        \addplot+[no marks, very thick, color=red] table[x
        expr=\thisrowno{0}*2, y expr=\thisrowno{1}, col sep=space] {Figures/surfaceD-a=1.0-lub.dat};
        \addplot+[no marks, very thick, color=cyan] table[x
        expr=\thisrowno{0}*2, y expr=\thisrowno{2}, col sep=space] {Figures/surfaceD-a=1.0-lub.dat};
        \addplot+[no marks, very thick, color=green!60!black] table[x
        expr=\thisrowno{0}*2, y expr=\thisrowno{3}, col sep=space] {Figures/surfaceD-a=1.0-lub.dat};
              \addlegendimage{empty legend}
        \addplot+[no marks, very thick, color=red, dashed] table[x
        expr=\thisrowno{0}, y expr=\thisrowno{4}, col sep=space] {Figures/surfaceD-a=1.0-lub.dat};
        \addplot+[no marks, very thick, color=cyan, dashed] table[x
        expr=\thisrowno{0}, y expr=\thisrowno{5}, col sep=space] {Figures/surfaceD-a=1.0-lub.dat};
        \addplot+[no marks, very thick, color=green!60!black, dashed] table[x
        expr=\thisrowno{0}, y expr=\thisrowno{6}, col sep=space] {Figures/surfaceD-a=1.0-lub.dat};
        \legend{
          {[text width=43pt,text depth=]Container:\; },
          {{\color{red} $D_{nn}'$}}\;\;,
          {{\color{cyan} $D_{nt}'=D_{tn}'$}}\;\;,
          {{\color{green!40!black} $D_{tt}'$}},
          {[text width=43pt,text depth=]\hspace*{0.1in}Object:\;},
          {{\color{red} $D_{nn}'$}}\;\;,
          {{\color{cyan} $D_{nt}'=D_{tn}'$}}\;\;,
          {{\color{green!40!black} $D_{tt}'$}}}
      \end{axis}
    \end{tikzpicture} \raisebox{0.5in}{\includegraphics[width=0.27\textwidth]{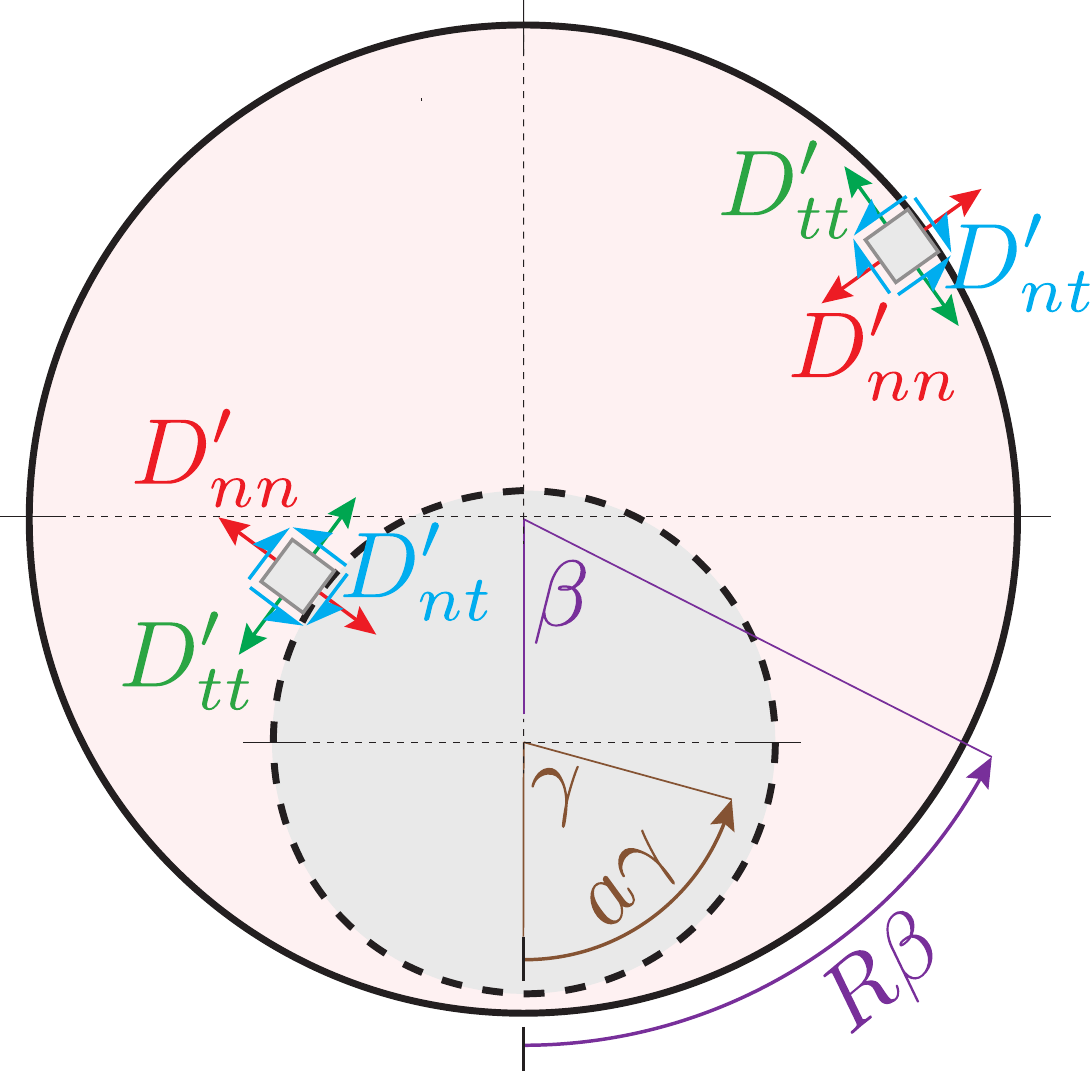}}
\caption{Deviatoric active stress components $\alpha\bD' = \alpha(\bD - \bI/2)$ of the active suspension along the container and object surfaces as indicated for $\alpha = -1.0$.  They are rotated to into local normal and tangential coordinates $(n,t)$. The circle's rotation and precession are both clockwise, so negative $\beta$ and $\gamma$ are ahead of its motion. }\label{fig:lubD}
\end{center}
\end{figure}
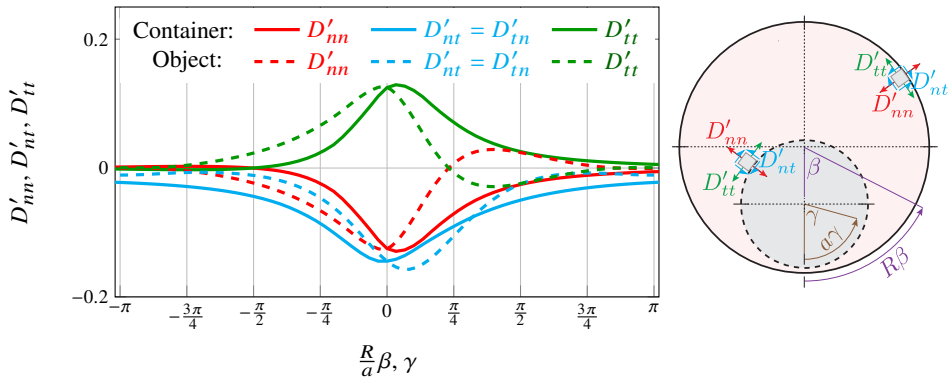

As it approaches, the object's motion toward the container wall is increasingly opposed by the usual large normal lubrication resistance.  However, active shear stress in the narrowing gap continues to cause counter-rolling rotation and continued precession along the container wall as it slowly approaches.  The deviatoric components of $\alpha\bD$ on the sufaces in this narrow lubrication-like gap are plotted in figure~\ref{fig:lubD}, rotated into local surface coordinates.  In the narrowest region, the wall-normal components are nearly the same across the gap, consistent with the lubrication limit.  They hold the object near the wall and for $\alpha = -1$ slowly pull it closer.  When a $\delta_c = 0.05$ standoff constraint is imposed, they hold it against this constraint as it precesses.  Upstream and downstream of the narrowest gap symmetry is broken with wall-normal stresses there significantly different. By $\gamma = \pm \pi/4$ they are compressive behind the object and tensile ahead of it, promoting the precession and shear strain rate in the smallest gap.  

\begin{figure}
  \begin{center}
  \subfigure[]{\includegraphics[width=0.35\textwidth]{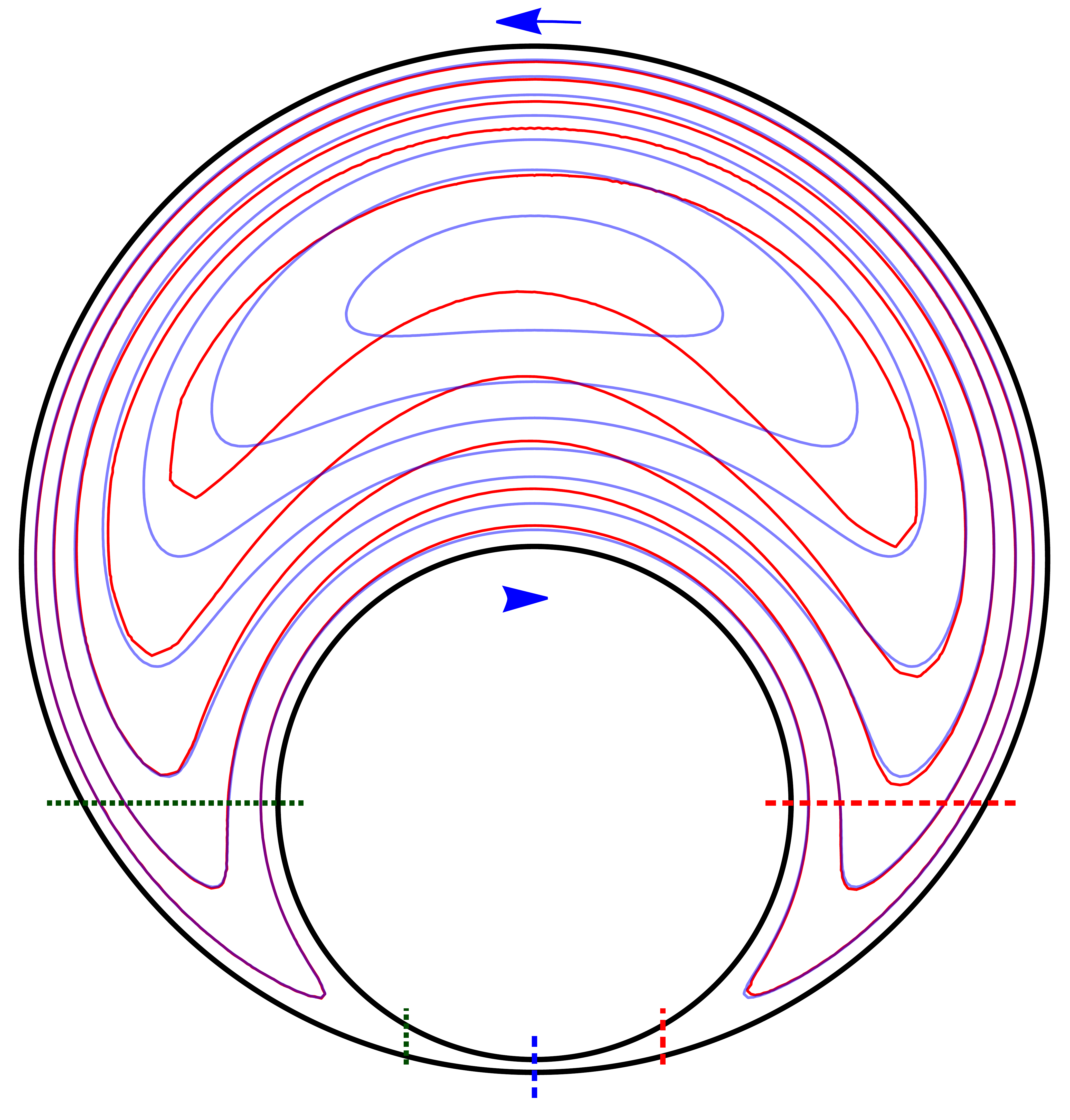}}
  \subfigure[]{
    \begin{tikzpicture}
      \begin{axis}
        [ 
        ymin = -0.006,
        ymax =  0.0085,
        xmin = 1,
        xmax = 1.8,
        ylabel={$|v(x,y_o)|$},
        xlabel={$|x|$},
        tick scale binop=\times,
        width=0.32\textwidth,
        height=0.35\textwidth,
        legend style={	at={(axis cs:-3,.19)},anchor=west,legend
         columns=4,draw=none,legend cell align=left},
      ]
        \addplot+[no marks, very thick, dashed, color=red] table[x
        expr=\thisrowno{0}, y expr=\thisrowno{1}, col sep=space] {Figures/Vline-a=1.0.dat};
        \addplot+[no marks, very thick, dotted, color=green!60!black] table[x
        expr=\thisrowno{0}, y expr=\thisrowno{3}, col sep=space] {Figures/Vline-a=1.0.dat};
        \addplot+[no marks, color=black] table[x
        expr=\thisrowno{0}, y expr=\thisrowno{4}, col sep=space] {Figures/Vline-a=1.0.dat};
      \end{axis}
    \end{tikzpicture}
  }
  \subfigure[]{
    \begin{tikzpicture}
      \begin{axis}
        [ 
        xmin = -0.005,
        xmax =  0.01,
        ymin = -1.95, 
        ymax = -1.81,
        axis y line*=left,
        xlabel={$u(x_o,y)$},
        ylabel={$y$},
        tick scale binop=\times,
        width=0.32\textwidth,
        height=0.35\textwidth,
        ]
        \addplot+[no marks, very thick, dashed, color=red] table[x
        expr=\thisrowno{1}, y expr=\thisrowno{0}, col sep=space] {Figures/Uline-a=1.0-xo=+0.5.dat};
        \addplot+[no marks, very thick, dotted, color=green!60!black] table[x
        expr=\thisrowno{1}, y expr=\thisrowno{0}, col sep=space] {Figures/Uline-a=1.0-xo=-0.5.dat};
        \addplot+[no marks, color=black] table[x
        expr=\thisrowno{2}, y expr=\thisrowno{0}, col sep=space] {Figures/Uline-a=1.0-xo=-0.5.dat};
      \end{axis}
      \begin{axis}[
        xmin = -0.005,
        xmax =  0.01,
        ymin = -2.01, 
        ymax = -1.95,
        yticklabel pos=right,
        axis y line*=right,
        axis x line=none,
        tick scale binop=\times,
        width=0.32\textwidth,
        height=0.35\textwidth,
        ]
        \addplot+[no marks, very thick, dashed, color=blue] table[x expr=\thisrowno{1}, y expr=\thisrowno{0}, col sep=space] {Figures/Uline-a=1.0-xo=0.0.dat};
        \addplot+[no marks, color=black] table[x expr=\thisrowno{2}, y expr=\thisrowno{0}, col sep=space] {Figures/Uline-a=1.0-xo=0.0.dat};
      \end{axis}
    \end{tikzpicture}
  }
  \end{center}
  \caption{ (a)  Streamfunction contours space by $\Delta \psi = 0.0005$ for $\alpha = -1$ comparing the active suspension (red) with the exact Newtonian fluid Stokes-flow solution (blue) for the same boundary velocities in a frame that tracks the object precession at fixed standoff $\delta_c = 0.05$.  The undetermined constant in $\psi$ is adjusted so that the contours algn in the lower left region of highest curvature.  The straight lines in (a) indicate where velocity profiles are compared with Stokes flow in (b) and (c), with the unbroken black curves shown the corresponding Stokes flow solution. }\label{fig:velNewt}
\end{figure}
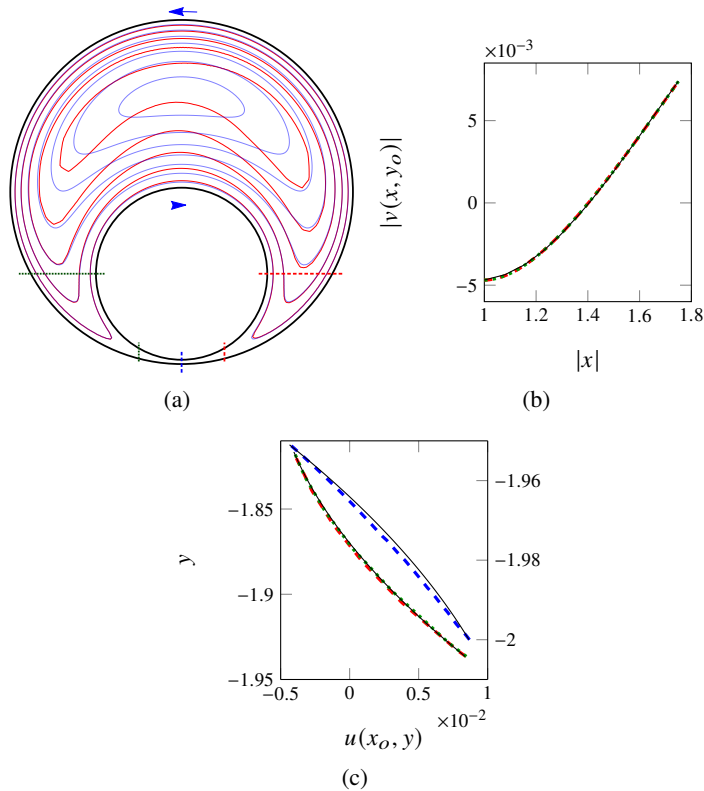

One notable feature of figure~\ref{fig:lubD} is that the active shear stresses are most significant in the lubrication layer near the wall of the container, activated there by the relatively high local strain rate. Two thirds (65.6 percent) of the net active-stress torque on the cylinder is on the lowest quarter of the circle ($|\gamma| \le \pi/4$), and 89 percent is on its lower half ($|\gamma| < \pi/2$).  Yet for such weak activity, even in these same regions, the flow remains nearly identical to a constant viscosity Newtonian fluid driven by the motion of the circle.  In the frame that fixes the precession angle, the streamfunction $\psi$  flow pattern (with $\bu = [\psi_y,-\psi_x]^T$) is compared with the exact Stokes flow solution~\citep{Wannier:1950} in figure~\ref{fig:velNewt} (a).  There is only a slight difference in the velocity profile in the region of highest strain rate (figure~\ref{fig:velNewt} c) and almost no difference at $|\gamma| = \pm \pi/2$ (figure~\ref{fig:velNewt} b).  This can be anticipated:  the flow is nearly Stokesian for low activity ($|\alpha|\to 0$), the mobility of the circle is required to (barely) maintain the flow, and the minimum dissipation flow must be a Stokes flow.  However, the activity, viewed as components of $\bD$, is not nearly so symmetric as the velocity field.  (The margination processess itself also obviously breaks the symmetry of a Stokes flow, although that is slower still than the velocities associated with precession and rotation.) 

Despite the flow pattern nearly matching that of a Newtonian fluid, applying lubrication theory is hindered by a lack of specificity in boundary conditions at either end of the nominal gap.  For example, figure~\ref{fig:lubD} shows that $D_{nt}$ changes continuously and significantly over at least $-\pi/2 < \gamma < \pi/2$, well beyond where the lubrication limit can be expected to be accurate.  However, the overall behavior does suggest that lubrication theory might afford a more useful description for still smaller gaps or other geometries with more extensive narrow regions than this circle-in-circle configuration.  Of course, if contact is truly close, the very character of the suspension must also be questioned, which is revisited in section~\ref{s:conclusions}.  For these reasons, we defer any further analysis contact.

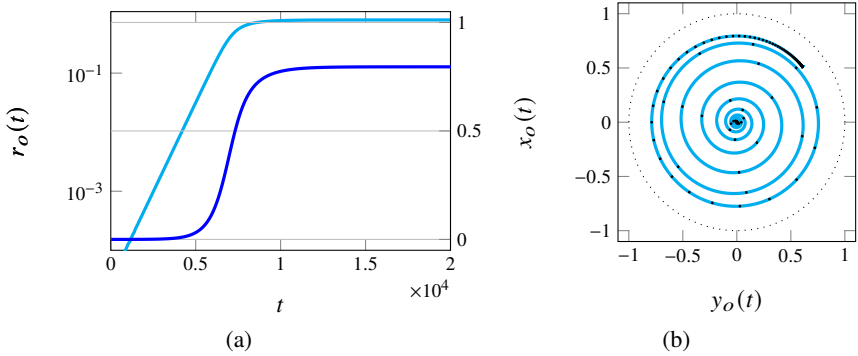
\begin{figure}
\begin{center}
  \subfigure[]{
    \begin{tikzpicture}
      \begin{semilogyaxis}
        [ 
        ymin = 1.e-4,
        ymax = 1.1e0,
        xmin = 0,
        xmax = 20000,
        ylabel={$r_o(t)$},
        xlabel={$t$},
        tick scale binop=\times,
        ytick pos=left,
        width=0.45\textwidth,
        height=0.35\textwidth,        
        xminorticks=true,
        ]
        \addplot+[no marks, very thick, color=cyan] table[x
        expr=\thisrowno{0}, y expr=\thisrowno{1}, col sep=space] {Figures/r-alpha=0.625-fine.dat};
      \end{semilogyaxis}
      \begin{axis}
        [ 
        ymin = -0.05,
        ymax = 1.05,
        xmin = 0,
        xmax = 20000,
        ylabel={$r_o(t)$},
        xlabel={$t$},
        axis y line*=right,
        axis x line = none,
        tick scale binop=\times,
        width=0.45\textwidth,
        height=0.35\textwidth,
        grid=major,
        max space between ticks=25,
        ]
        \addplot+[no marks, very thick, color=blue] table[x
        expr=\thisrowno{0}, y expr=\thisrowno{1}, col sep=space] {Figures/r-alpha=0.625-fine.dat};
      \end{axis}
    \end{tikzpicture}
  }
  \subfigure[]{
    \begin{tikzpicture}
      \begin{axis}
        [ 
        ymin = -1.1,
        ymax = 1.1,
        xmin = -1.1,
        xmax = 1.1,
        ylabel={$x_o(t)$},
        xlabel={$y_o(t)$},
        tick scale binop=\times,
        width=0.35\textwidth,
        height=0.35\textwidth,
        ]
        \addplot+[no marks, very thick, color=cyan] table[x
        expr=\thisrowno{0}, y expr=\thisrowno{1}, col sep=space] {Figures/xx-alpha=0.625-fine.dat};
        \addplot+[mark=*, black, mark options={fill=black,draw=none},
        draw=none, only marks, mark size = 0.3pt] table[x
        expr=\thisrowno{0}, y expr=\thisrowno{1}, col sep=space]
        {Figures/xx-alpha=0.625-coarse.dat};
        \addplot [samples=100, domain=0:2*pi,dotted] ( {cos(deg(x))}, {sin(deg(x))} );
      \end{axis}
    \end{tikzpicture}
  }
\caption{Migration from center point $\bx_o(0) = (0,0)$, $r_o(0) = 0$ outward for $\alpha = -0.625$:  (a) radial distance in log (light blue) and linear (blue) scales, and (b) the trajectory over this same time period.  The dotted circle in (b) indicates the radius of contact.  The black dots on the trajectory are equally spaced in time, with $\Delta t = 200$.  See also animated visualization in suplemental movie~1.}\label{fig:migrate0625}
\end{center}

\end{figure}

For the small $|\alpha|$ cases considered thus far, immobilizing the object at any $r_o$ stabilizes the suspension, which confirms that the entire flow is intimately linked to the mobility of the object.   We can estimate the viscous resistance that the object must overcome to remain rotating (and hence slowly marginating) based on the Newtonian fluid limit.  The required torque to maintain rotation without suspension activity ($\alpha=0$) increases rapidly near the point of contact.  Assuming that the angular precession rate equals the angular rotation rate of the object, it increases by only a factor of 1.14 from $r_o=0$ to $0.8$, but doubles by $r_o = 0.956$ and quadruples by $r_o = 0.990$.  
It follows that weaker activity, or other external resistance on the
object, should lead to arrest at smaller $r_o$.  For $\alpha =
-0.75$ the flow indeed ceases at $r_o = 0.949$ ($\delta = 0.051$), rather than the $r_o = 0.986$ ($\delta = 0.014$) for $\alpha = -1$.  For $\alpha = -0.625$, which is near the limit of any fluid instability for this configuration, it stops at $r_o = 0.80$, after following a similar spiral trajectory (figure~\ref{fig:migrate0625}).  In this case, the migration rate is smaller relative to the precession rate, leading to a more tightly spaced spiral than for $\alpha = -1$ (figure~\ref{fig:migrate}).   When a rotational resistance torque of $T_r = -8\Omega$ is added for $\alpha = -1$, as discussed in section~\ref{s:susstab}, the additional dissipation arrests the circle further from the container wall at $\delta= 0.020$.

The dependence of suspension instability on the mobility of the object introduces the question of what changes for activity that is self-sustaining independently of the mobility of the circle.  This is the case for $\alpha \le -10$ in the following section~\ref{s:highactivity}.

\section{Strong activity ($-80 \le \alpha \le -10$)}
\label{s:highactivity}

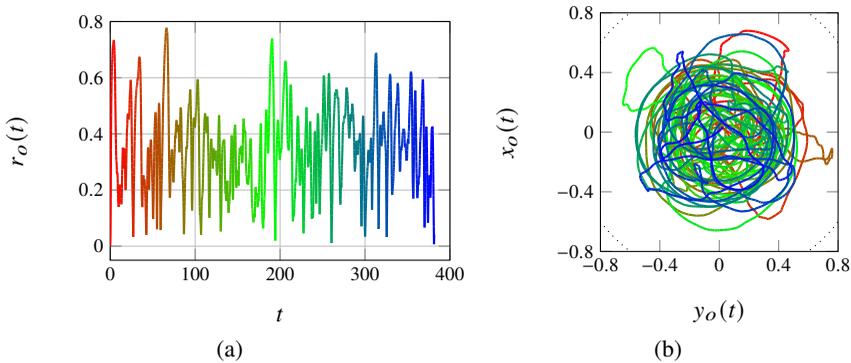
\begin{figure}
\begin{center}
  \subfigure[]{
    \begin{tikzpicture}
      \begin{axis}
        [ 
        ymin = -0.05,
        ymax = 0.8,
        xmin = 0,
        xmax = 400,
        ylabel={$r_o(t)$},
        xlabel={$t$},
        tick scale binop=\times,
        width=0.45\textwidth,
        height=0.35\textwidth,
        grid=major,
        max space between ticks=25,
        ]
        \addplot+[no marks, thick, line join=round, , mesh, point meta=\thisrowno{0}, colormap/bluered,
        colormap={}{ 
           [1cm] color(0cm)=(red) color(1cm)=(green) color(2cm)=(blue)
        }] table[x
        expr=\thisrowno{0}, y expr=\thisrowno{1}, col sep=space] {Figures/r-alpha=20p0.dat};
      \end{axis}
    \end{tikzpicture}
  }
  \subfigure[]{
    \begin{tikzpicture}
      \begin{axis}
        [ 
        ymin = -0.8,
        ymax = 0.8,
        xmin = -0.8,
        xmax = 0.8,
        ylabel={$x_o(t)$},
        xlabel={$y_o(t)$},
        tick scale binop=\times,
        width=0.35\textwidth,
        height=0.35\textwidth,
        xtick={-0.8,-0.4,0.0,0.4,0.8},
        ytick={-0.8,-0.4,0.0,0.4,0.8},
        ]
        \addplot+[no marks, thick, mesh, point meta=\thisrowno{2}, colormap/bluered,
        colormap={}{ 
           [1cm] color(0cm)=(red) color(1cm)=(green) color(2cm)=(blue)
        }] table[x
        expr=\thisrowno{0}, y expr=\thisrowno{1}, col sep=space] {Figures/xx.dat};

        \addplot [samples=100, domain=0:2*pi,dotted] ( {cos(deg(x))}, {sin(deg(x))} );
      \end{axis}
    \end{tikzpicture}
  }
\caption{Trajectory from  $\bx_o(0) = (0,0)$, $r_o(0) = 0$
  for $\alpha = -20$:  (a) radial distance from the container
  center, and (b) the trajectory over this same time
  period.  The visible portions of the dotted circle in (b) indicates
  the radius that would correspond to contact.  The same color pattern tracks evolution in
  time in both (a) and (b).}\label{fig:a20trajectory}
\end{center}
\end{figure}

\begin{figure}
  \begin{center}
    \subfigure[$\alpha = -10$ (and $-5$), $a = 1$]{
      \begin{tikzpicture}
        \node[anchor=south west, inner sep=0] (image) at (0,0)
        {\includegraphics[width=0.27\textwidth]{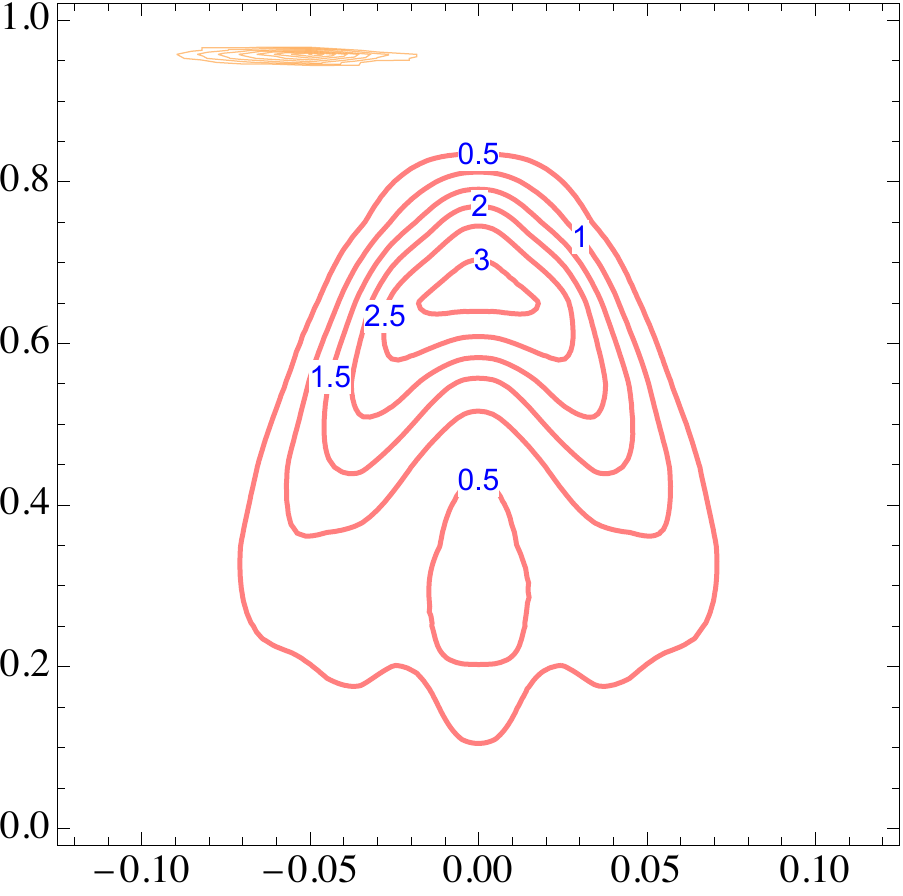}};
        \begin{scope}[x={(image.south east)},y={(image.north west)}]    
          \node at (0.5,-0.07) {$\Omega/\alpha$}; 
          \node[rotate=90] at (-0.025,0.5) {$r_o$};
          \node at (0.25, 0.88) {{\footnotesize \color{orange} $\alpha = -5$}};
          \draw[blue!40!white, thick] (0.85,0.85) circle [radius=0.105];
          \draw[blue,fill=blue!40!white,draw=none] (0.85,0.85) circle [radius=0.05];
        \end{scope}
      \end{tikzpicture}
}
    \subfigure[$\alpha =-20$, $a = 1$]{\axislabledfigureC{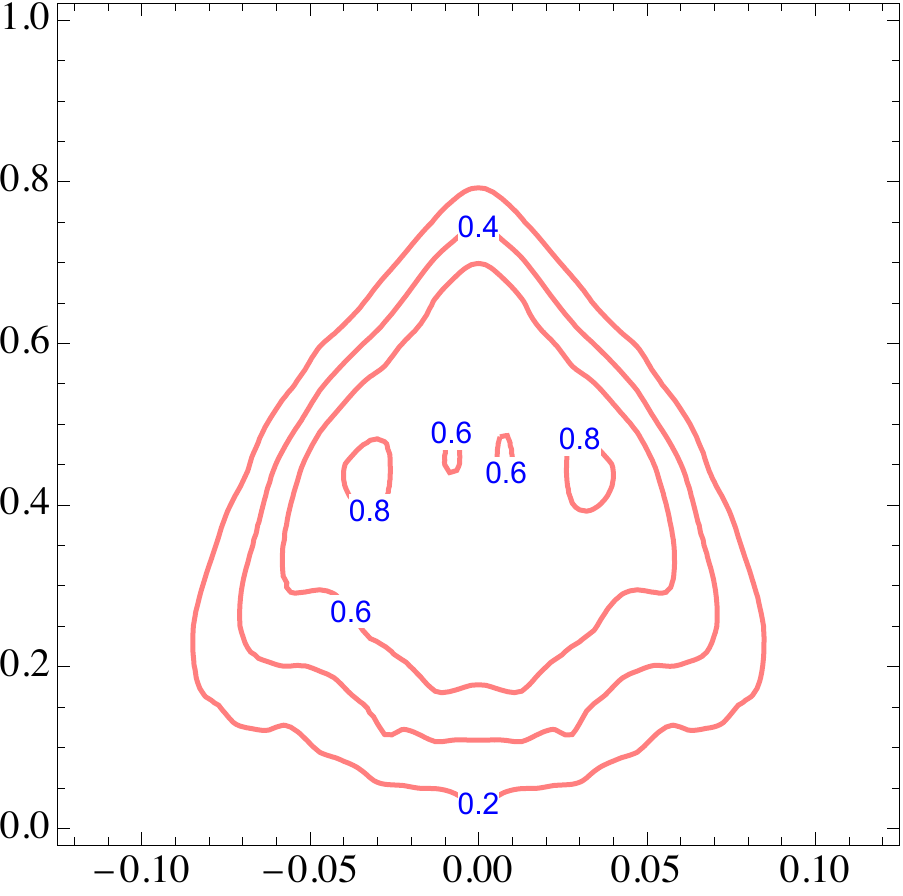}{$\Omega/\alpha$}{$r_o$}{0.27}{0.05}}
    \subfigure[$\alpha =-80$, $a = 1$]{\axislabledfigureC{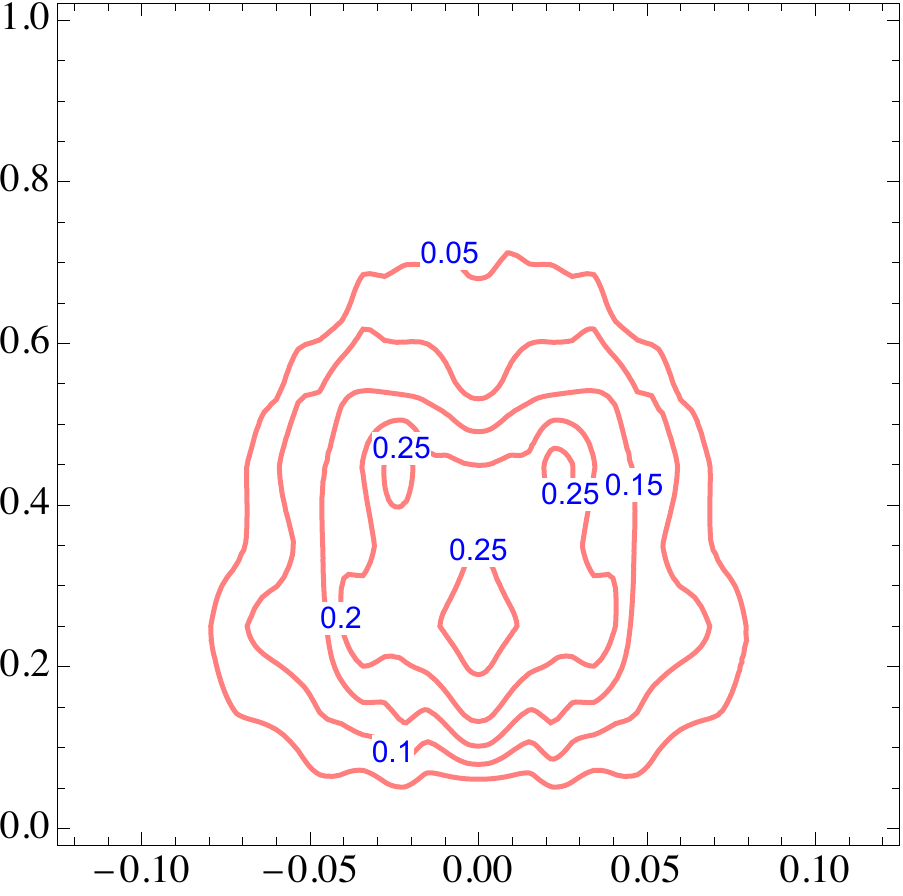}{$\Omega/\alpha$}{$r_o$}{0.27}{0.05}}
    \subfigure[$\alpha =-40$, $a = 0.5$]{\axislabledfigureC{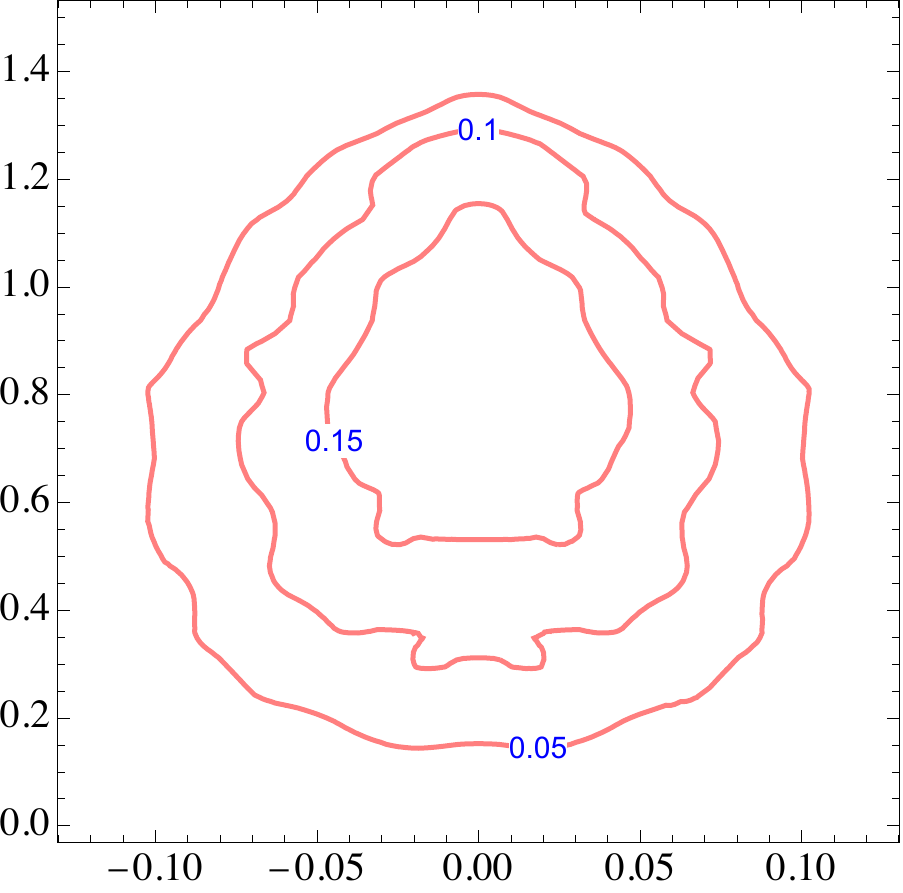}{$\Omega/\alpha$}{$r_o$}{0.27}{0.025}}
    \subfigure[$\alpha =-40$, $a = 1$]{\axislabledfigureC{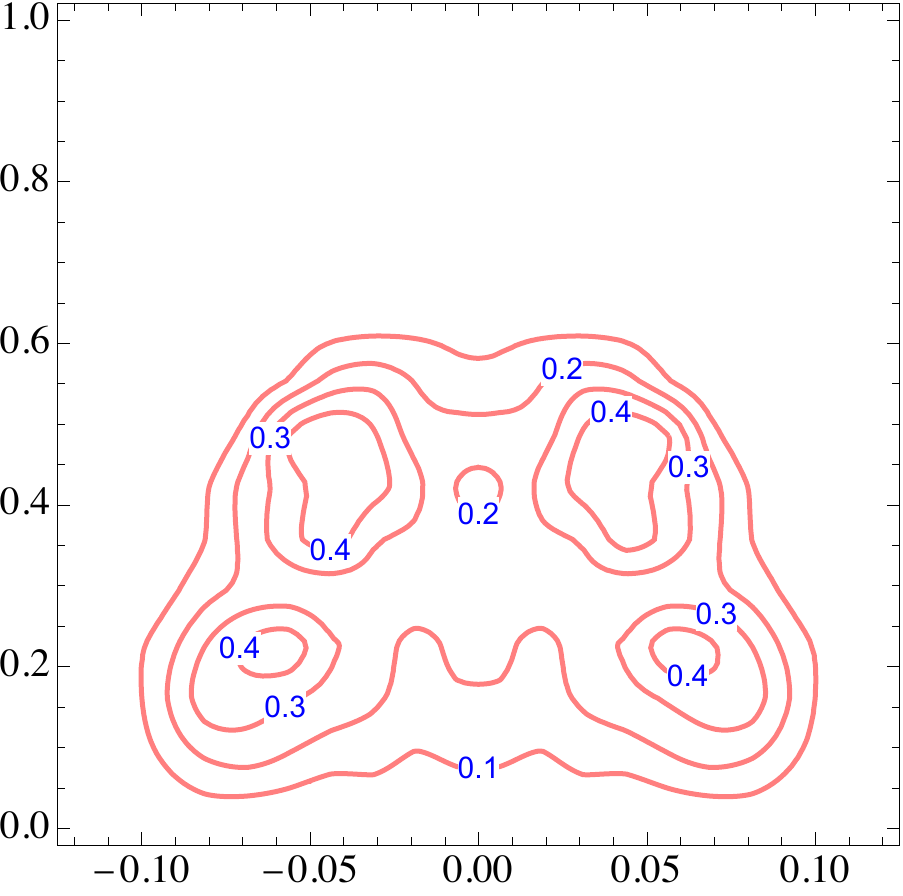}{$\Omega/\alpha$}{$r_o$}{0.27}{0.05}}
    \subfigure[$\alpha =-40$, $a = 1.5$]{\axislabledfigureC{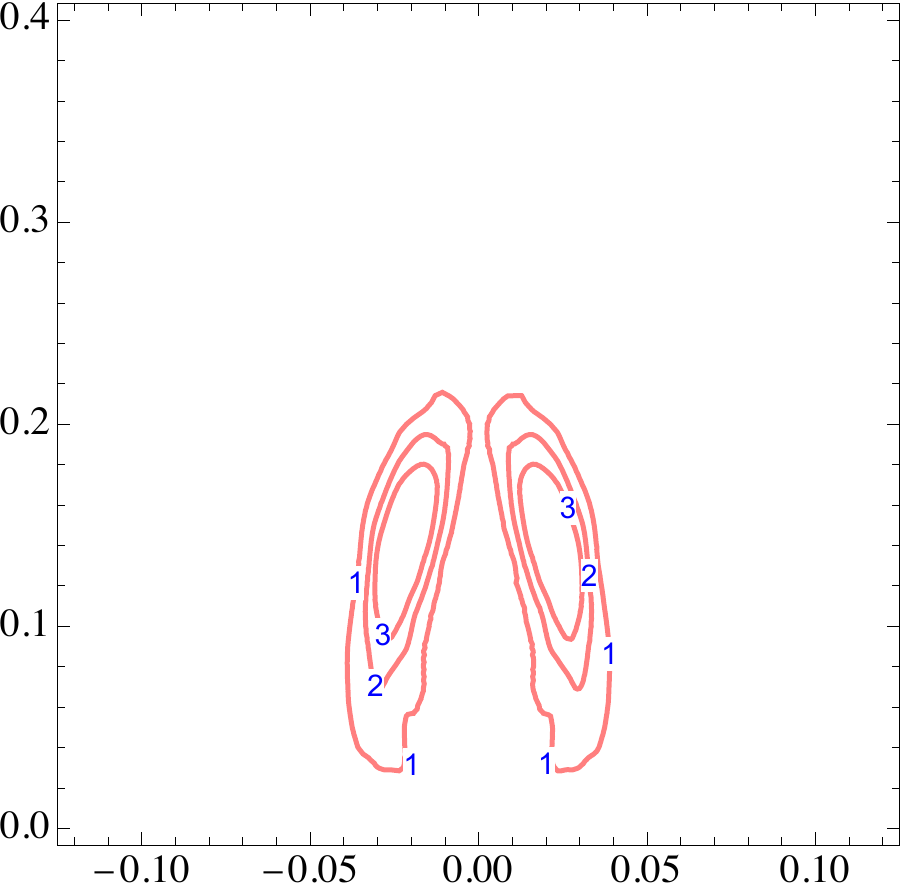}{$\Omega/\alpha$}{$r_o$}{0.27}{0.075}}

  \end{center}
  \caption{(a--f) Joint probability density function (p.d.f.) of radial position $r_o$ and angular rotation rate $\Omega$ for the cases as labeled.  In (a), also shown in orange is the corresponding narrow p.d.f.\ for the $\alpha=-5$ case, which follows a relatively deterministic path (see section~\ref{s:transition}).  Note the changing vertical scale for $r_o$ for the larger and smaller radius circle cases (d) and (f).  Aside from the $\alpha=-5$ inset in (a), all cases were observed to change rotation sense multiple times and were thus averaged for $\pm \Omega$ symmetry.  Animated visualizations of these cases are available in supplemental movies~4--10.}\label{fig:rOmegaPDF}
\end{figure}
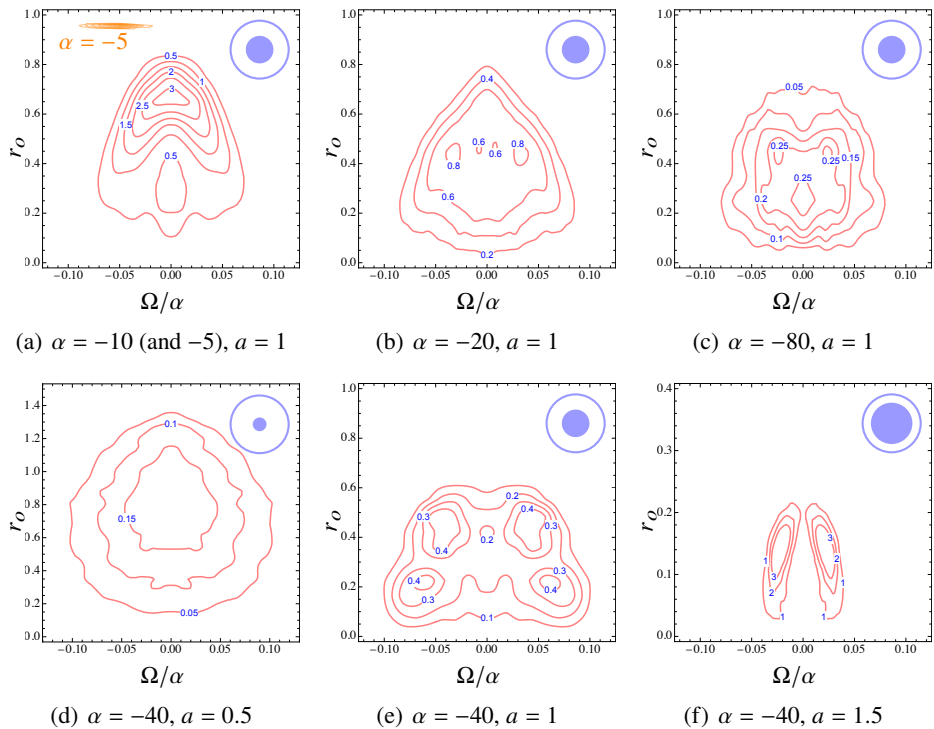

Figure~\ref{fig:a20trajectory} shows the apparently chaotic trajectory
of the same $a=1$ circle in a $R = 2$ container, now for $\alpha=-20$.  For the course of this
simulation, or any similar simulation for $\alpha \lesssim -10$, the object never
approaches closer than $\delta = 0.1$ to the wall, and rarely approaches even this close.  Figure~\ref{fig:rOmegaPDF} quantifies this along with the rotation rate of the circle with joint probability distributions. For $\alpha = -10$, it is most likely rotating slowly and at a relatively large $r_o \approx 0.7$ ($r_o = 1$ would be contact).  With stronger activity, the circle is increasingly likely to be closer to the center of the container at $r_o=0$, and it experiences a broader range of rotation rates $\Omega$ (figures~\ref{fig:rOmegaPDF} a--c and e).  The width of the $\Omega$ distribution scales approximately with $\alpha$.  None of the distributions are simple, neither in $r_o$ nor in $\Omega$, suggesting that the dynamics of this configuration, with its finite-sized object, are too complex to represent as a simple statistical process.  

The dependence on the radius of the free-floating circle is a striking example of geometry dependent statistics.  Figures~\ref{fig:rOmegaPDF} (d,e,f) compare the $r_o$--$\Omega$ joint probability distributions for $\alpha = -40$ and $a = 0.5$, $1$ and $1.5$.  The small $a=0.5$ circle is the closest to following a simple random-walk-like process, though only in some regions.  For $r_o \lesssim 0.8$, it has a broad flat distribution in $\Omega$ and an approximately linear increase with $r_o$, as would be commensurate with a random sampling in this geometry.  However, this simple trend ends for $r_o\gtrsim 0.8$, still well away from the $r_o=1.5$ radius of contact.  For radius $a = 1$, non-zero rotation is the most likely, and the object rarely leaves the $r_o < 0.6$ region of the container.  The $a=1.5$ cases is most striking.  The circle nearly always rotates, though slowly relative to the other case.  Its changes of direction, which are not uncommon, are sudden, contributing little to the  $\Omega \approx 0$ portion of the distributions.  Even this larger circle never approaches close to the container walls, rarely reaching beyond $r_o= 0.2$, which leaves a $\delta = 0.3$ gap before contact.

\begin{figure}
  \begin{center}
\subfigure[$t=456.5$]{\includegraphics[width=0.32\textwidth]{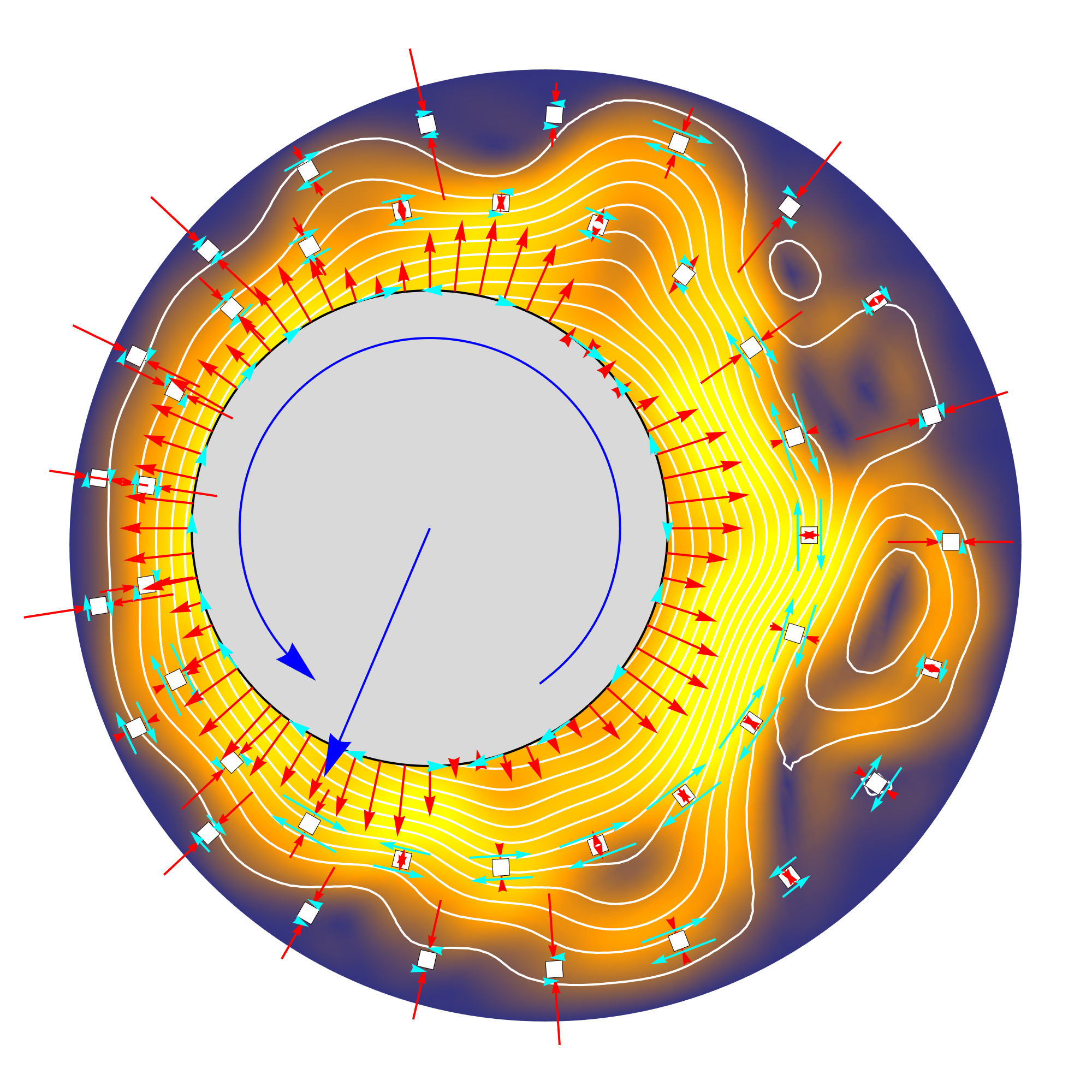}}
\subfigure[$t=457.5$]{\includegraphics[width=0.32\textwidth]{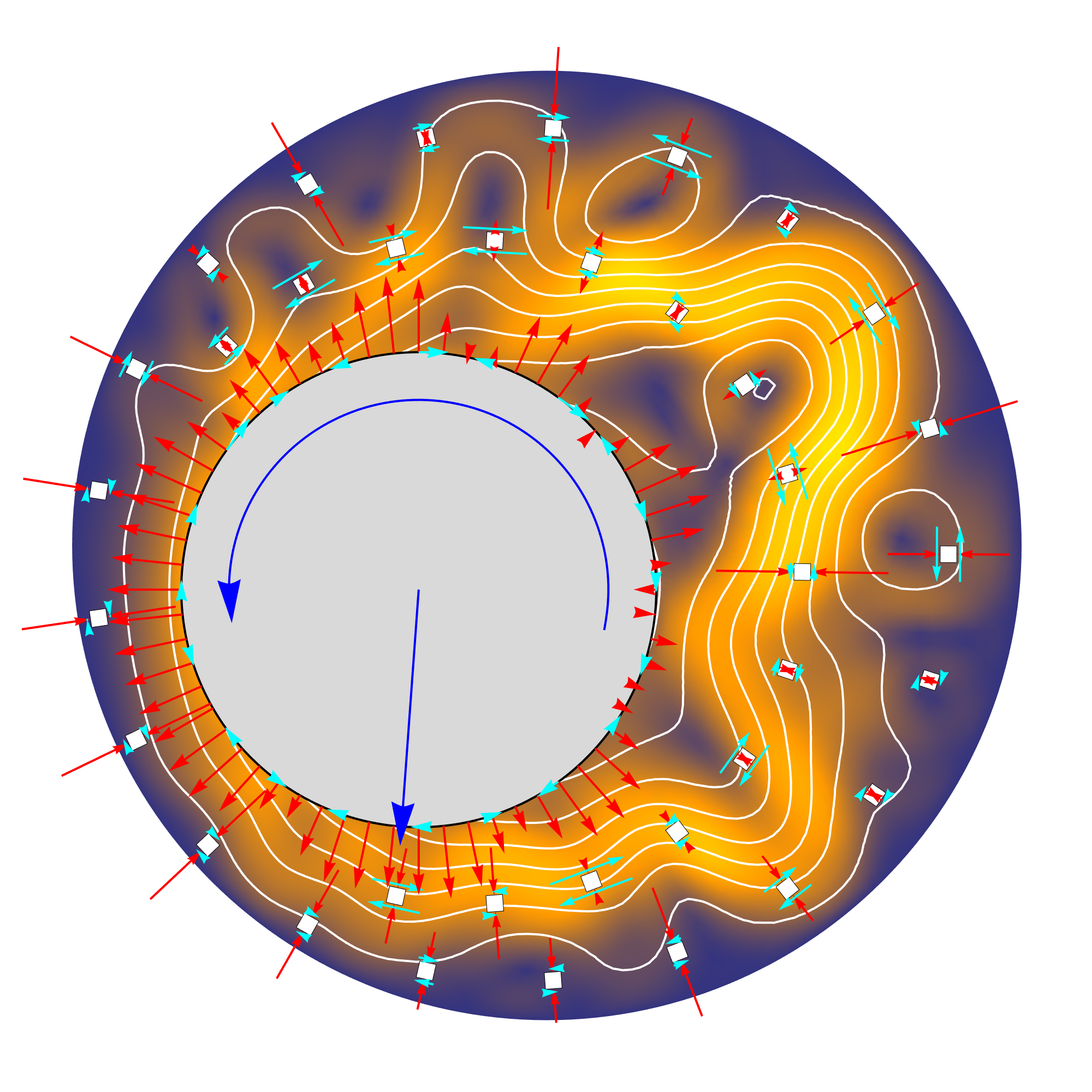}}
\subfigure[$t=458.5$]{\includegraphics[width=0.32\textwidth]{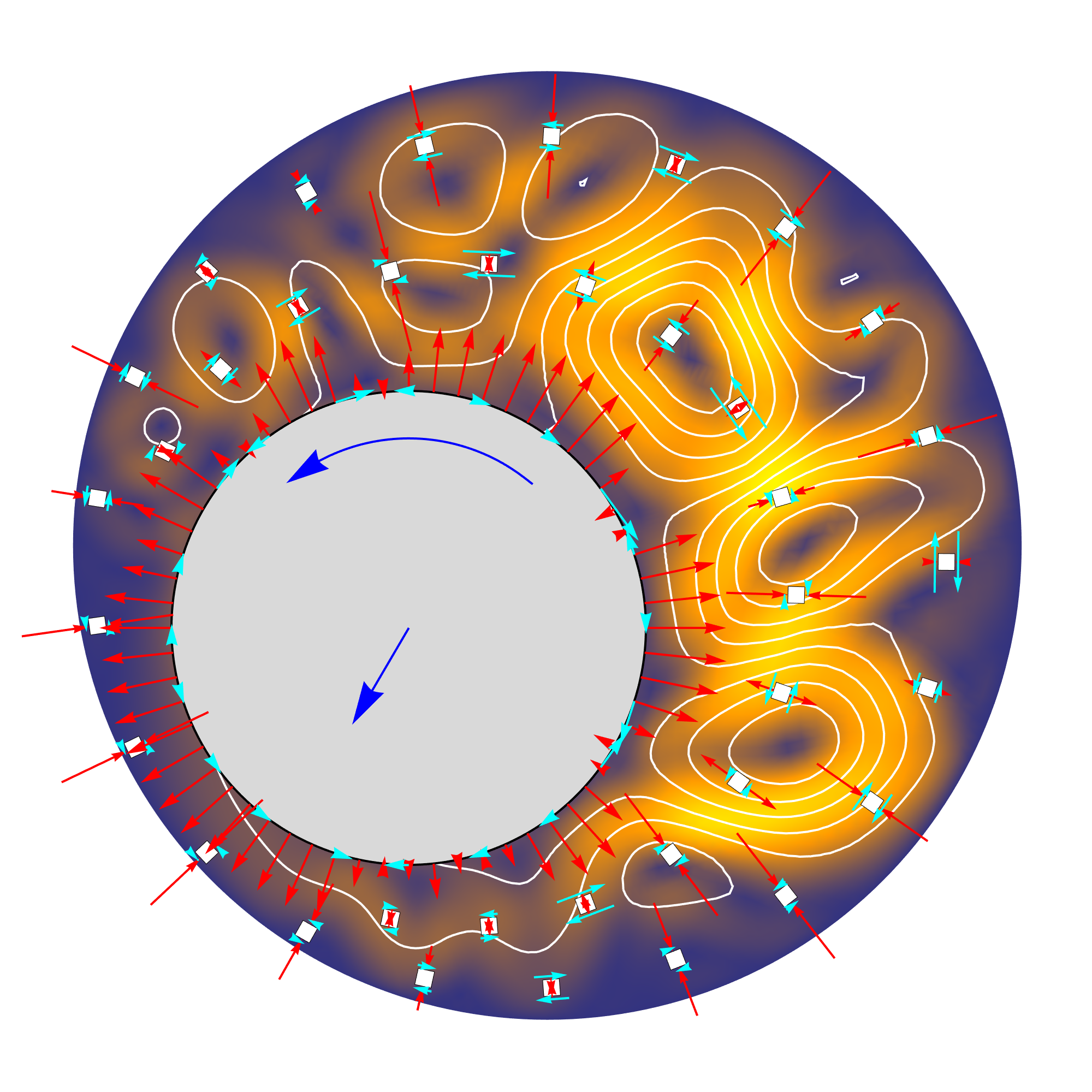}}
\subfigure[$t=459.5$]{\includegraphics[width=0.32\textwidth]{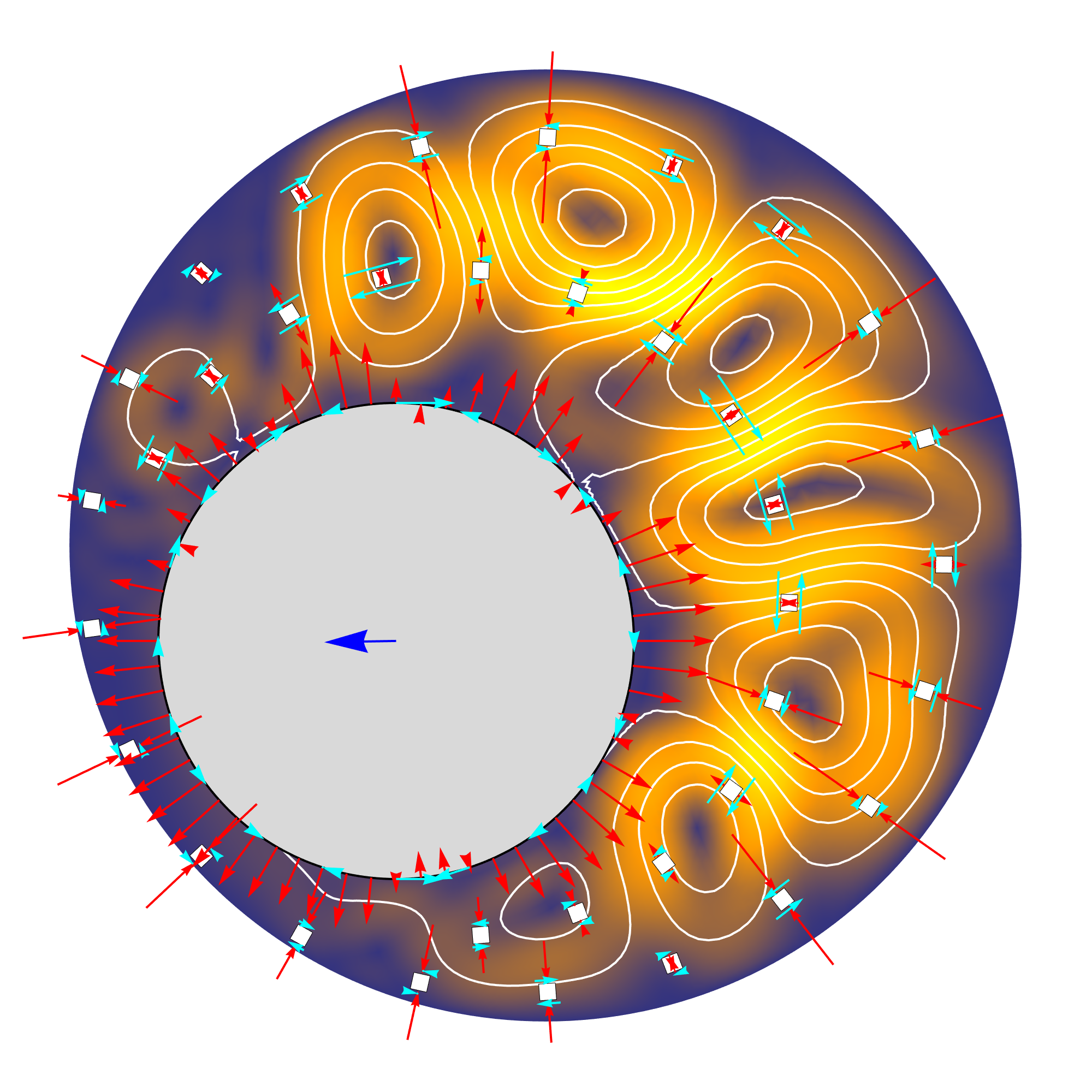}}
\subfigure[$t=460.5$]{\includegraphics[width=0.32\textwidth]{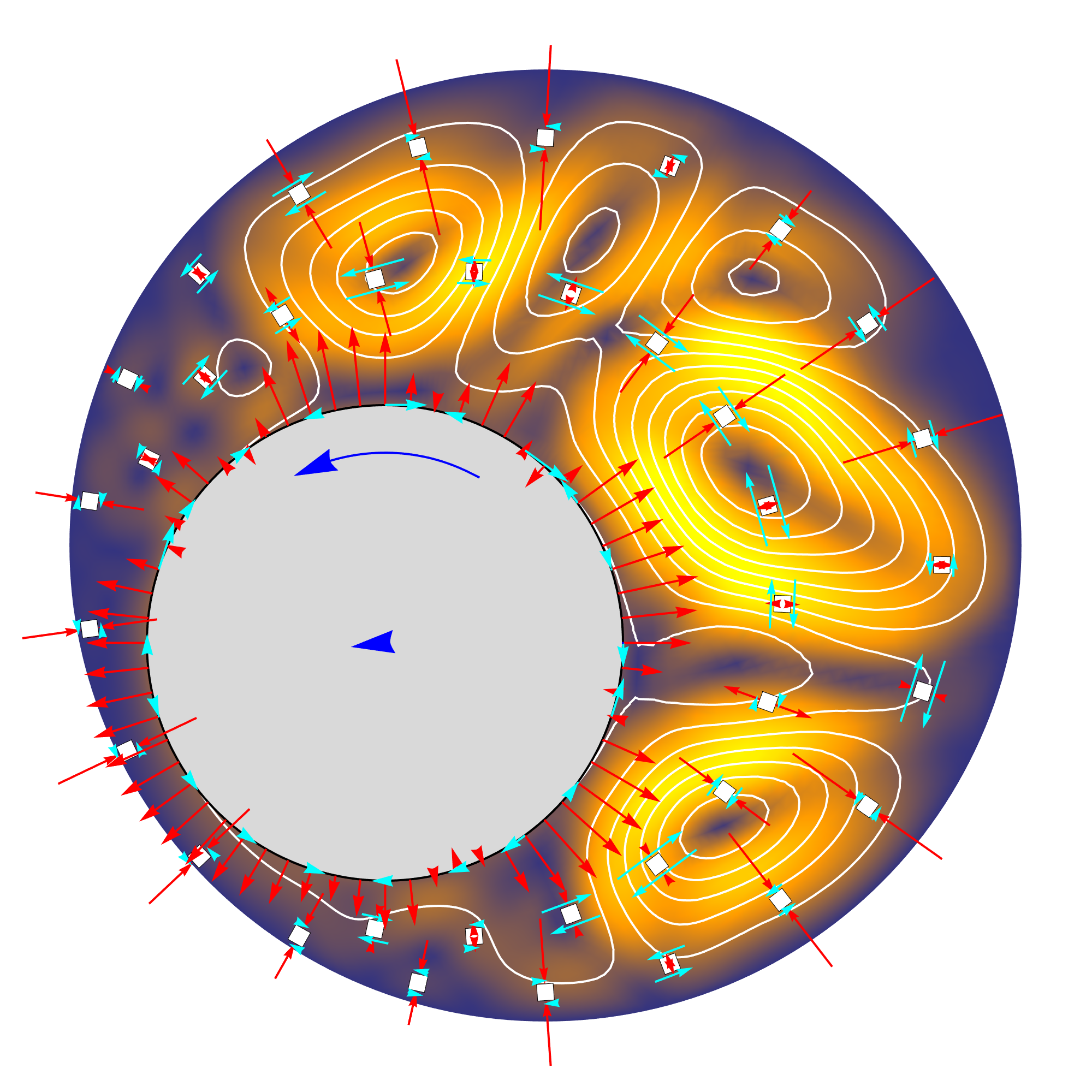}}
\subfigure[$t=461.5$]{\includegraphics[width=0.32\textwidth]{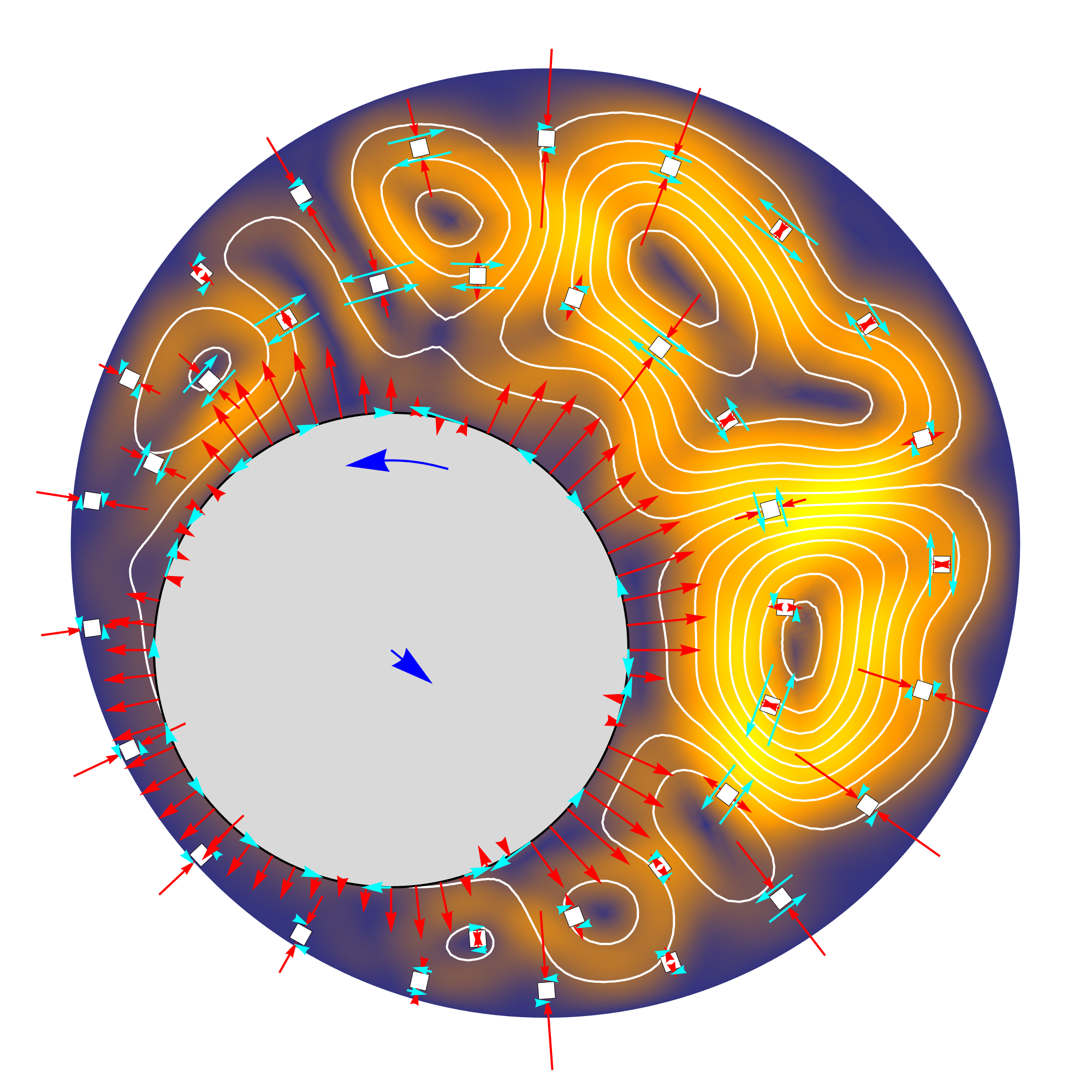}}
\subfigure[$t=462.5$]{\includegraphics[width=0.32\textwidth]{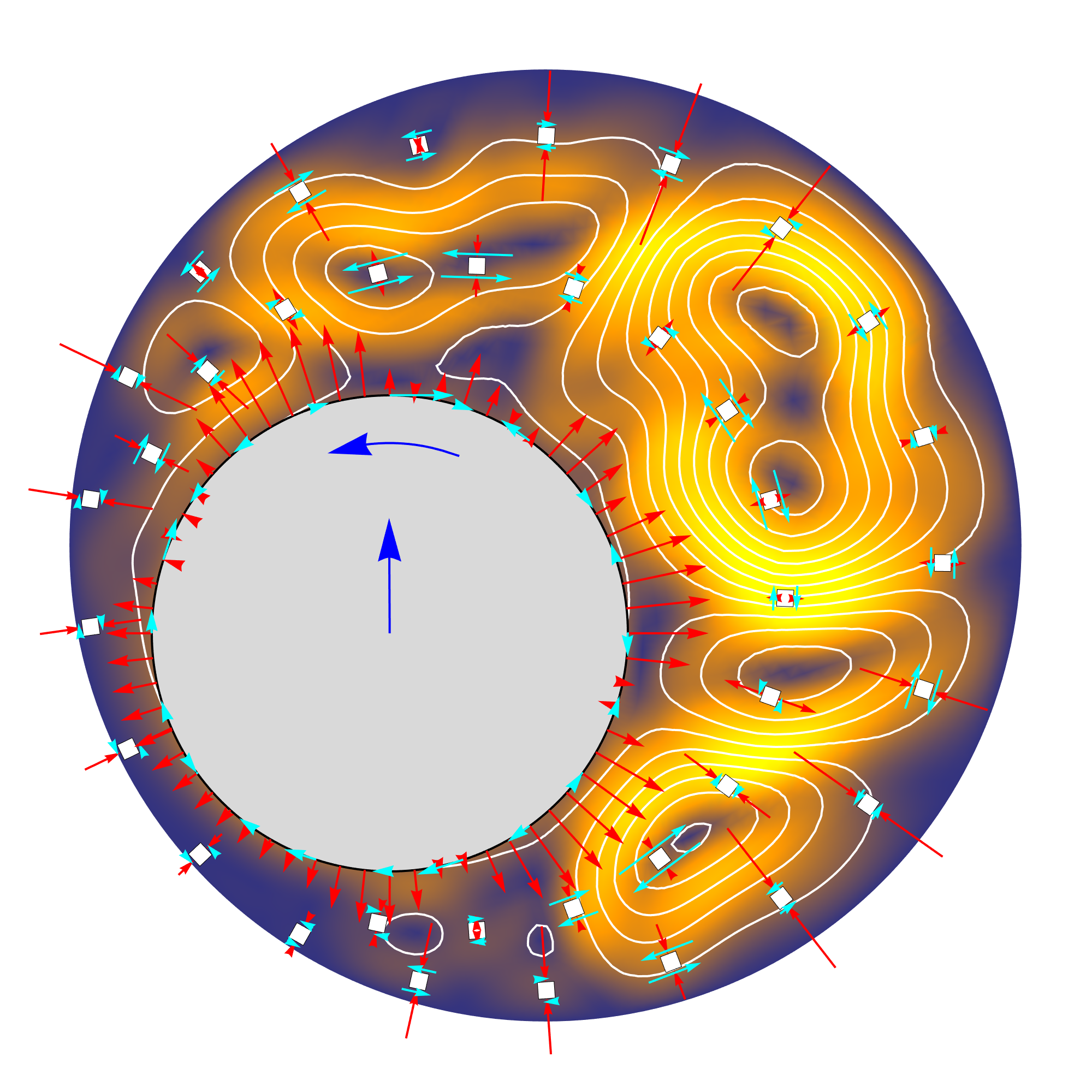}}
\subfigure[$t=463.5$]{\includegraphics[width=0.32\textwidth]{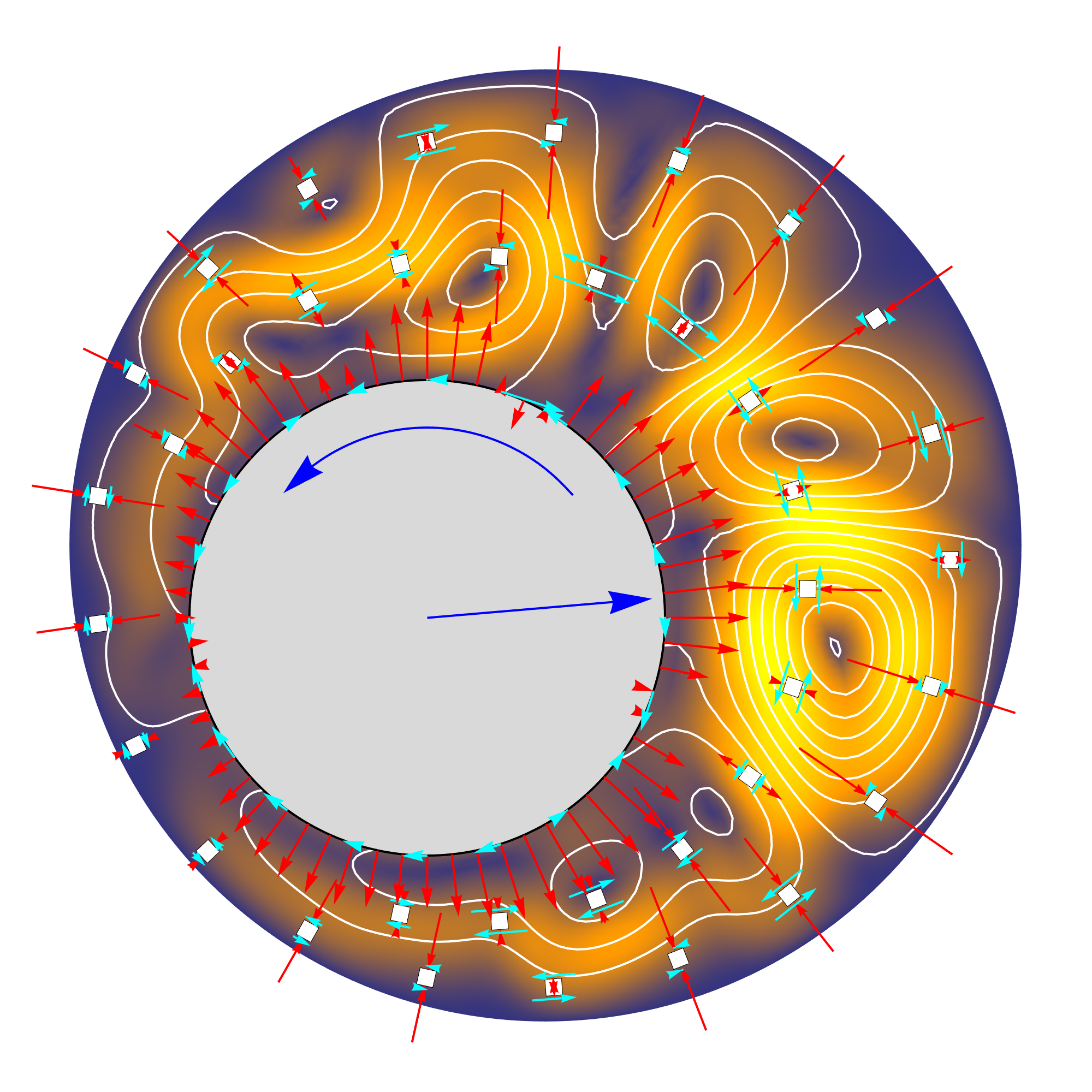}}
\subfigure[$t=464.5$]{\includegraphics[width=0.32\textwidth]{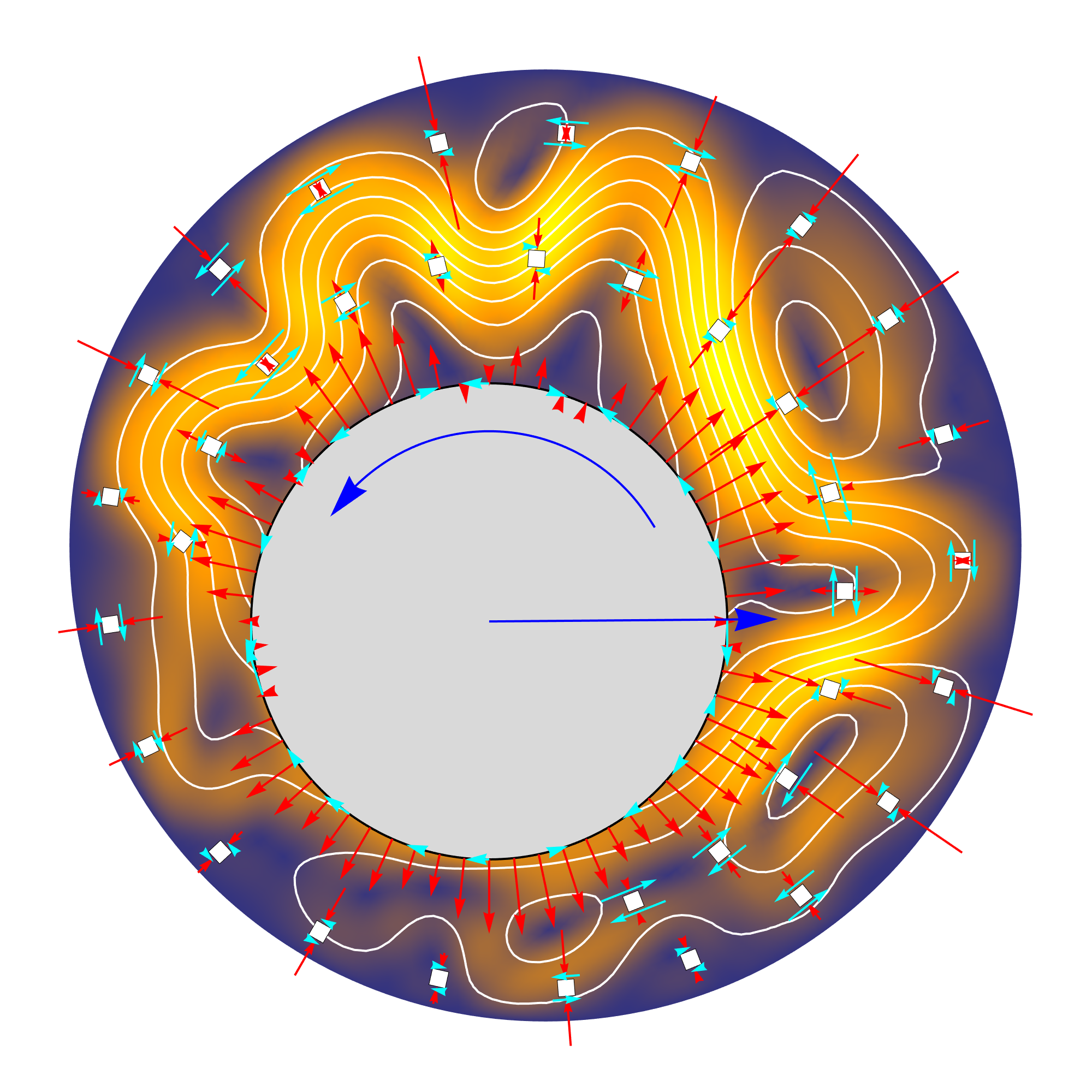}}
\caption{Visualization of a near approach to the container wall and repulsion for $\alpha = -20$ at the times labeled, with blue arrows visualizing the velocity $\bU$ and rotation rate $\Omega$ of the immersed circle.  The red arrows show the normal component $D_{nn}$ of $\bD$ directed toward the immersed object and outer wall, with its specific angle linearly interpolated between closest points on each circle.   The cyan arrows visualize the shear component $D_{nt}$ in this same orientation.  The white contours are streamfunction $\Psi$ contours with spacing $\Delta\psi = 0.005$ in this rotating frame and the same color levels visualize the velocity magnitude $|\bu|$ in this same frame.}\label{fig:viz20}
\end{center}
\end{figure}

The mechanism that counters contact is visualized in figure~\ref{fig:viz20} for a particularly close approach and subsequent repulsion for $\alpha = -20$.  The sequence starts in a period of fast rotation and correlated flow around the circumference of the circle.  The overall flow at this time loosely resembles the uniform circulation flow of the $\alpha = -0.6$ case in figure~\ref{fig:lowalphainitial}, though with additional waviness reflecting its additional instabilities.  In this state, the circle is pulled toward the container wall by shear-driven tensile normal stress, following the same basic margination mechanism of the geometric instability for weak activity discussed in section~\ref{ss:geominstab}.   However, as it approaches, there is a sharp decrease in this circle's rotation.  The overall circulation rapidly fails as the gap narrows, and the strain rate in the gap similarly drops.  The circulating flow is replaced by distinct vorticies in the larger region, which resemble those associated with the traveling-wave instability observed in a fixed-wall annular geometry~\citep{Chen:2018}.  No flow structures appear in the smaller gap.  Both the viscous resistance of this configuration and de-correlation of the azimuthal flow structure lead to slower rotation of the circle; compared to the approach phase, its rotation essentially ceases.  Without sustaining shear strain, the net active stresses decay in the narrow region, as seen in figures~\ref{fig:viz20} (c)--(g).  In this same period, the array of vortex-like structures strengthen and fill the wider gap.  For each oppositely rotating vortex, the active shear component $D_{nt}$ on the circle changes sign, so they do not apply a significant net torque.  In contrast, the normal stress component $D_{nn}$ for all of them act in the same tensile sense to draw the circle away from the container wall, which happens rapidly.  All close approaches observed for all $\alpha \le -10$ cases showed a similar evolution.

\begin{figure}
  \begin{center}
    \subfigure[]{\includegraphics[width=0.38\textwidth]{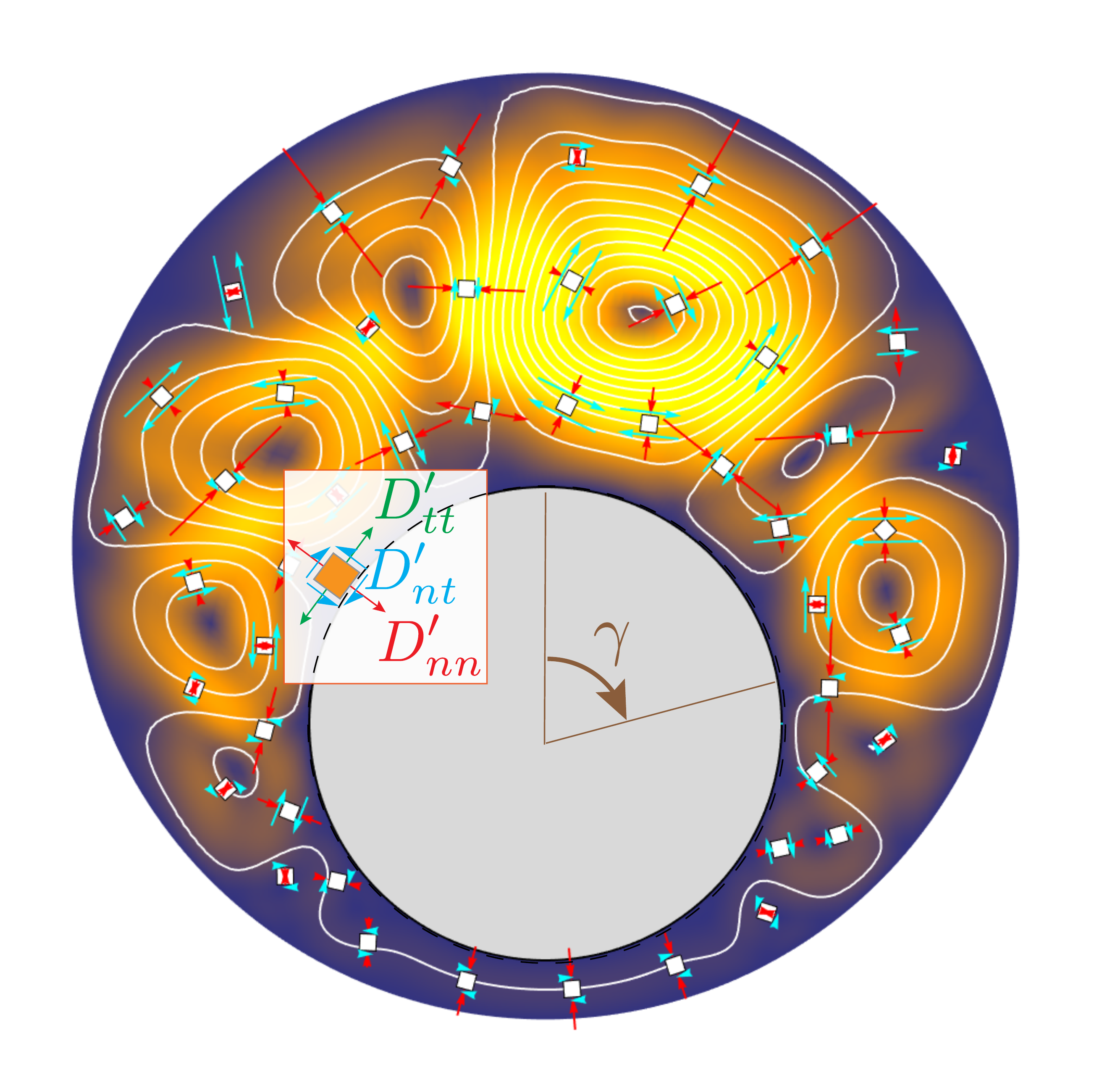}}
    \subfigure[]{
    \begin{tikzpicture}
      \begin{axis}
        [ 
        ymin = -0.5,
        ymax = 0.81,
        xmin = -3.2,
        xmax = 3.2,
        ylabel={$D_{nn}'$, $D_{nt}'$, $D_{tt}'$},
        xlabel={$\gamma$},
        tick scale binop=\times,
        width=0.58\textwidth,
        height=0.4\textwidth,
        grid=major,
        legend style={	at={(axis cs:-3.1,.64)},anchor=west,legend
        columns=4,draw=none,legend cell align=left},
         xtick={-3.14159,-2.35619,-1.5708,-0.785398,0.,0.785398,1.5
   708,2.35619,3.14159},
    xticklabels={$-\pi$ ,$-\frac{3 \pi }{4}$,$-\frac{\pi
   }{2}$,$-\frac{\pi }{4}$,$0$,$\frac{\pi }{4}$,$\frac{\pi
   }{2}$,$\frac{3 \pi }{4}$,$\pi$ }
      ]
      \addlegendimage{empty legend}
        \addplot+[no marks, very thick, color=red] table[x
        expr=\thisrowno{0}, y expr=\thisrowno{1}, col sep=space] {Figures/surfaceD-a=20-U0.dat};
        \addplot+[no marks, very thick, color=cyan] table[x
        expr=\thisrowno{0}, y expr=\thisrowno{2}, col sep=space] {Figures/surfaceD-a=20-U0.dat};
        \addplot+[no marks, very thick, color=green!60!black] table[x
        expr=\thisrowno{0}, y expr=\thisrowno{3}, col sep=space] {Figures/surfaceD-a=20-U0.dat};
              \addlegendimage{empty legend}
        \addplot+[no marks, very thick, color=red, dashed] table[x
        expr=\thisrowno{0}, y expr=\thisrowno{4}, col sep=space] {Figures/surfaceD-a=20-U0.dat};
        \addplot+[no marks, very thick, color=cyan, dashed] table[x
        expr=\thisrowno{0}, y expr=\thisrowno{5}, col sep=space] {Figures/surfaceD-a=20-U0.dat};
        \addplot+[no marks, very thick, color=green!60!black, dashed] table[x
        expr=\thisrowno{0}, y expr=\thisrowno{6}, col sep=space] {Figures/surfaceD-a=20-U0.dat};
        \legend{
          {[text width=40pt,text depth=] Instant.:\;\; },
          {{\color{red} $D_{nn}'$}}\;\;,
          {{\color{cyan} $D_{nt}'$}}\;\;,
          {{\color{green!40!black} $D_{tt}'$}},
          {[text width=40pt,text depth=]\hspace*{0.0in}Average:\;\;},
          {{\color{red} $D_{nn}'$}}\;\;,
          {{\color{cyan} $D_{nt}'$}}\;\;,
          {{\color{green!40!black} $D_{tt}'$}}}
      \end{axis}
    \end{tikzpicture}
    }
  \end{center}
  \caption{Case with $\alpha = -20$ and the object fixed and not
    rotating ($\bU=0$, $\Omega = 0$) at $r_o =
    0.75$:  (a) contours of the streamfunction $\psi$ with $\Delta
    \psi = 0.005$ spacing and showing components of local $\bD$ as in figure~\ref{fig:viz20}, and $|\bu|$ (flood colors), and (b) components of
    $\bD$ refrenced to the object-normal direction, show both an
    example instantaneous profile and a long-time average.} \label{fig:fixed}
\end{figure}

To more explicitly illustrate the lift forces, figure~\ref{fig:fixed} (a) visualizes a corresponding case with the circle held fixed ($\bU=0$ and $\Omega = 0$) at $r_o = 0.75$.  Unlike the small $\alpha$ cases, the suspension activity does not depend on the motion of the object to sustain it, and a vigorous flow persists in the larger space.  The alternating vortex structures are distinct, and there is only weak flow in the narrow gap.   Active stresses on the circle (figure~\ref{fig:fixed} b) reflect the alternating vortex structure, leading to a significant normal stress generally pulling the circle away from the container wall.  The corresponding shear stresses alternate sign with near-zero mean.

\begin{figure}
  \begin{center}
    \subfigure[$\alpha = -20$]{\begin{minipage}{0.45\textwidth}
        \axislabledfigure{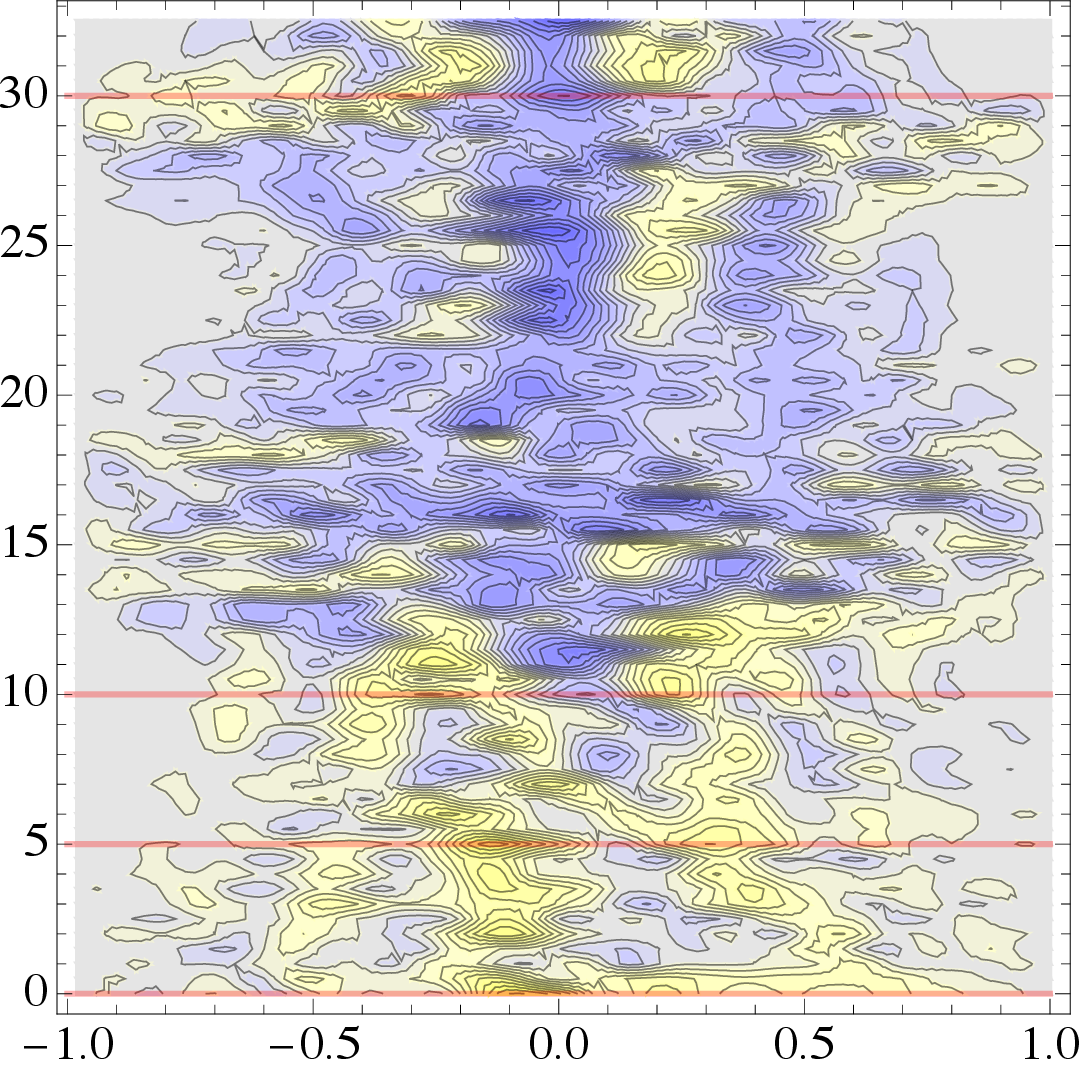}{$\gamma/\pi$}{$t$}{1.0}
      \end{minipage}
        \vspace*{-0.3in}
      \hspace*{0.25in}
      \raisebox{0.3in}{\begin{minipage}{0.45\textwidth}
        \includegraphics[width=0.48\textwidth]{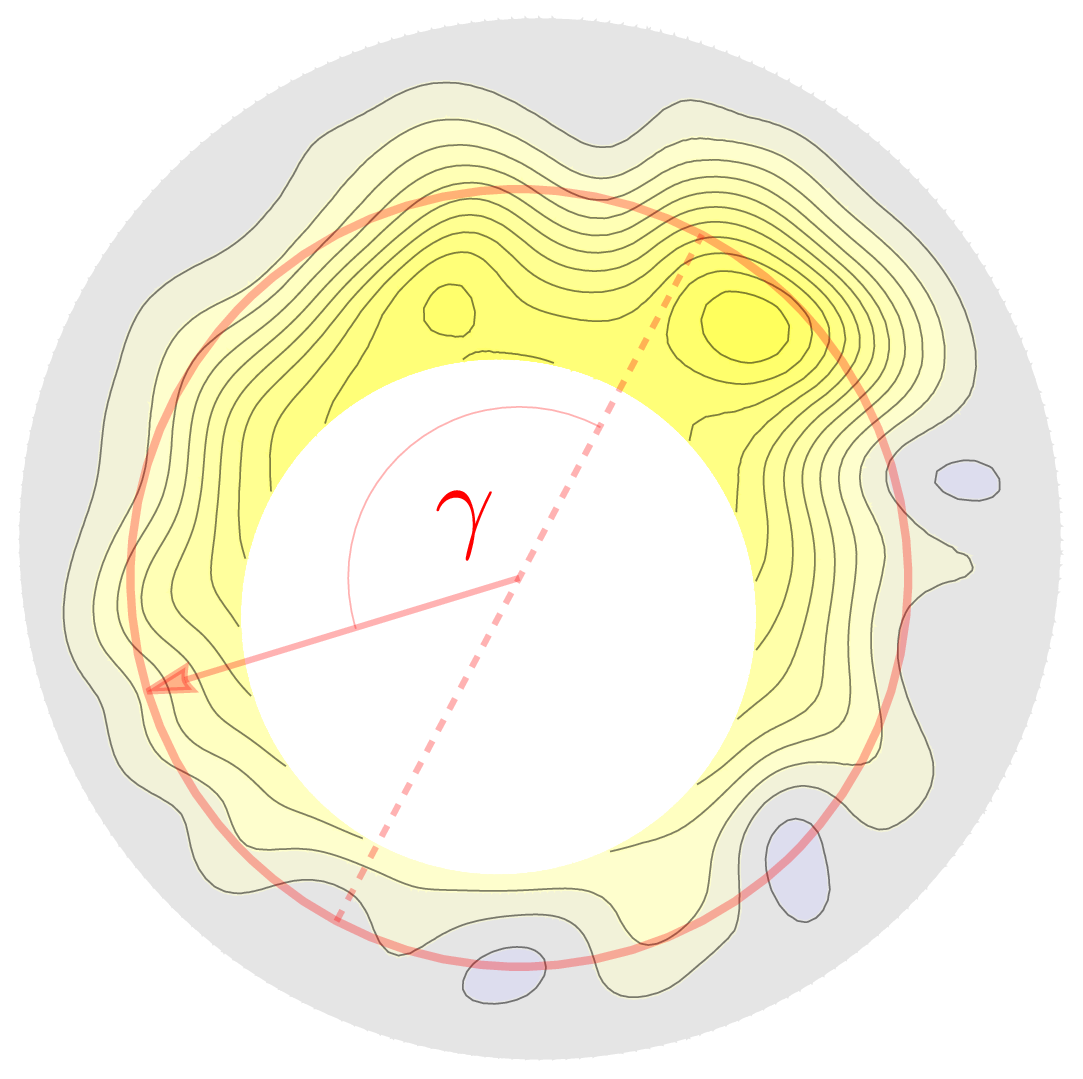}
        \includegraphics[width=0.48\textwidth]{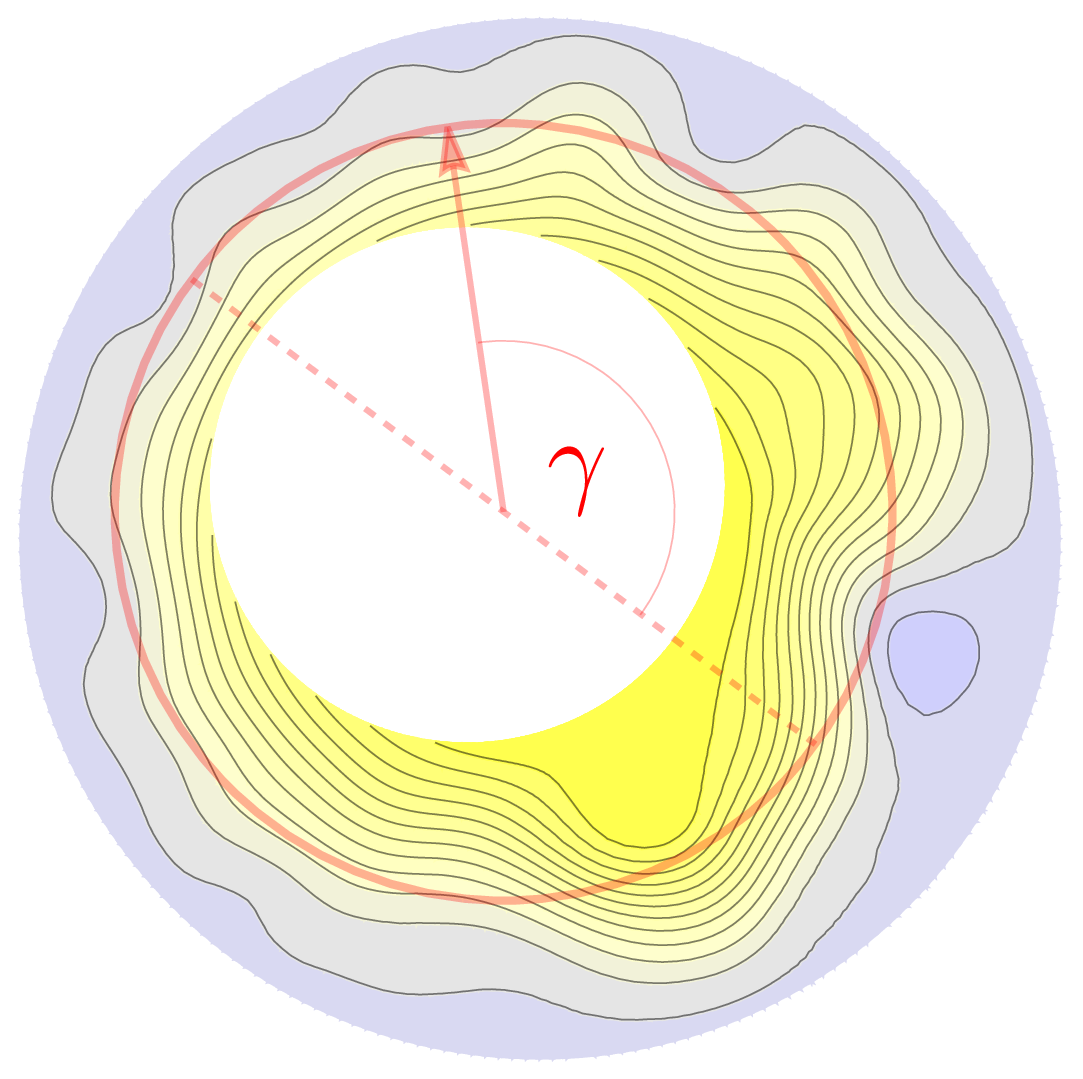}
        \includegraphics[width=0.48\textwidth]{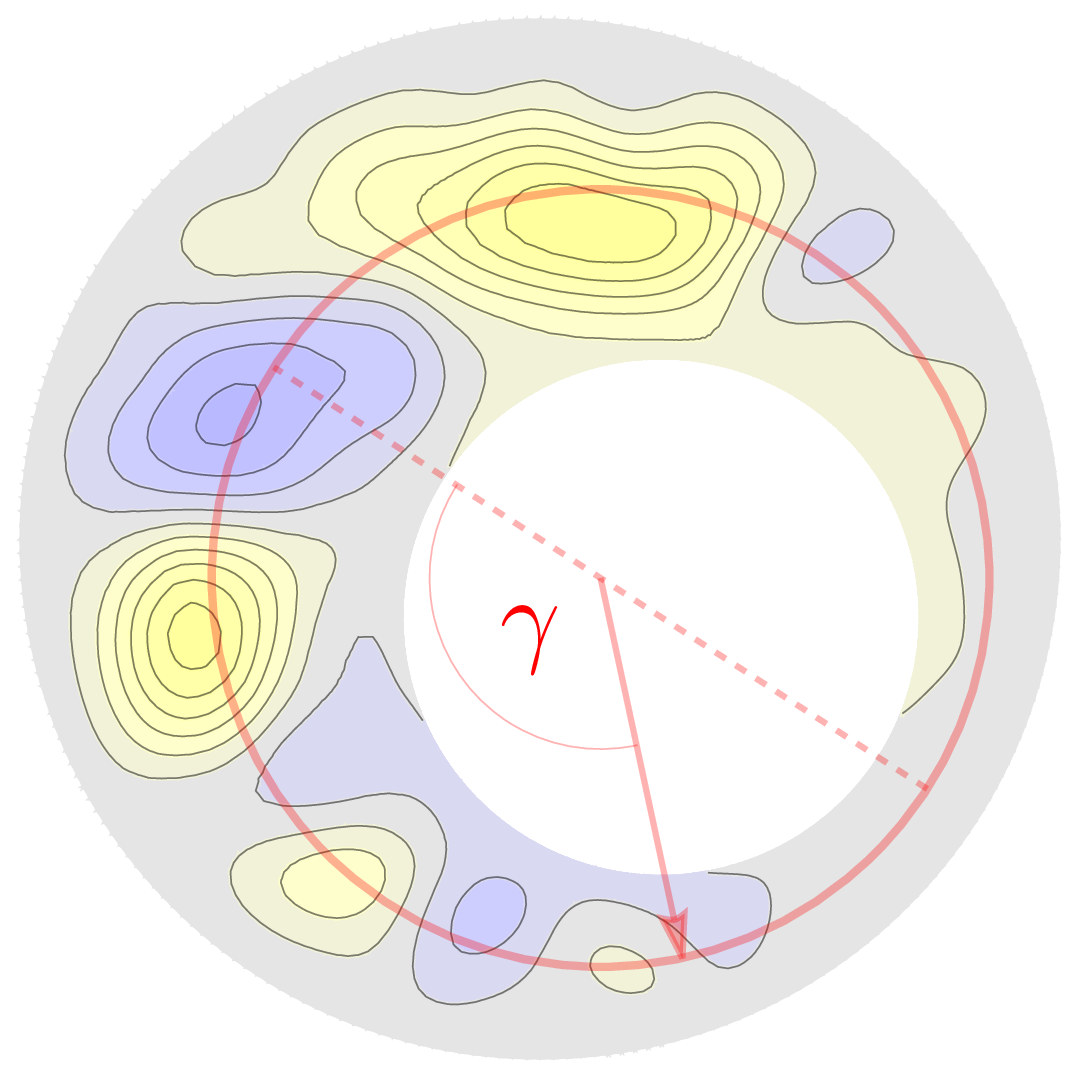}
        \includegraphics[width=0.48\textwidth]{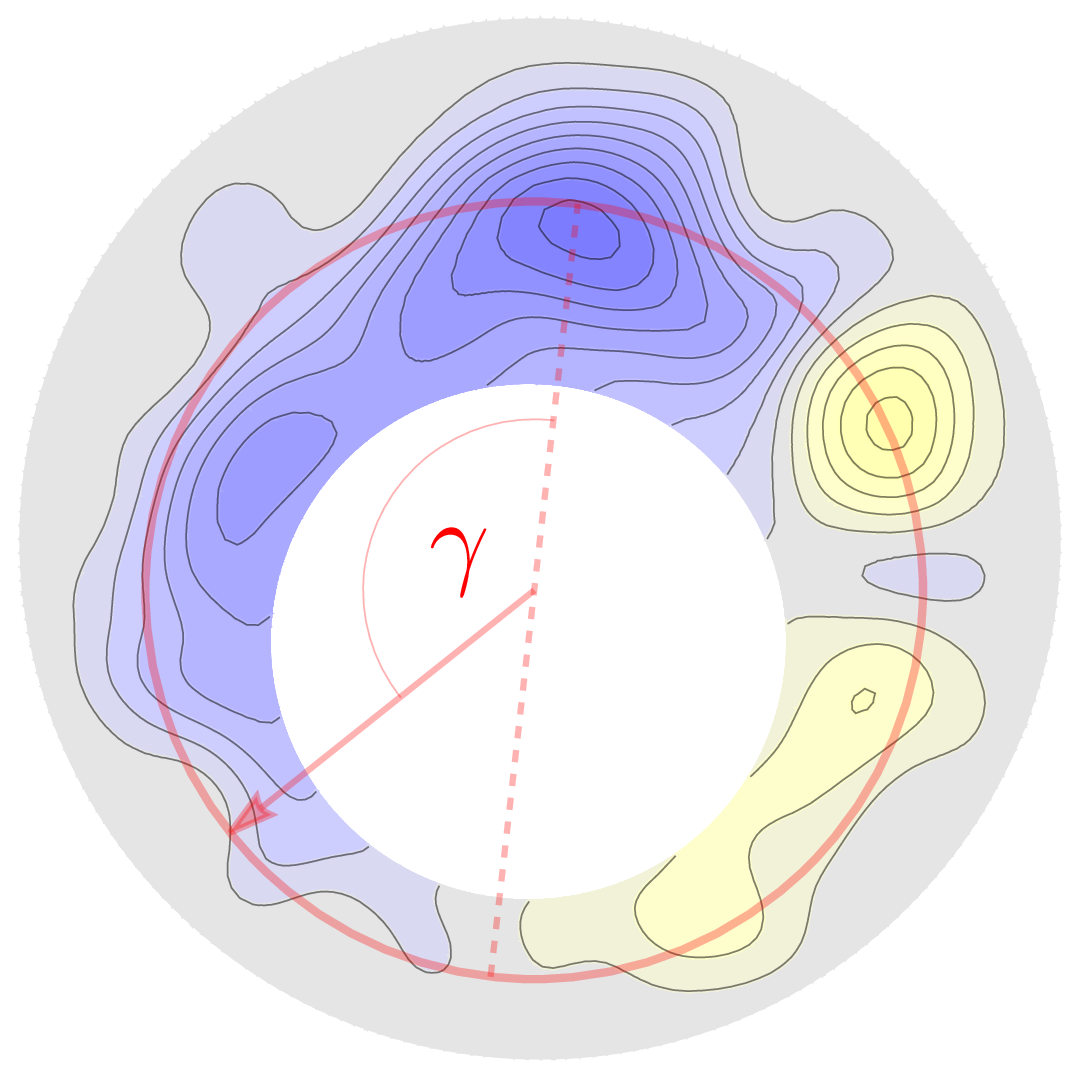}
      \end{minipage}}
  }
  \vspace*{0.2in}
    \subfigure[$\alpha = -80$]{\begin{minipage}{0.45\textwidth}
        \axislabledfigure{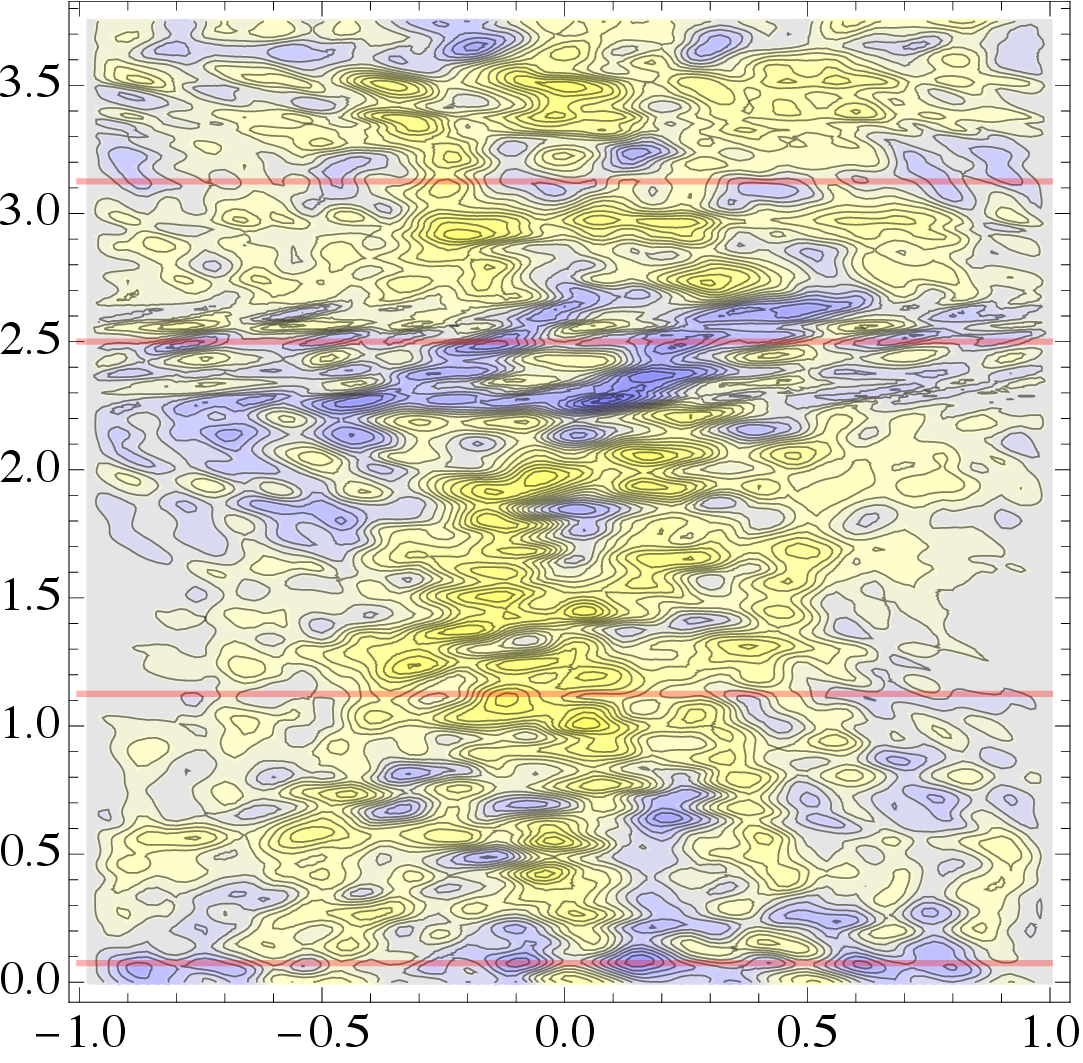}{$\gamma/\pi$}{$t$}{1.0}
      \end{minipage}
        \vspace*{-0.3in}
      \hspace*{0.25in}
      \raisebox{0.3in}{\begin{minipage}{0.45\textwidth}
        \includegraphics[width=0.48\textwidth]{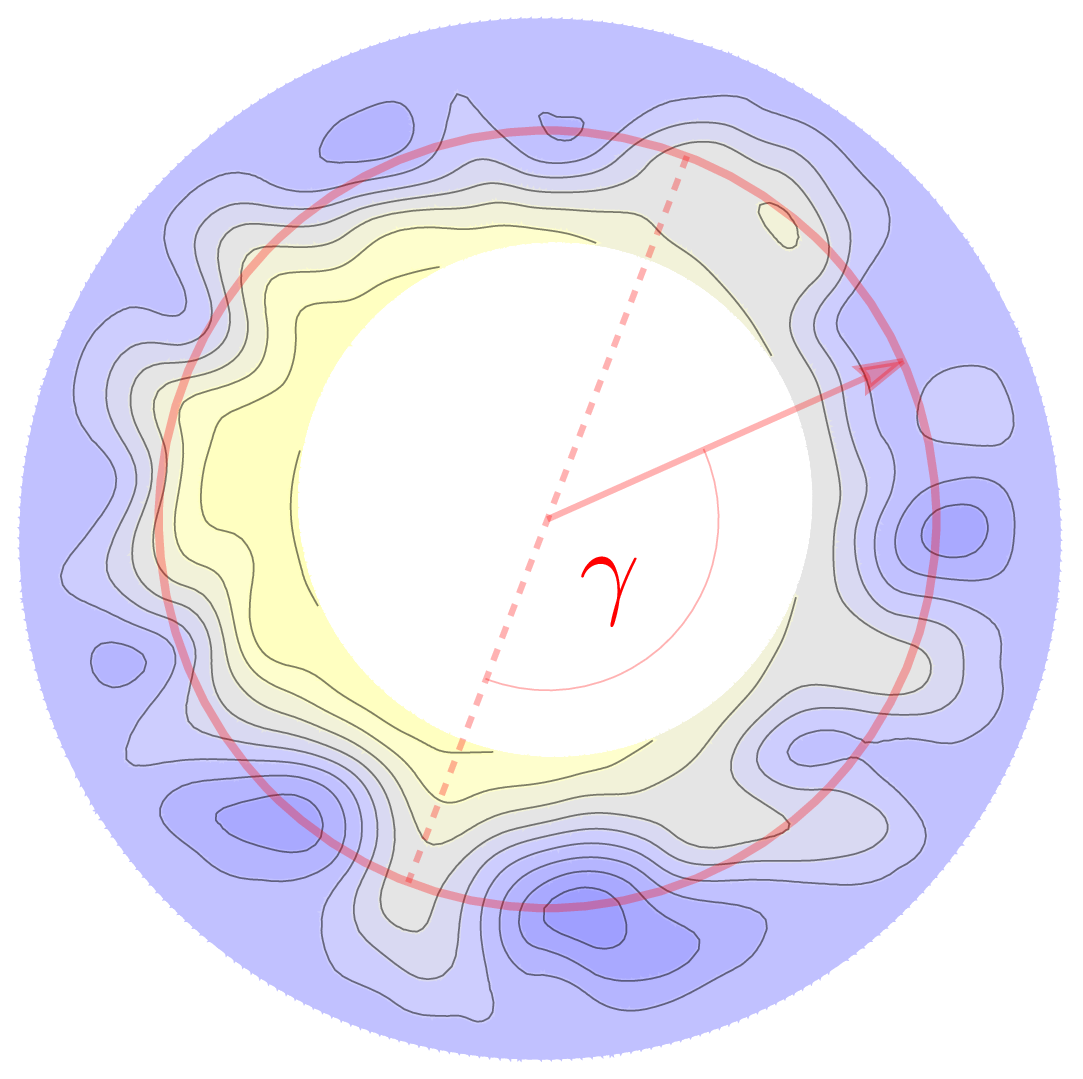}
        \includegraphics[width=0.48\textwidth]{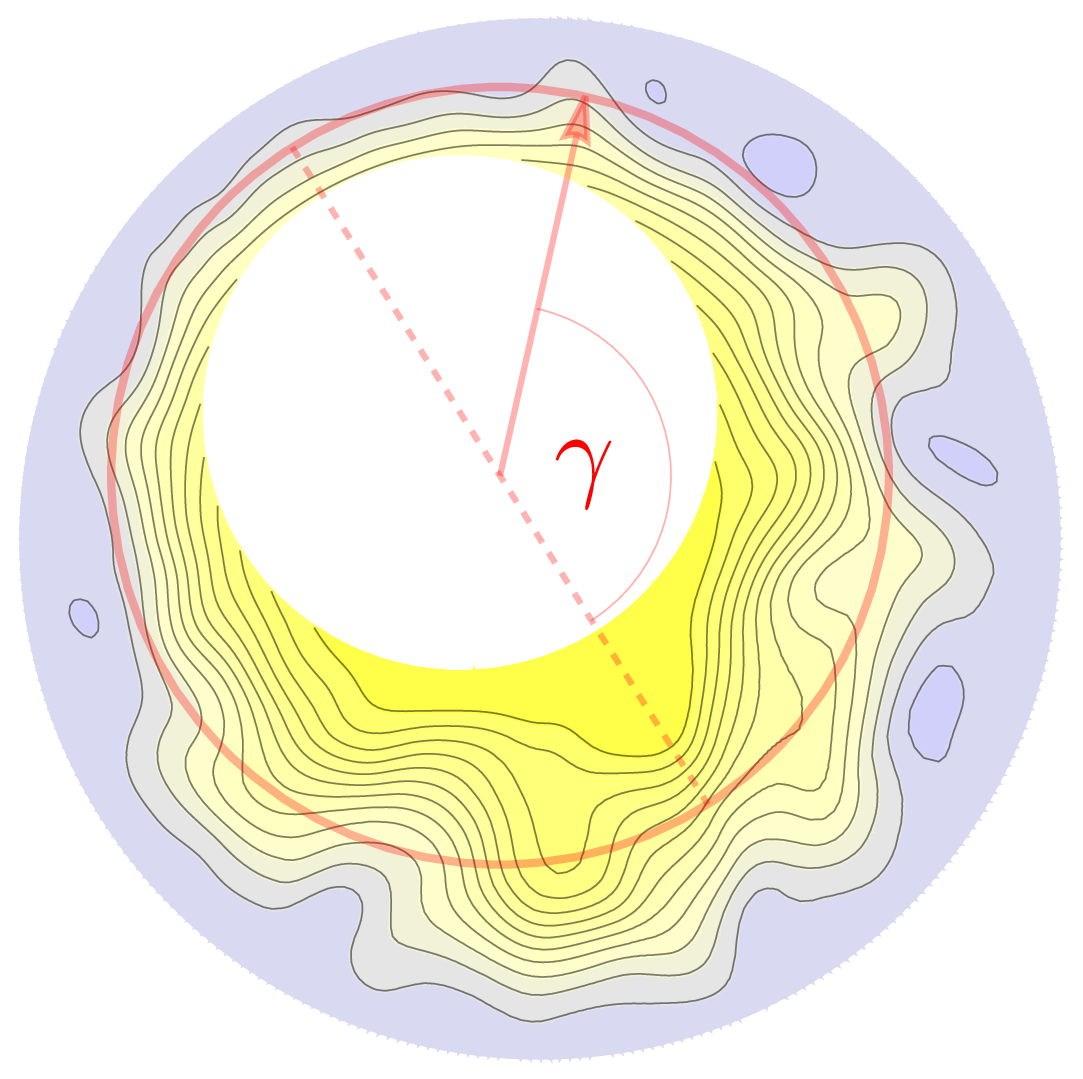}
        \includegraphics[width=0.48\textwidth]{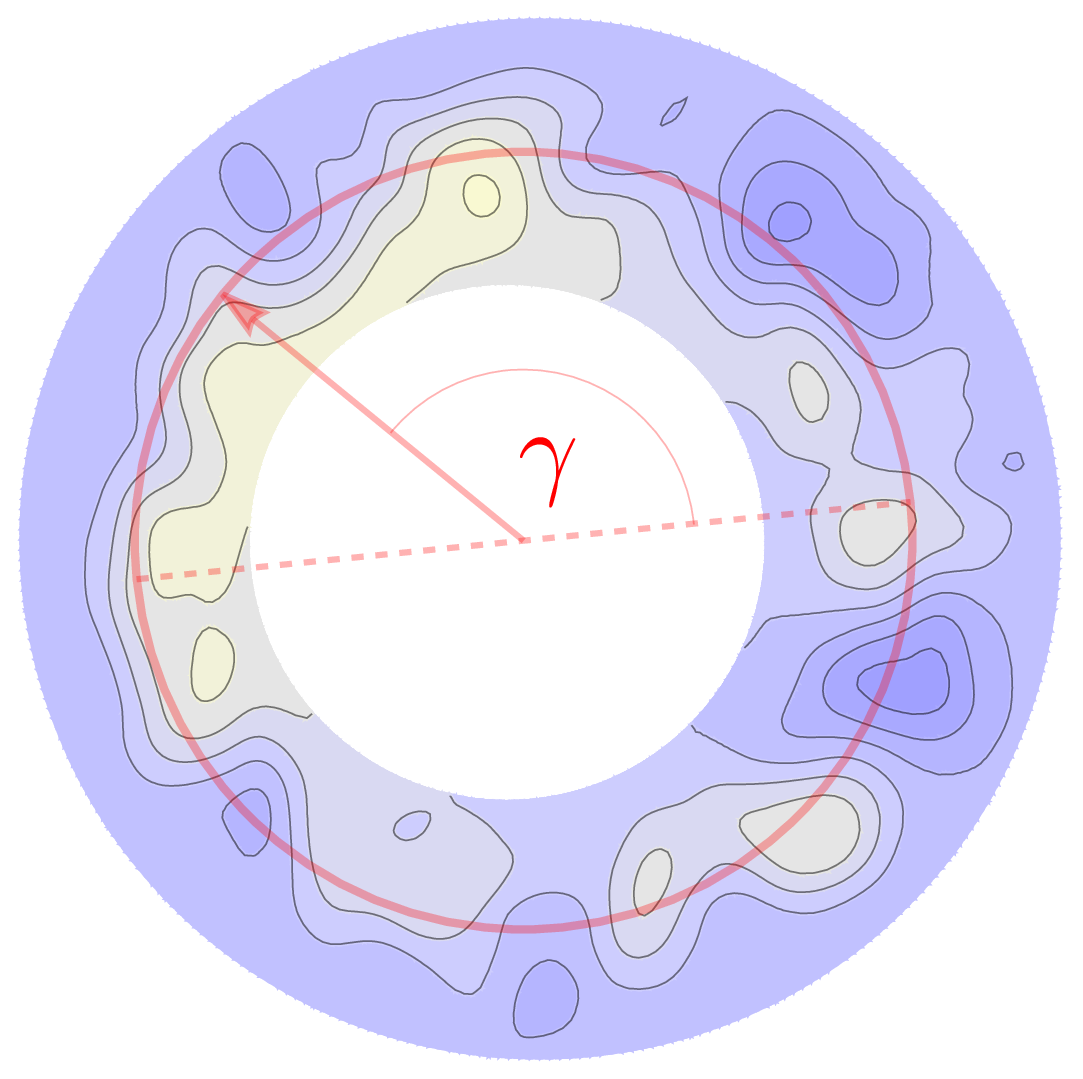}
        \includegraphics[width=0.48\textwidth]{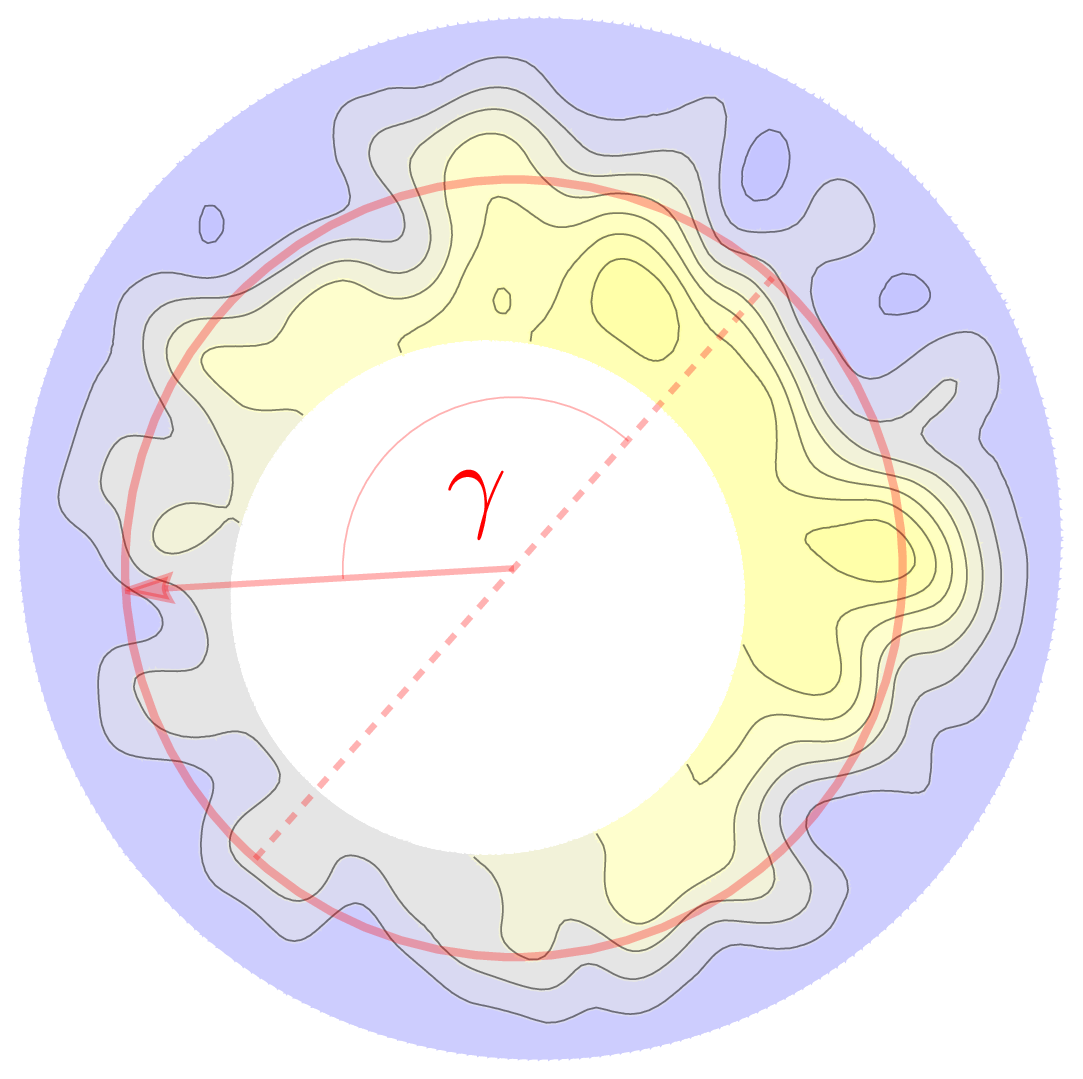}
      \end{minipage}}
  }
\end{center}
\caption{Visualized streamfunction $\psi$ for cases (a) $\alpha = -20$ and (b) $\alpha = -80$.  The horizontal lines in the $\gamma$--$t$ plots indicate the selected instances visualized to the right of each (with time increasing left-to-right, top-to-bottom); the $\gamma$--$t$ data are taken from the circle radius $(R+a)/2 = 1.5$ that passes through the midpoint at the smallest and largest container--object separation.  For each time, $\psi = 0$ is set at $\gamma = \pm \pi$, there are 20 equally space contours between $\pm |\psi|_{\text{max}}$.  Animated visualizations of these cases are available in supplemental movies~6 and 10.}\label{fig:XTstrong}
\end{figure}

The $\alpha=-20$ case was selected for the above examination because the structures can be more clearly identified than for stronger activity levels.  However, the same behavior is observed for $\alpha = -80$.  Figure~\ref{fig:XTstrong} visualizes the evolution of the flow structure in both cases for longer periods.  The $\alpha=-80$ case (figure~\ref{fig:XTstrong} b) shows more random and smaller structures than $\alpha = -20$ (figure~\ref{fig:XTstrong} b), but both also show periods of relatively correlated circulating flow and periods with relatively distinct arrays of vorticies of alternating sense.  For $\alpha=-80$, these arrays are less distinct, due to smaller scale instabilities available for this $\alpha$.  For both cases, rotation of the circle remains primarily in a single direction for periods.  Still, even for $\alpha \le -10$, there are many reversals of rotation direction.  It should be noted that the time period visualized in the figures of this section are a small fraction of the total times simulated and analyzed.

\section{Transition behavior ($-5 \le \alpha \le -1$)}
\label{s:transition}

In section~\ref{s:lowactivity}, where mobility of the object was essential for instability of the suspension, it was moved toward the container wall until viscous resistance limited its mobility to a degree that the suspension stabilized and flow ceased.  In contrast, in section~\ref{s:highactivity}, the suspension remained active, even if the object were held fixed.  When the object approaches the container wall, organized activity in the gap was significantly suppressed, and the persistent activity in the bulk drew it away from the container wall.  The transition between these behaviors is complex.

\begin{figure}
  \begin{center}
    \subfigure[$\alpha = -5$; $\bar\delta = 0.0445$ with $\sigma = 0.005$]{\begin{minipage}{0.38\textwidth}
        \axislabledfigure{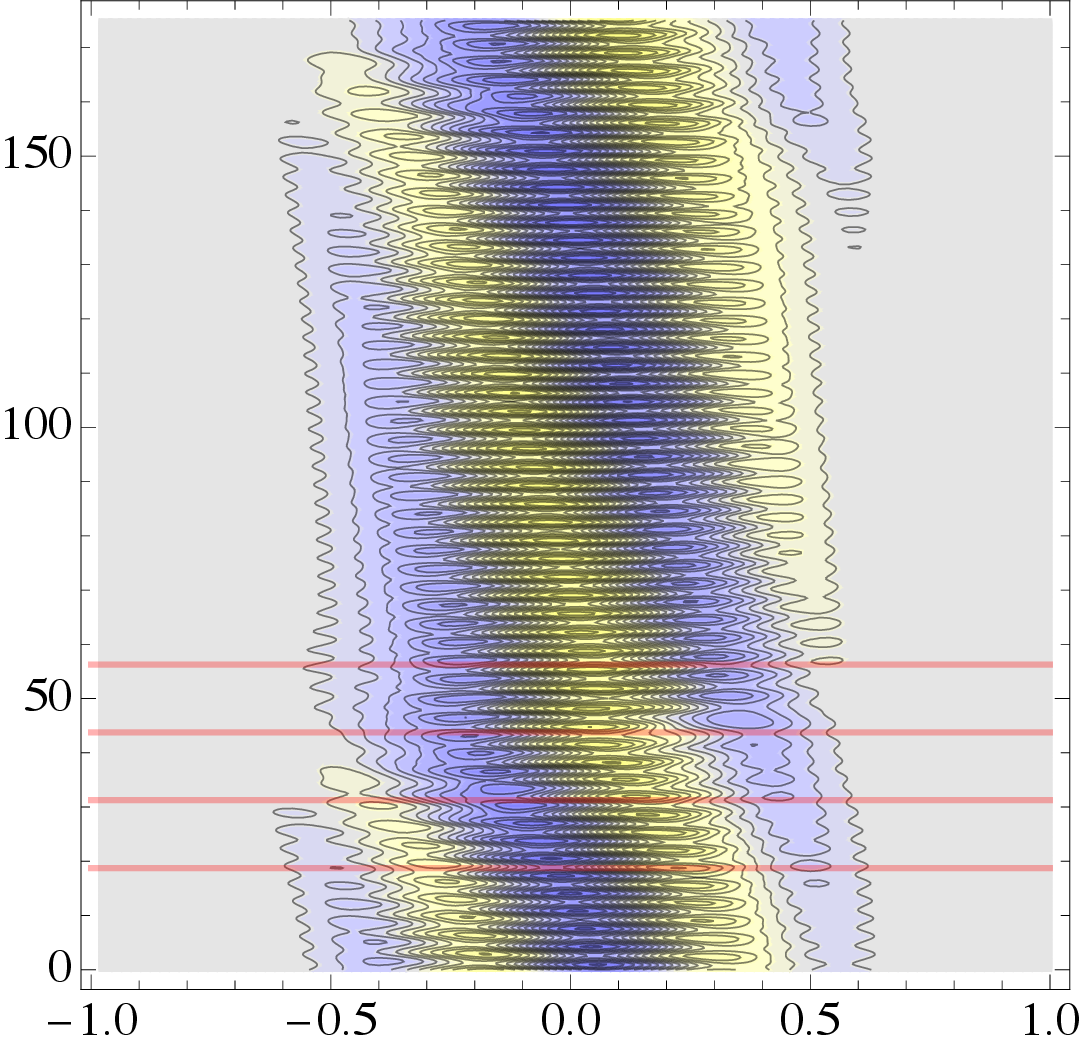}{$\gamma/\pi$}{$t$}{1.0}
      \end{minipage}
        \vspace*{-0.3in}
      \hspace*{0.25in}
      \raisebox{0.3in}{\begin{minipage}{0.38\textwidth}
        \includegraphics[width=0.48\textwidth]{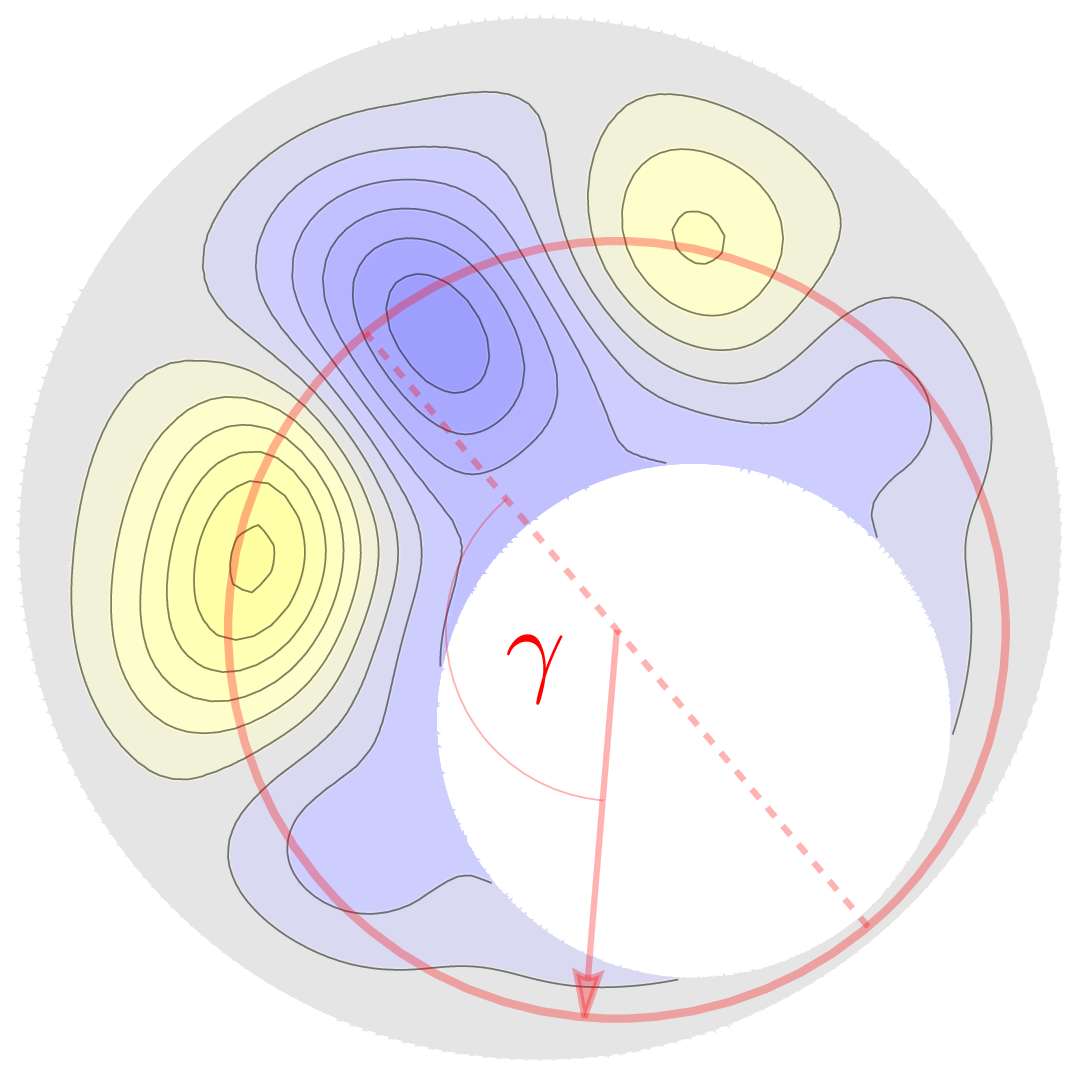}
        \includegraphics[width=0.48\textwidth]{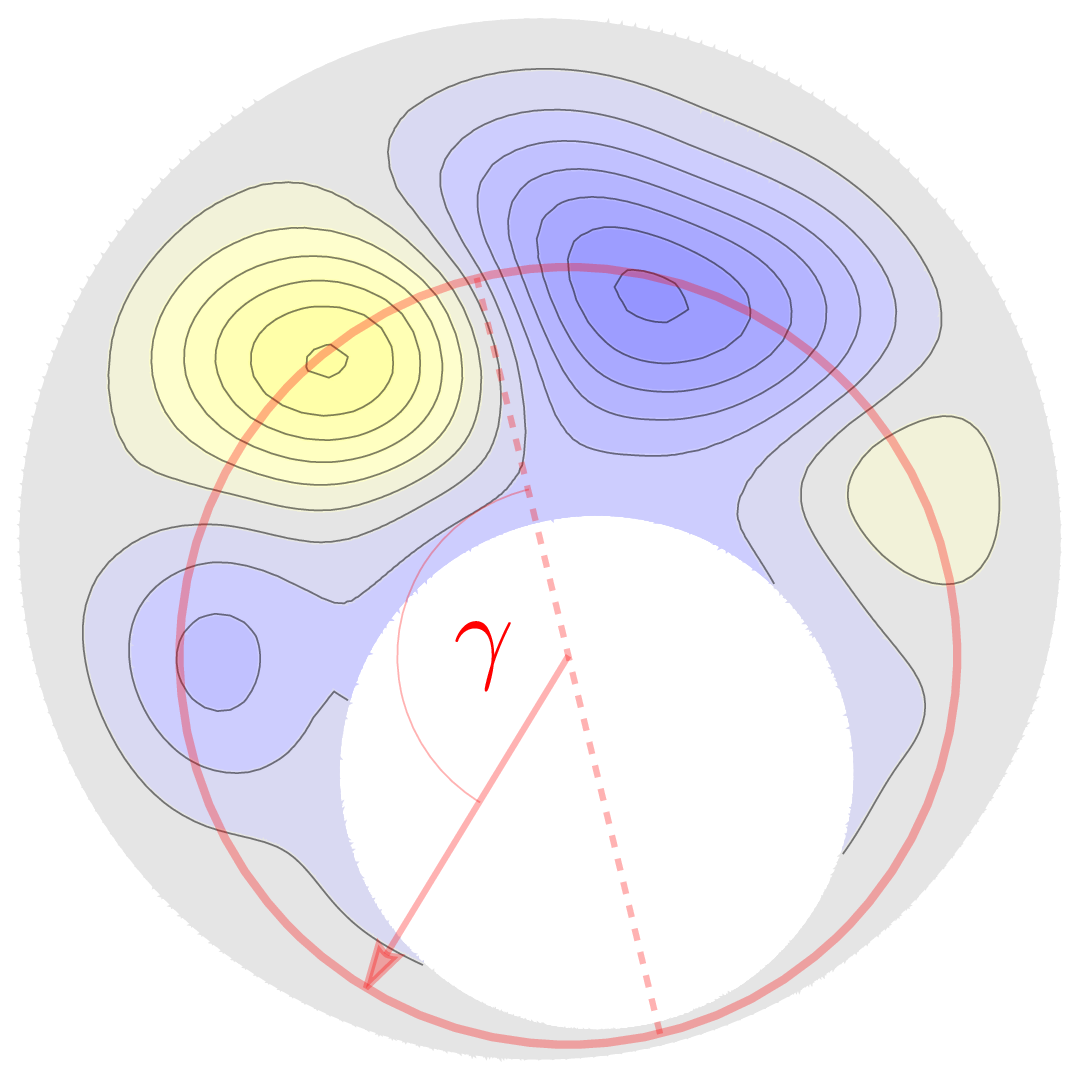}
        \includegraphics[width=0.48\textwidth]{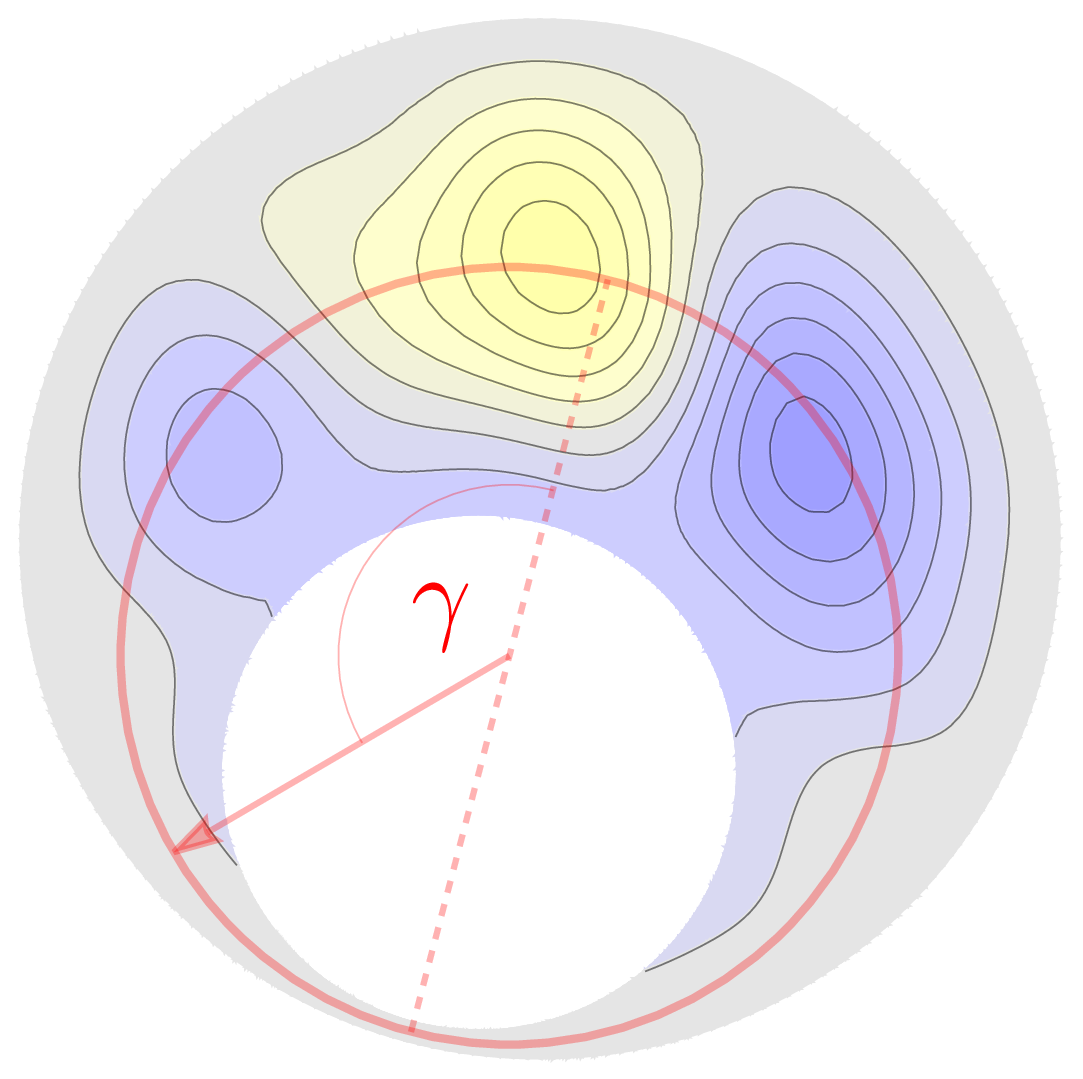}
        \includegraphics[width=0.48\textwidth]{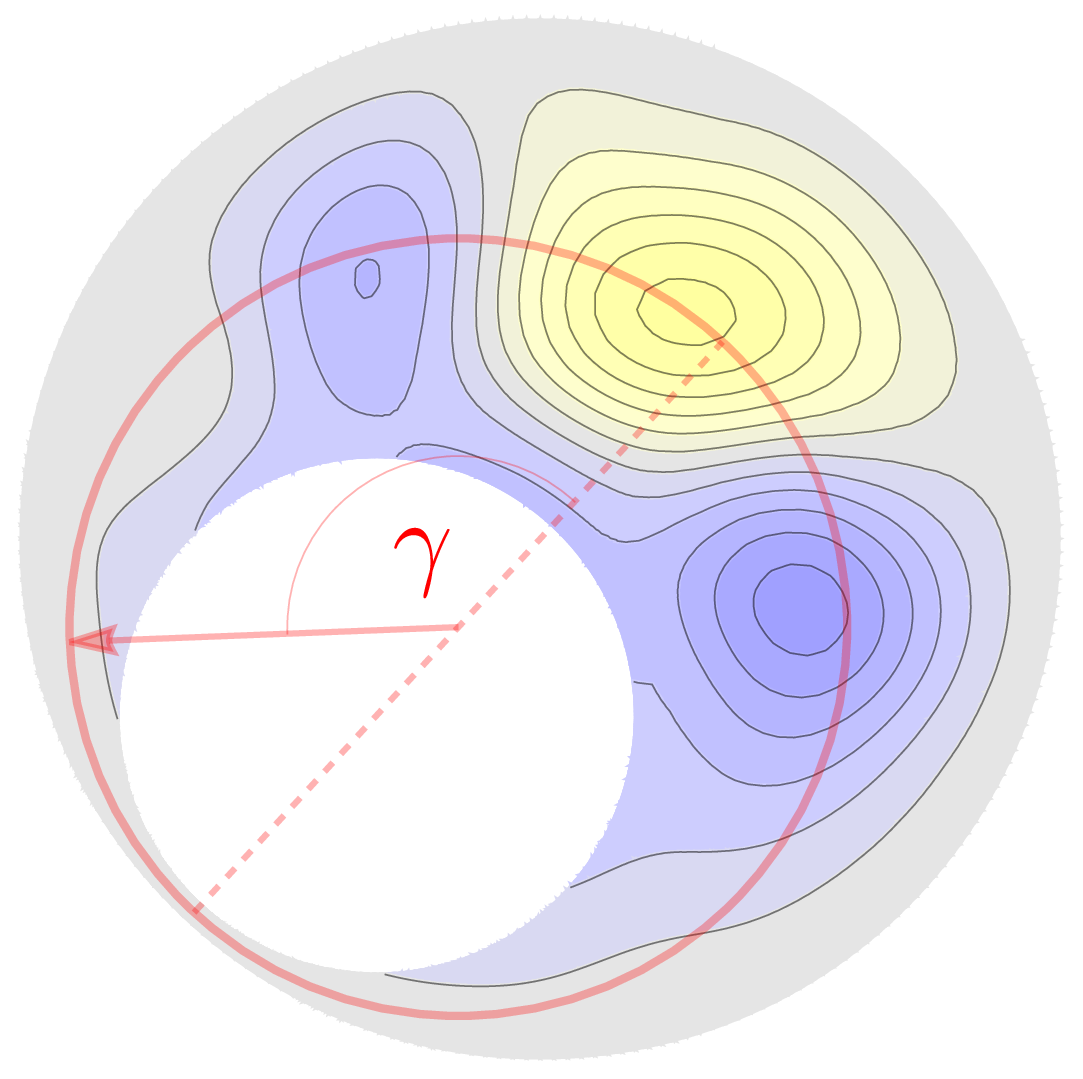}
      \end{minipage}}
  }
      \subfigure[$\alpha = -5$; $\delta_c=0.04$]{\begin{minipage}{0.26\textwidth}
        \axislabledfigure{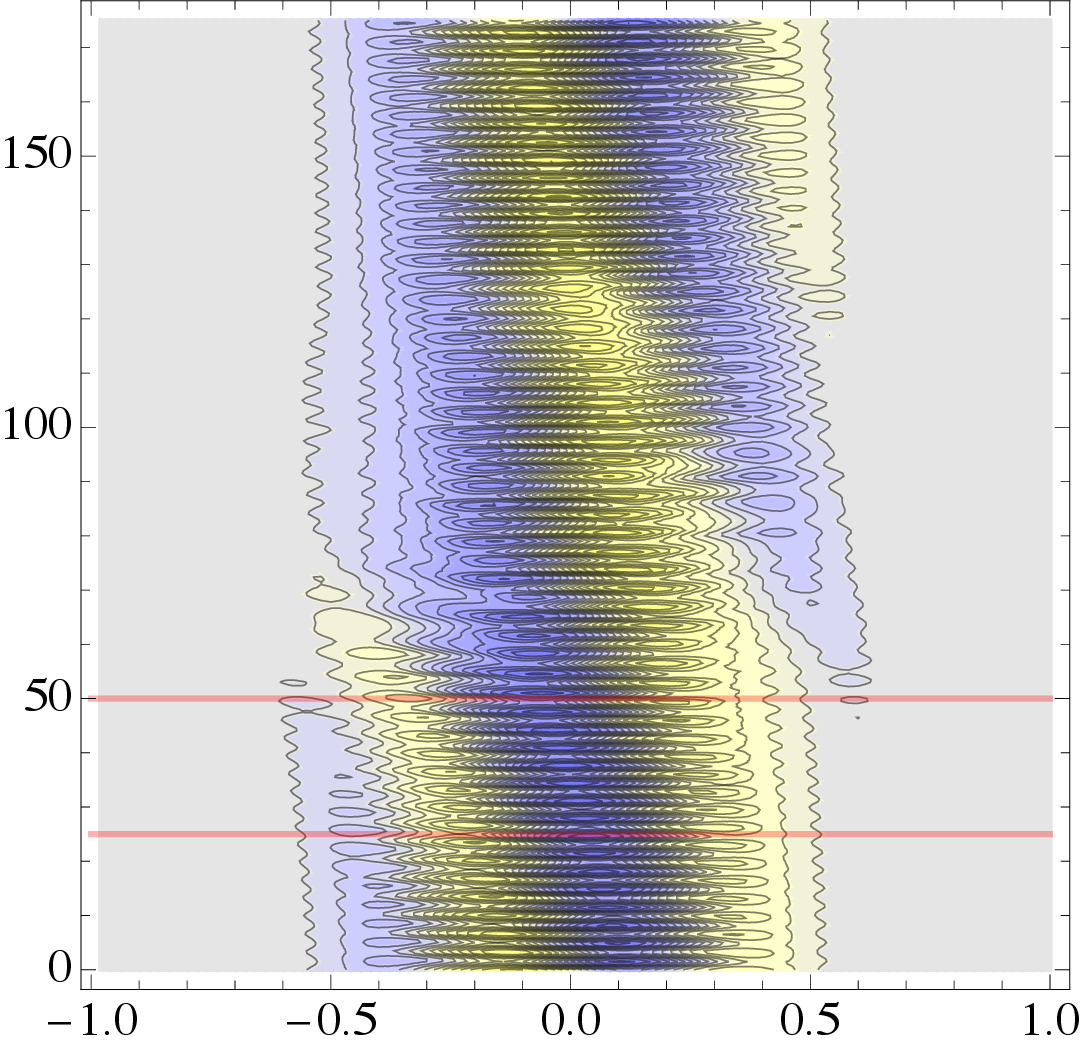}{$\gamma/\pi$}{$t$}{1.0}
      \end{minipage}
      \hspace*{0.25in}
      \raisebox{0.3in}{\begin{minipage}{0.12\textwidth}
        \includegraphics[width=\textwidth]{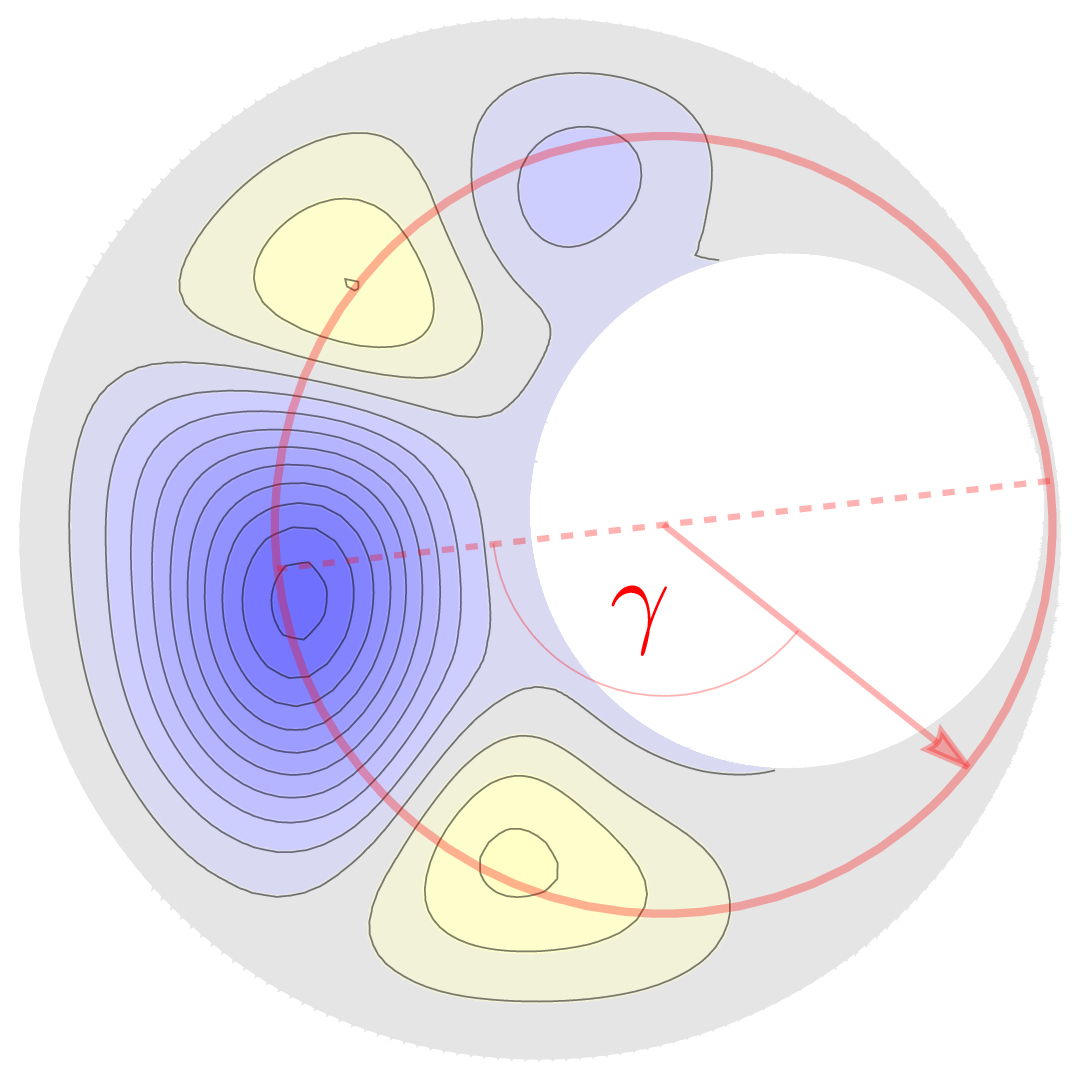}
        \includegraphics[width=\textwidth]{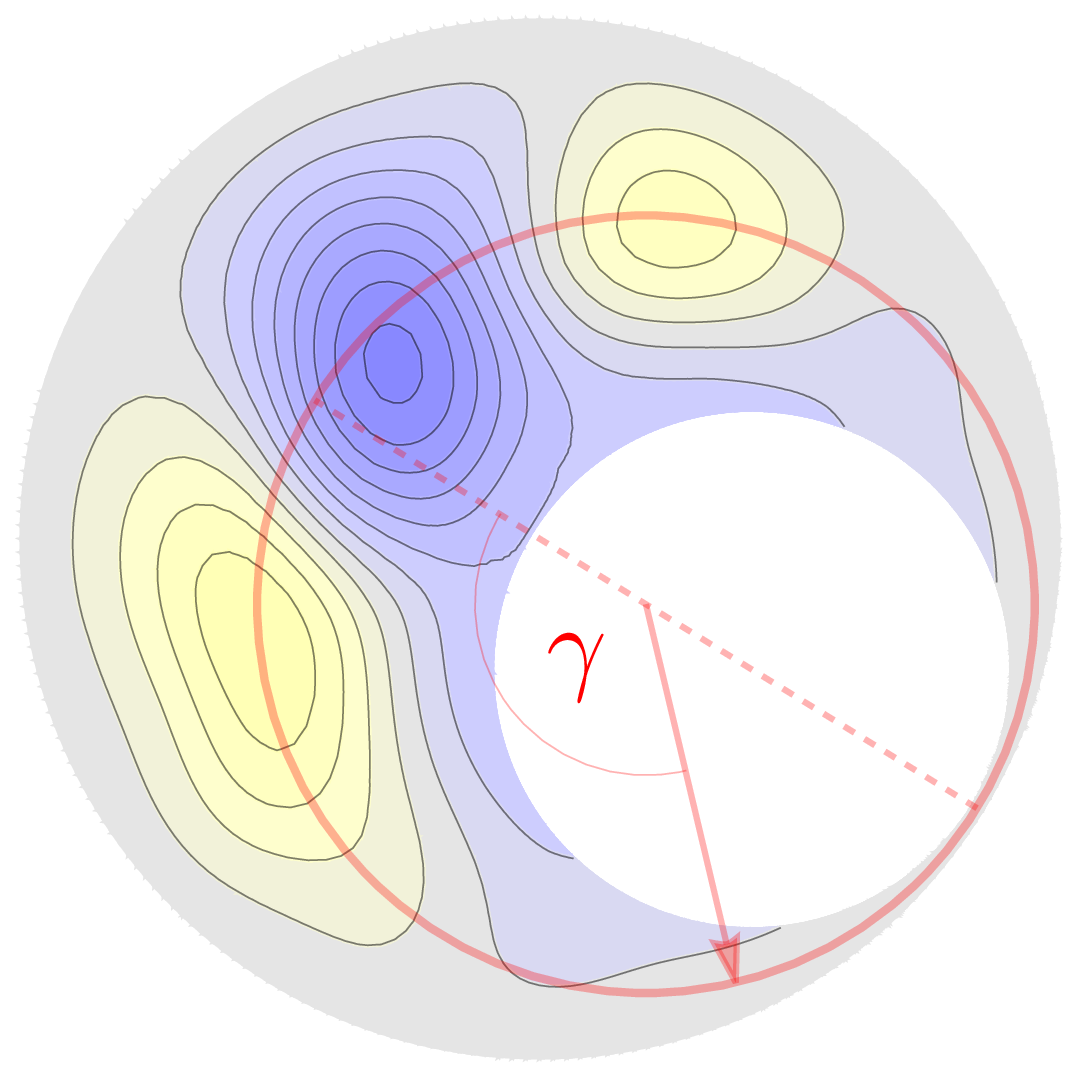}
      \end{minipage}}
  }\hfill
      \subfigure[$\alpha = -2$; $\delta_c = 0.10$]{\begin{minipage}{0.26\textwidth}
        \axislabledfigure{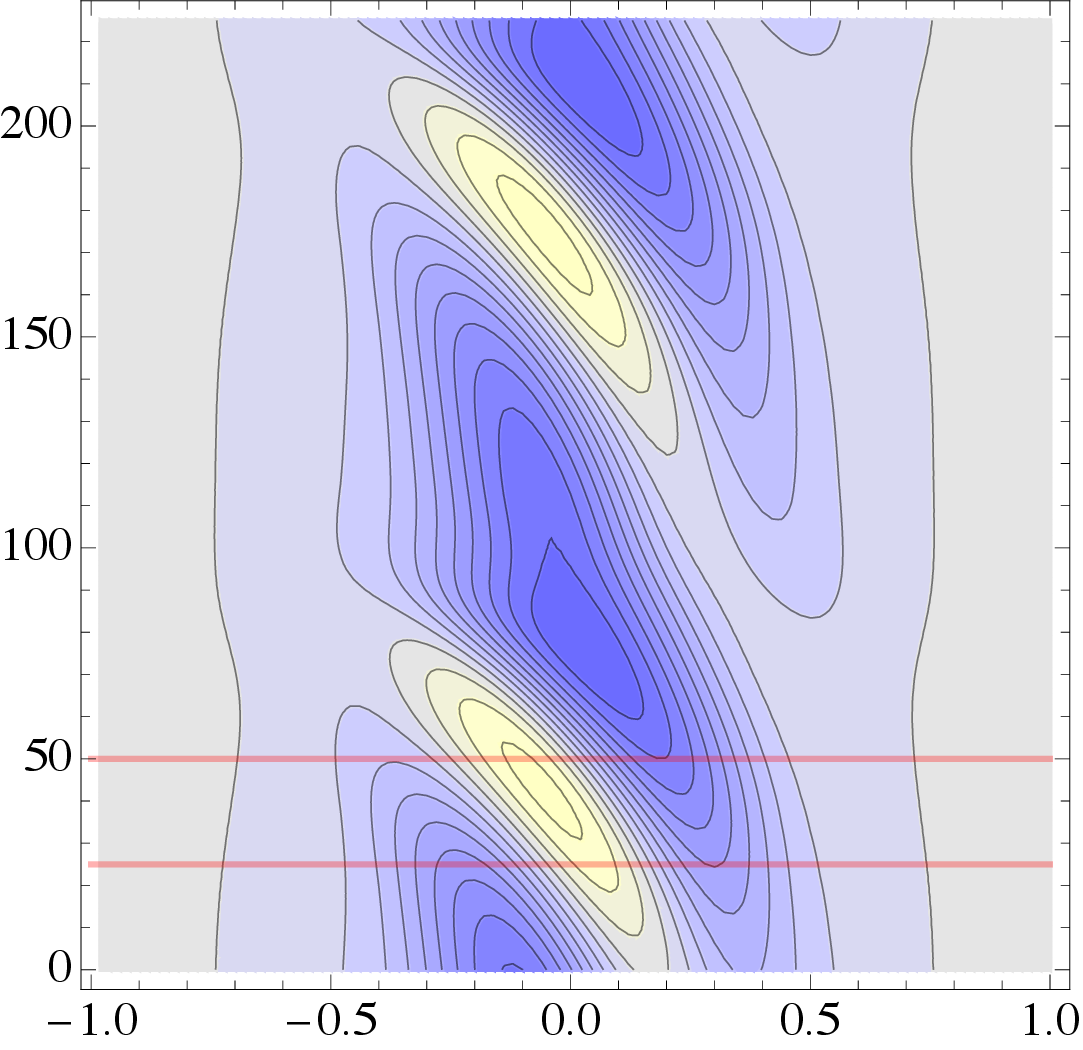}{$\gamma/\pi$}{$t$}{1.0}
      \end{minipage}
      \hspace*{0.25in}
      \raisebox{0.3in}{\begin{minipage}{0.12\textwidth}
        \includegraphics[width=\textwidth]{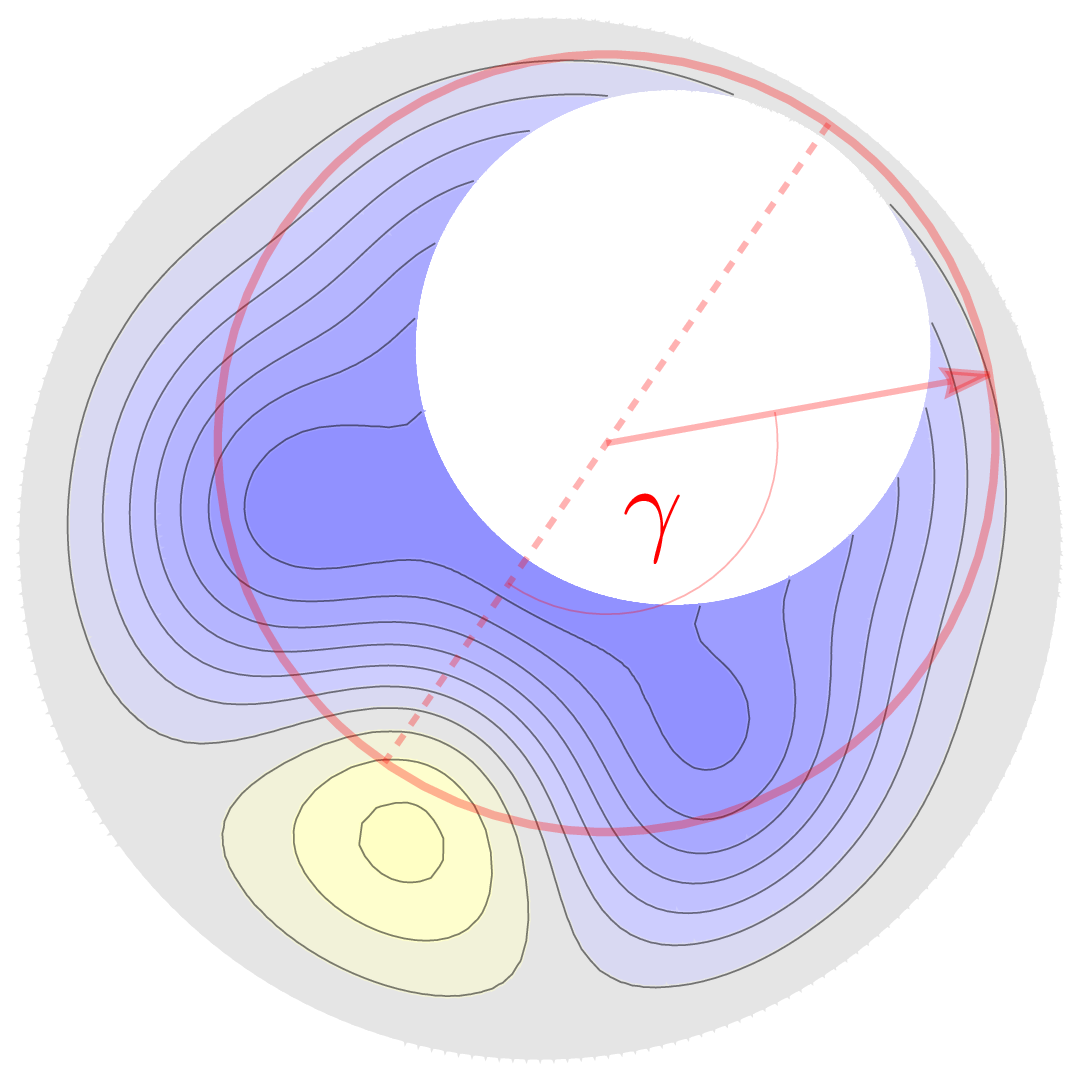}
        \includegraphics[width=\textwidth]{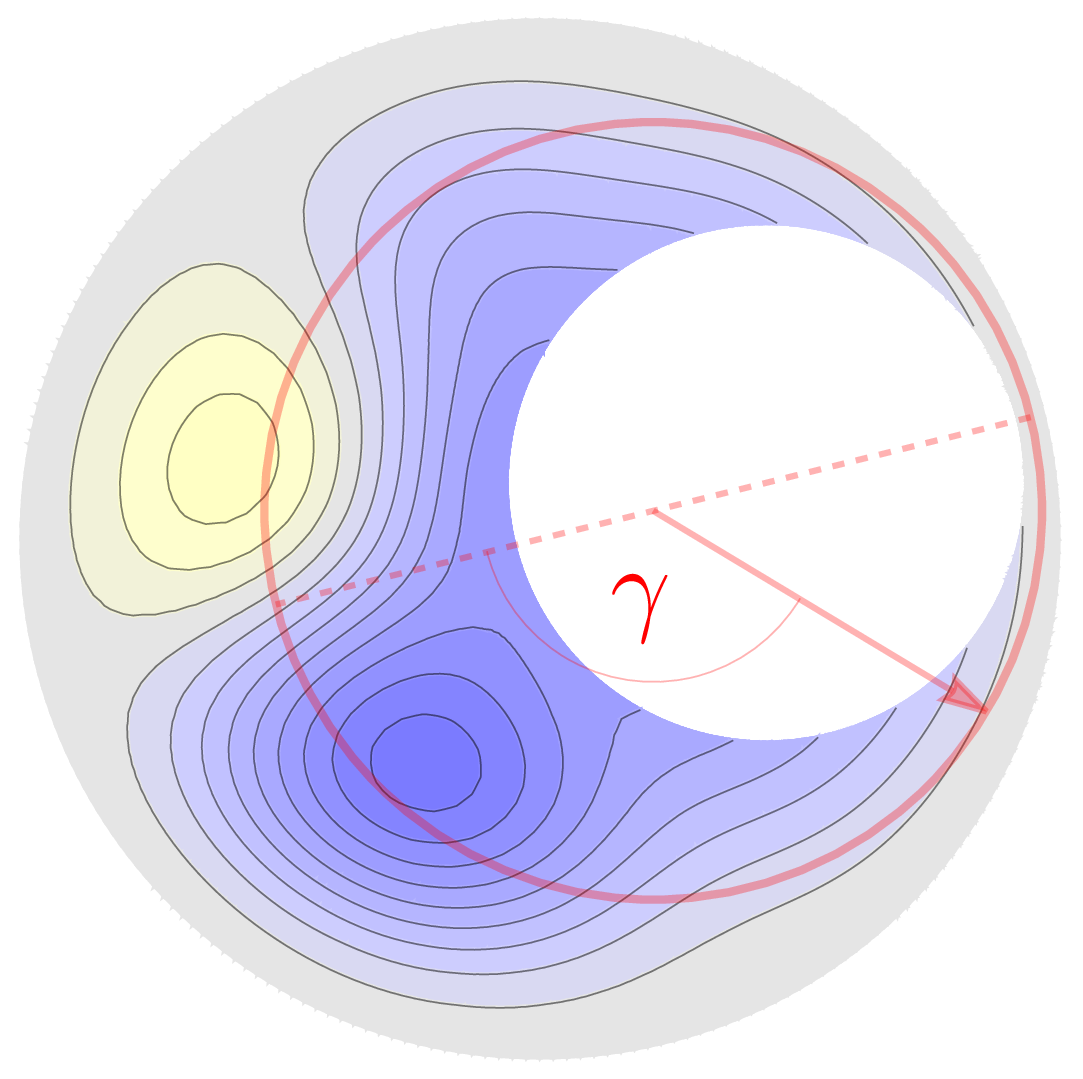}
      \end{minipage}}
  }
    \subfigure[$\alpha = -2.5$; $\bar\delta = 0.0124$ with $\sigma =8\times 10^{-5}$]{\begin{minipage}{0.26\textwidth}
        \axislabledfigure{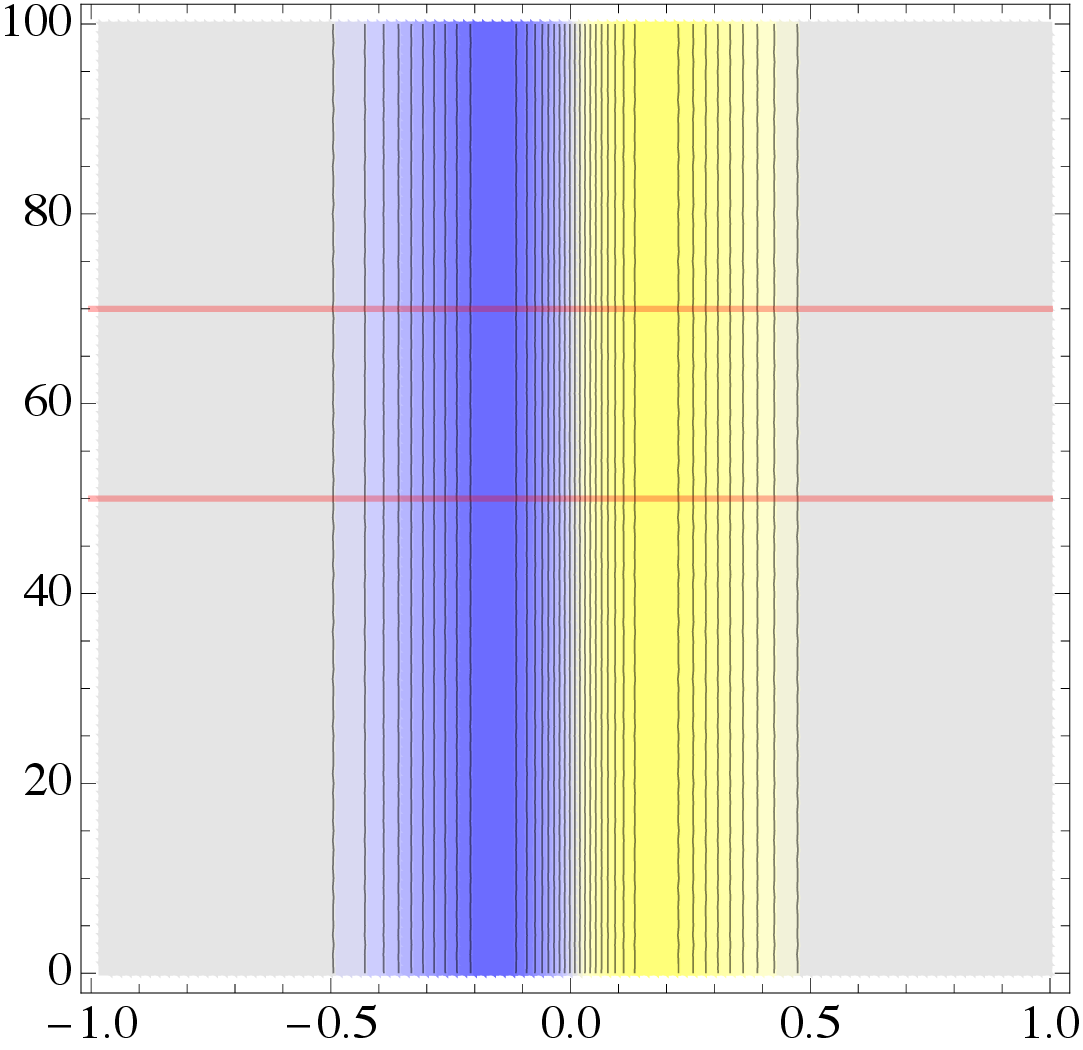}{$\gamma/\pi$}{$t$}{1.0}
      \end{minipage}
      \hspace*{0.25in}
      \raisebox{0.3in}{\begin{minipage}{0.12\textwidth}
        \includegraphics[width=\textwidth]{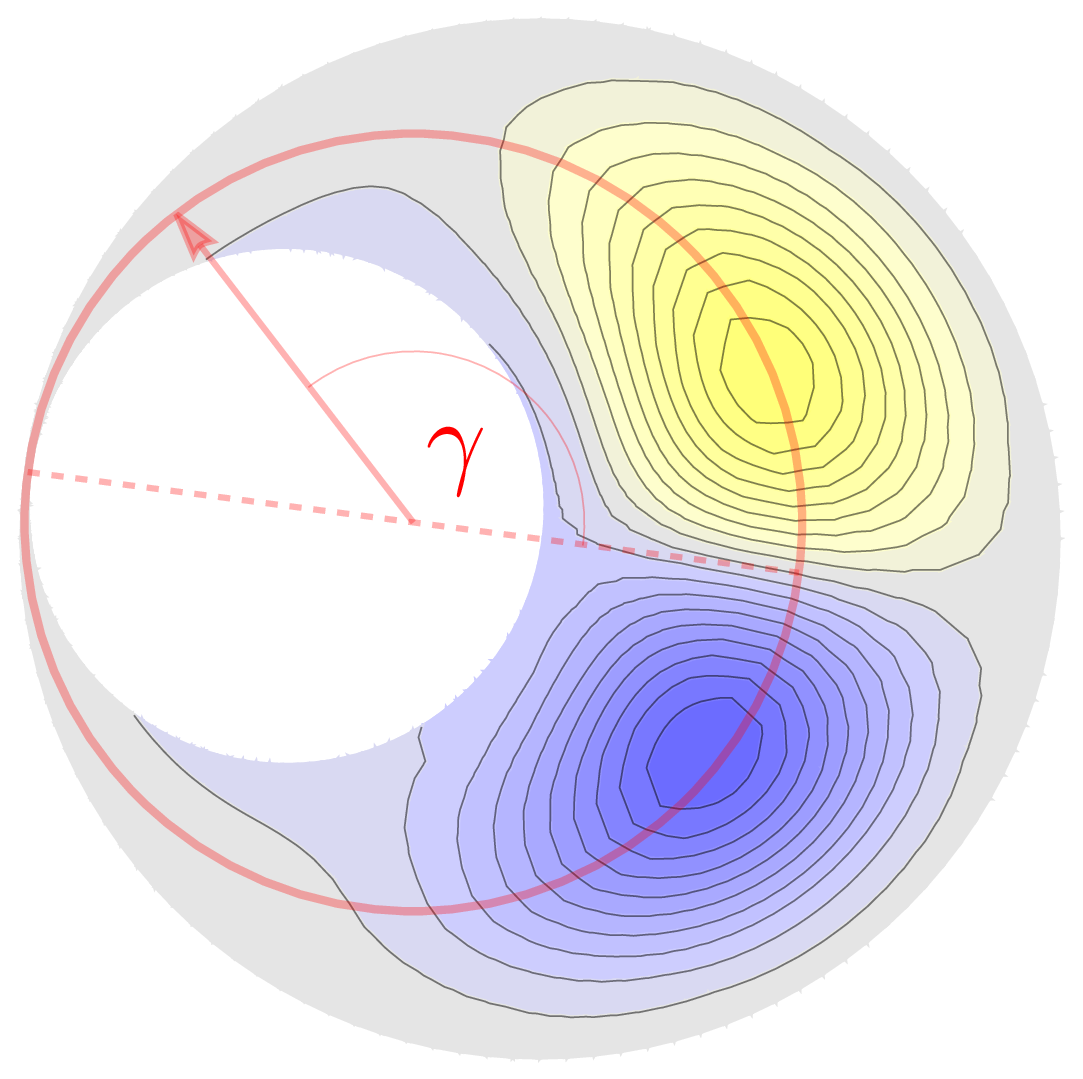}
        \includegraphics[width=\textwidth]{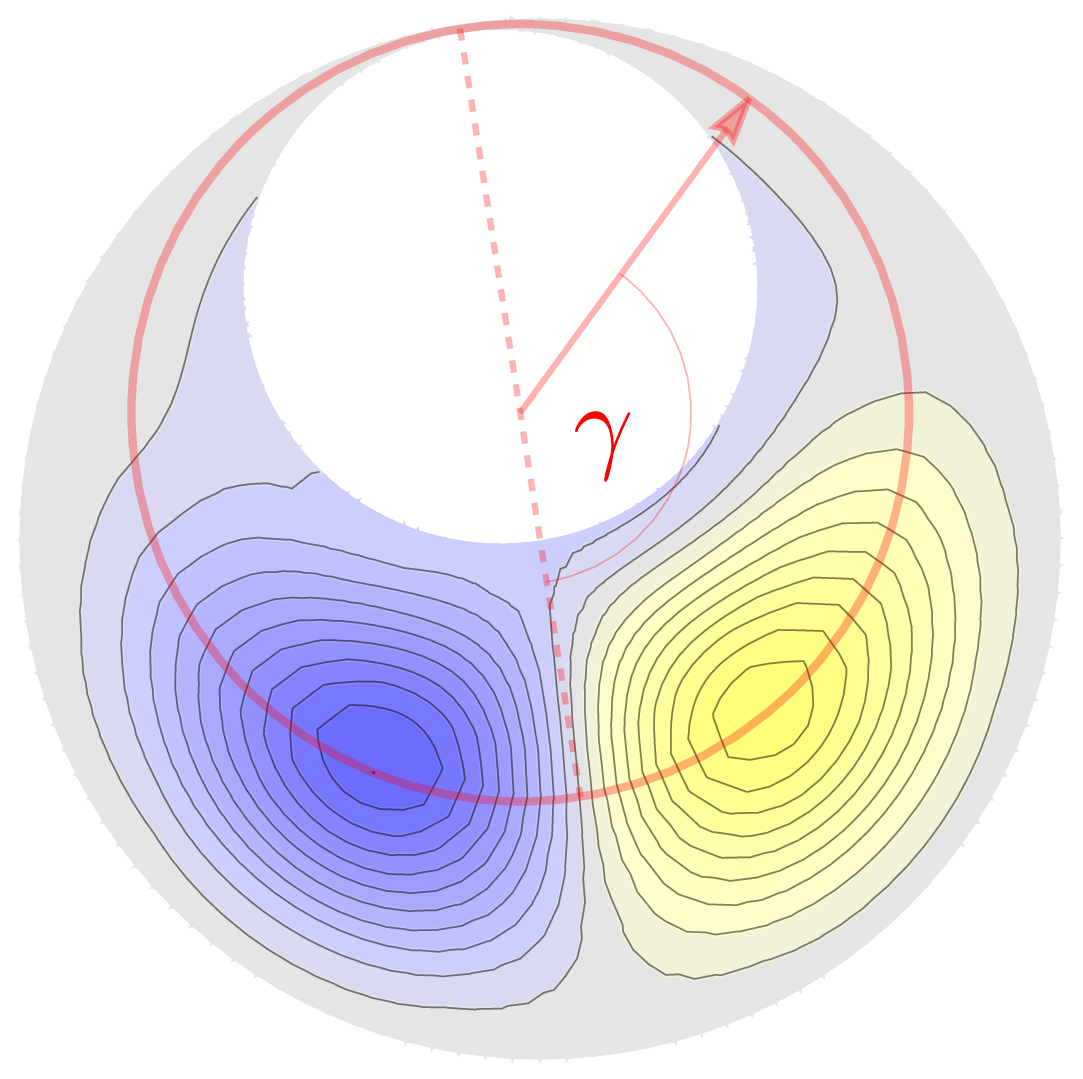}
      \end{minipage}}
    }\hfill
        \subfigure[$\alpha = -4$; $\delta_c = 0.03$]{\vspace*{-0.2in}\begin{minipage}{0.26\textwidth}
        \axislabledfigure{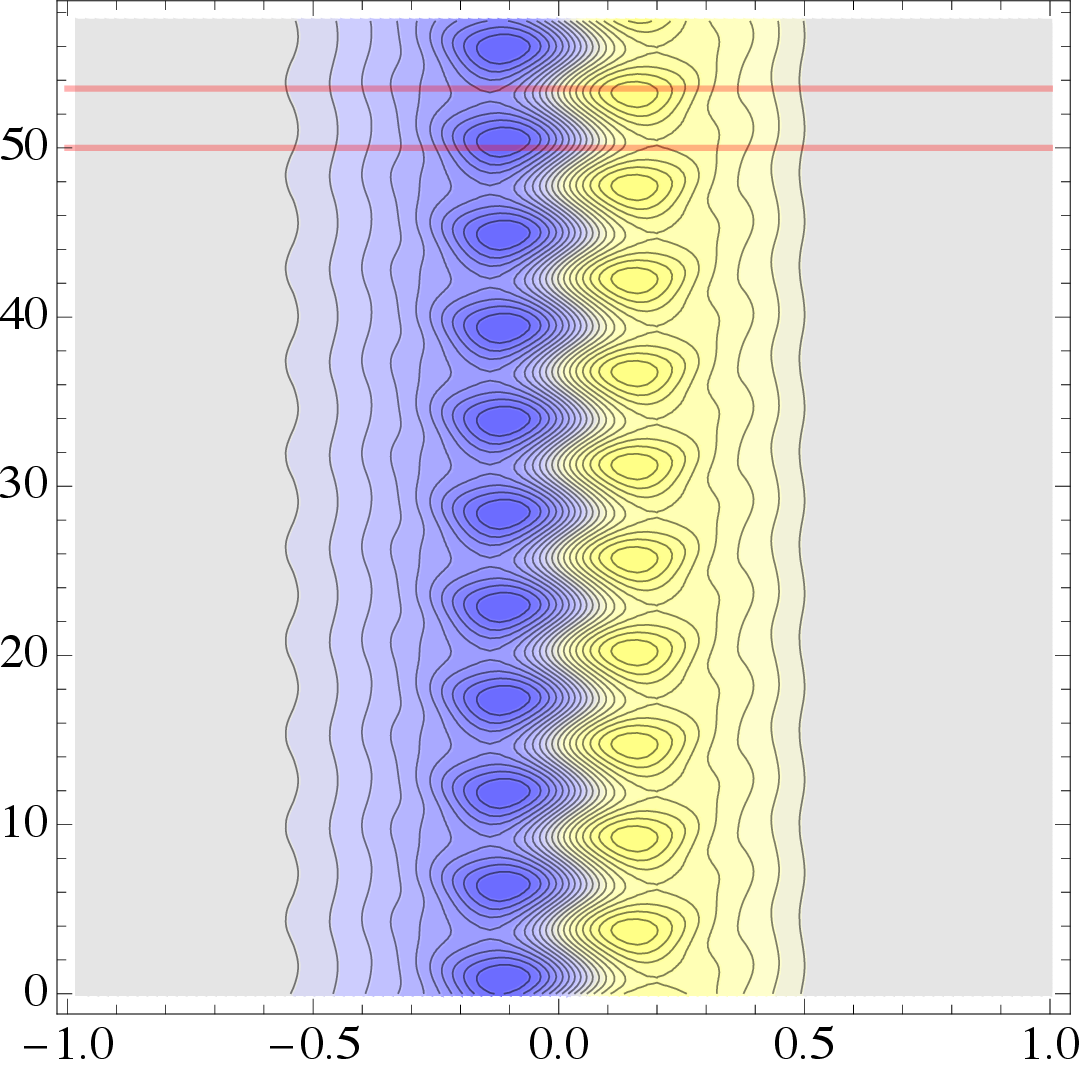}{$\gamma/\pi$}{$t$}{1.0}
      \end{minipage}
      \hspace*{0.25in}
      \raisebox{0.3in}{\begin{minipage}{0.12\textwidth}
        \includegraphics[width=\textwidth]{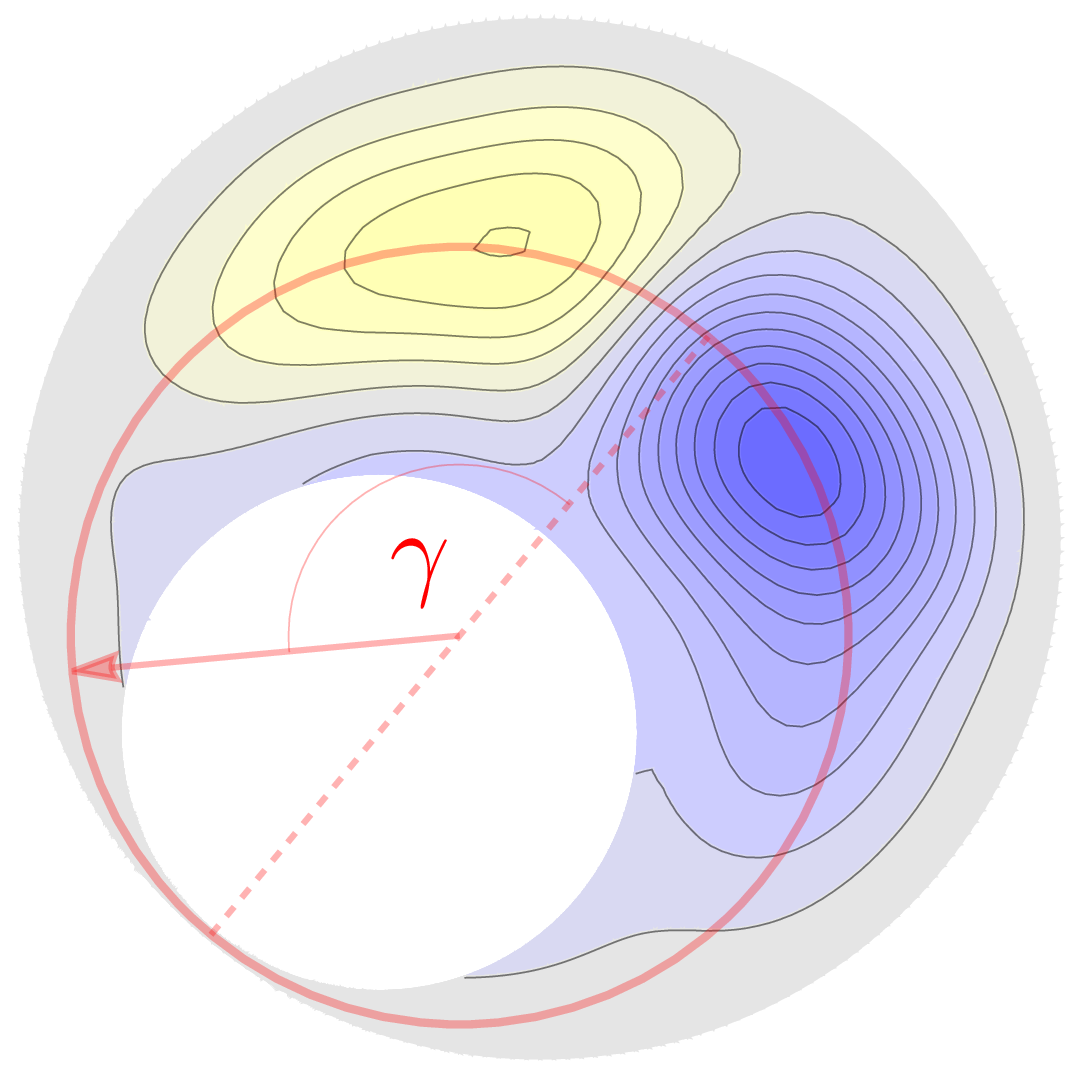}
        \includegraphics[width=\textwidth]{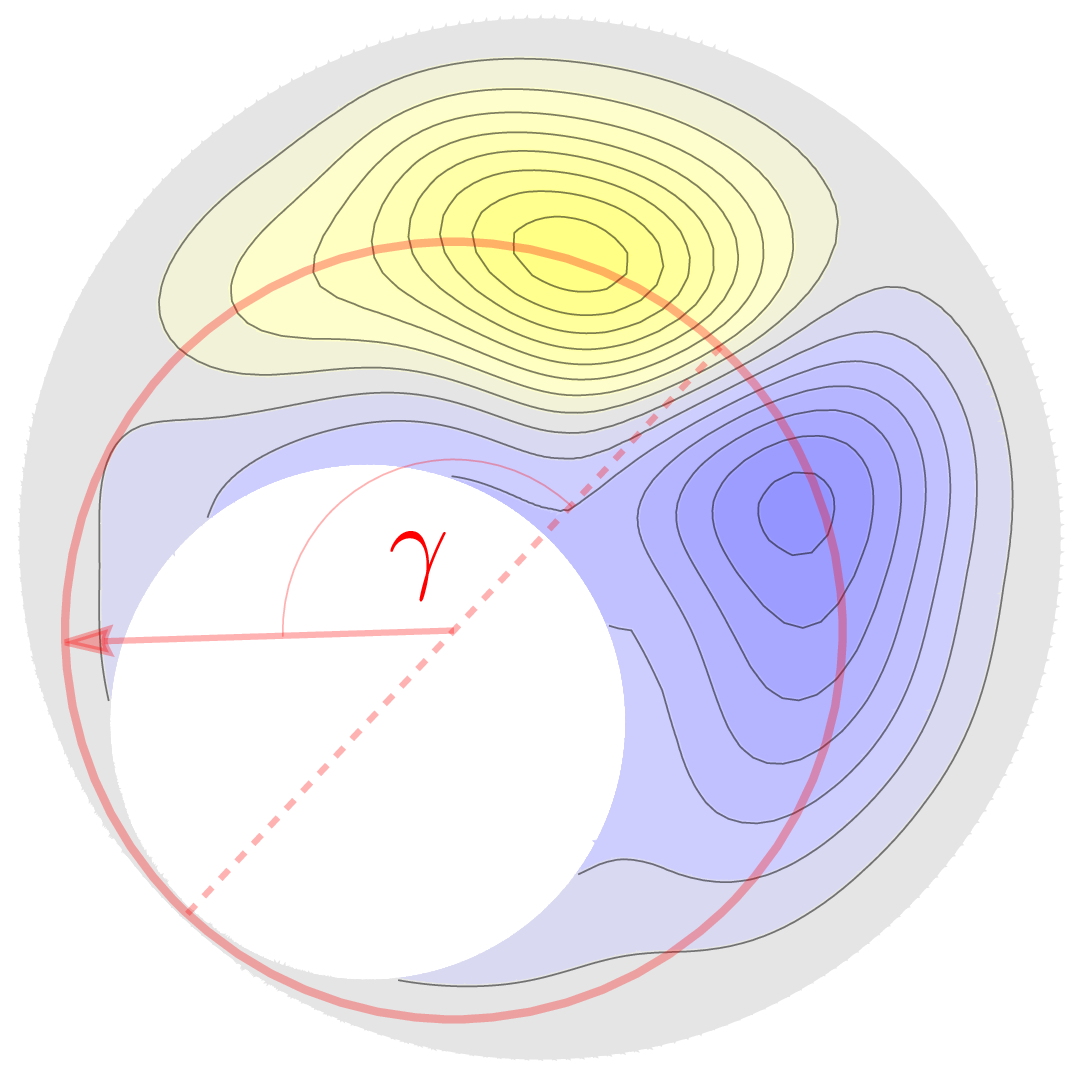}
      \end{minipage}
    }
  }
\end{center}
  \caption{(a)--(e) Visualized streamfunction $\psi$ for cases as labeled, with either a constrained wall-separation height $\delta_c$ or unconstrained mean separation $\bar \delta$ with r.m.s.\ fluctuations $\sigma$.  The horizontal lines in the $\gamma$--$t$ plots indicate the instances visualized to the right of each (with time increasing left-to-right, top-to-bottom); the $\gamma$--$t$ data are taken from the circle of radius $(R+a)/2 = 1.5$ that passes through the midpoint at smallest and largest container--object separation.  For each time, $\psi = 0$ is set at $\gamma = \pm \pi$ with 20 equally space contours between the overall $\pm |\psi|_{\text{max}}$. In all cases, the circle and precession are clockwise due to the initial perturbations.  Animated visualizations for (a) and (d) are available in supplemental movies~3 and~4.}\label{fig:XTmed}
\end{figure}

In cases, its standoff distance is nearly constant.  For example, for $\alpha = -5$ it remains about $\delta \approx 0.045$ above the wall with root-mean-square (r.m.s.) fluctuations of only $\sigma_s=0.005$, though its rotation rate fluctuates as it interacts with the unsteady vortical structures that persist in the flow (as will be seen in figure~\ref{fig:XTmed} a).  This stability was seen in figure~\ref{fig:rOmegaPDF} (a): its radial position is stable and it never changes its sense of rotation or precession.  This reflects a balance between the sustained activity in the high-strain-rate narrow gap, also a feature of the weaker activity cases of section~\ref{s:lowactivity}, and the tensile normal stresses of the persistent structures in the wider region that counter close approaches to the container wall in section~\ref{s:highactivity}. However, for somewhat weaker activity, such as $\alpha = -2.5$, the object seems to approach indefinitely toward contact, though increasingly slowly due to increasing lubrication resistance.  To study these dynamics in an analogous statistically stationary flow that avoids the discretization challenges and physical complexity of actual contact, we constrain the circle to maintain a constant distance $\delta_c$ from the wall.  The corresponding Lagrange multiplier force is interpretable as countering the net hydrodynamic lift force $L$, which is negative toward and positive away from the nearby container wall.  In corresponding unconstrained free-flowing precessing cases, fluctuating forces cause variations in $\delta$, but motion normal to the wall is so strongly resisted by lubrication effects that the corresponding changes in $\delta$ are slow, allowing at most small $\delta$ fluctuation.

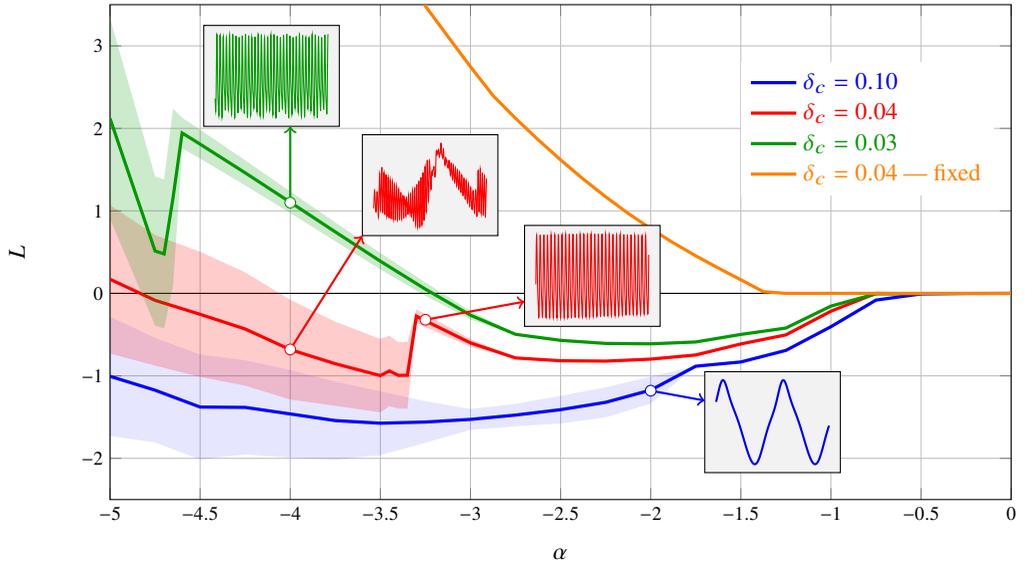
\begin{figure}
  \begin{center}
    \begin{tikzpicture}
      \begin{axis}
        [ 
        ymin = -2.5,
        ymax = 3.5,
        xmin = -5.0,
        xmax = 0.0,
        ylabel={$L$},
        xlabel={$\alpha$},
        tick scale binop=\times,
        width=\textwidth,
        height=0.6\textwidth,
        grid=major,
        fill between/on layer={main},
        legend style={	at={(axis cs:-1.5,2)},anchor=west,legend
        columns=1,draw=none,legend cell align=left},
   %
      ]
      \addplot+[no marks, black, solid, forget plot] coordinates
      {(-5,0) (0,0)};
        \addplot+[no marks, name path=A, blue, solid, very thick] table[x
        expr=\thisrowno{0}, y expr=\thisrowno{1}, col sep=space]
        {Figures/Dstats0.90.out};
        \addplot+[no marks, name path=A, draw=none, forget plot] table[x
        expr=\thisrowno{0}, y expr=\thisrowno{2}, col sep=space]
        {Figures/Dstats0.90.out};
        \addplot+[no marks,name path=B,draw=none, forget plot] table[x
        expr=\thisrowno{0}, y expr=\thisrowno{3}, col sep=space,
  forget plot]
        {Figures/Dstats0.90.out};
        \addplot[blue, fill opacity=0.1,
  forget plot] fill between[of=A and B];
        \addplot+[no marks, name path=A, red, solid, very thick] table[x
        expr=\thisrowno{0}, y expr=\thisrowno{1}, col sep=space]
        {Figures/Dstats0.96.out};
        \addplot+[no marks, name path=A, draw=none, forget plot] table[x
        expr=\thisrowno{0}, y expr=\thisrowno{2}, col sep=space]
        {Figures/Dstats0.96.out};
        \addplot+[no marks,name path=B, draw=none, forget plot] table[x
        expr=\thisrowno{0}, y expr=\thisrowno{3}, col sep=space]
        {Figures/Dstats0.96.out};
        \addplot[red, fill opacity=0.2,
  forget plot] fill between[of=A and B];
        \addplot+[no marks, green!60!black, solid, very thick] table[x
        expr=\thisrowno{0}, y expr=\thisrowno{1}, col sep=space]
        {Figures/Dstats0.97.out};
        \addplot+[no marks, name path=A, draw=none, forget plot] table[x
        expr=\thisrowno{0}, y expr=\thisrowno{2}, col sep=space]
        {Figures/Dstats0.97.out};
        \addplot+[no marks,name path=B, draw=none, forget plot] table[x
        expr=\thisrowno{0}, y expr=\thisrowno{3}, col sep=space]
        {Figures/Dstats0.97.out};
        \addplot[green!60!black, fill opacity=0.2,
  forget plot] fill between[of=A and B];
        \addplot+[no marks, orange, solid, very thick] table[x
        expr=\thisrowno{0}, y expr=\thisrowno{1}, col sep=space]
        {Figures/Dstats0.96-U0.out};
        \addplot+[no marks, name path=A, draw=none, forget plot] table[x
        expr=\thisrowno{0}, y expr=\thisrowno{2}, col sep=space]
        {Figures/Dstats0.96-U0.out};
        \addplot+[no marks,name path=B, draw=none,
  forget plot] table[x
        expr=\thisrowno{0}, y expr=\thisrowno{3}, col sep=space]
        {Figures/Dstats0.96-U0.out};
        \addplot[orange, fill opacity=0.2,
        forget plot] fill between[of=A and B];
        \legend{{\color{blue} $\delta_c = 0.10$}, {\color{red} $\delta_c =
          0.04$}, {\color{green!60!black} $\delta_c = 0.03$}, 
        {\color{orange} $\delta_c = 0.04$ --- fixed}}
          \coordinate (inset90) at (axis cs:-1.7,-0.95);
          \coordinate (mark90) at (axis cs:-2,-1.18); 
          \coordinate (target90) at (axis cs:-1.7,-1.3); 
          \coordinate (inset96) at (axis cs:-3.6,0.7);
          \coordinate (mark96) at (axis cs:-4,-0.68);
          \coordinate (target96) at (axis cs:-3.6,0.7);
          \coordinate (inset96a) at (axis cs:-2.7,-0.4);
          \coordinate (mark96a) at (axis cs:-3.25,-0.32);
          \coordinate (target96a) at (axis cs:-2.7,-0.1);
          \coordinate (inset97) at (axis cs:-4.48,2.025);
          \coordinate (mark97) at (axis cs:-4,1.1); 
          \coordinate (target97) at (axis cs:-4,2.025);
        \end{axis} 
  \begin{axis}[
    at={(inset90)},
    anchor=north west,
    name=inset,
    width=0.25\textwidth,
    ticks=none,
    axis background/.style={fill=gray!10}
    ] 
    \addplot[no marks, blue, thick] table [x expr=\thisrowno{0}, y
    expr=\thisrowno{2}, col sep=space] {Figures/tseries90.out};
  \end{axis}
  \draw[->,color=blue,thick] (mark90) -- (target90);
  \fill[draw=blue,fill=white] (mark90) circle [radius=2pt];
  \begin{axis}[
    at={(inset96)},
    anchor=south west,
    name=inset,
    width=0.25\textwidth,
    ticks=none,
    axis background/.style={fill=gray!10}
    ] 
    \addplot[no marks, red, thin] table [x expr=\thisrowno{0}, y
    expr=\thisrowno{2}, col sep=space] {Figures/tseries96.out};
  \end{axis}
  \draw[->,color=red,thick] (mark96) -- (target96);
  \fill[draw=red,fill=white] (mark96) circle [radius=2pt];
  \begin{axis}[
    at={(inset96a)},
    anchor=south west,
    name=inset,
    width=0.25\textwidth,
    ticks=none,
    axis background/.style={fill=gray!10}
    ] 
    \addplot[no marks, red, thin] table [x expr=\thisrowno{0}, y
    expr=\thisrowno{2}, col sep=space] {Figures/tseries96-3p25.out};
  \end{axis}
  \draw[->,color=red,thick] (mark96a) -- (target96a);
  \fill[draw=red,fill=white] (mark96a) circle [radius=2pt];
    \begin{axis}[
    at={(inset97)},
    anchor=south west,
    name=inset,
    width=0.25\textwidth,
    ticks=none,
    axis background/.style={fill=gray!10}
    ] 
    \addplot[no marks, green!60!black, thin] table [x expr=\thisrowno{0}, y
    expr=\thisrowno{2}, col sep=space] {Figures/tseries97.out};
  \end{axis}
  \draw[->,color=green!60!black,thick] (mark97) -- (target97);
  \fill[draw=green!60!black,fill=white] (mark97) circle [radius=2pt];
    \end{tikzpicture}
  \end{center}
  \caption{The net lift force $L$ normal to the container wall for different stand-off distance constraints $\delta_c$, plus one case with a completely fixed object with $\delta = 0.04$.  The lines indicate time averages (for time 1000 each), and the height of the corrsponding shaded regions indicate the $\pm\sigma$ r.m.s.\ fluctuations.   The insets show $L(t)$ time series of length $\Delta t = 250$ with amplitudes normalized by their respective $\sigma$.} \label{fig:lift}
\end{figure}

The lift $L$ is plotted in figure~\ref{fig:lift} for wall-distance constraints $\delta_c = 0.1$, $0.04$ and $0.03$ for $\alpha\in[-5,0]$.  All these $\delta_c$ show a cessation of motion, zero lift, and indeed absence of flow for small enough $|\alpha|$, consistent with the $\delta$ and $\alpha$ values needed for sustained activity in section~\ref{s:lowactivity}.  For stronger $|\alpha|$ and smaller $\delta_c$, the lift switches from attraction ($L<0$) to repulsion ($L>0$).  This is consistent with the observed persistent precession for $\alpha = -5$.  The lift $L$ is steady for smaller $|\alpha|$, but this changes in two ways when $|\alpha|$ increases.  A high-frequency instability appears, seemingly independently of $\delta_c$, for $\alpha \lesssim -2.5$.  This corresponds to the traveling-wave instability observed in the corresponding fixed-wall annular geometry and analyzed in a planar configuration~\citep{Gao:2017}.  The near-independence of $\delta_c$ is anticipated because it is expected to depend primarily on the voritices in the larger space, which barely changes here for the range of $\delta_c$ considered.  In all cases, the high-frequency component amplitude increases with stronger activity.

For the fixed $\bU=0$, $\Omega = 0$, $\delta_c = 0.04$ case, also shown in figure~\ref{fig:lift}, $L$ fluctuations also appear $\alpha \lesssim -2.5$, although their amplitude is smaller.  For $\alpha = -5$, $L = 8.9 \pm 0.1$ (mean $\pm$ r.m.s.). The biggest difference, however, for this fixed case is that without significant shear strain between the fixed object and the container wall, there is no induced active stress attraction.  Only the self-sustaining flow above the object is active, and it induces positive lift.

In some cases there is also a distinct low frequency, with time scale comparable to the inverse precession rate.  When the circle is further from the wall, this longer period behavior appears for smaller $|\alpha|$.  For $\delta_c = 0.9$, it is distinct for $\alpha \lesssim -1.75$ as shown in the inset of figure~\ref{fig:XTmed}, which is before the onset of the high-frequency instability.  However,
it only appears, apparently suddenly, for $\alpha \lesssim -3.35$ for
$\delta_c = 0.96$ and $\alpha \lesssim -4.60$ for $\delta_c = 0.97$.
Under some conditions, it is higher amplitude than the traveling wave instability
and always much lower frequency.  It is never observed if the circle is held fixed with $\bU = 0$ and $\Omega = 0$.  Its mechanism is straightforward to explain, but the complexity of the overall configuration hampers a full quantitative description.  Recall that the circle rotates with a sense opposite of what it would be were it rolling along the container wall.  It thus pumps fluid ahead of it through the narrow gap between it and the wall, which in turn drives an additional slow circulation of fluid in the container, so the overall circulating flow rate modestly exceeds that of the precession rate of the circle.  When this is strong enough, it advects the basic traveling-wave instability structures faster than the precession rate.  This does not happen for weaker activity, when the structures exist typically as a large pair filling in the largest region of the fluid. It also is suppressed for smaller $\delta_c$ cases because there is less of this circulation flow driven by the smaller gap spacing, which is insufficient to overcome the preferred arrangement of the vortex array.  This is consistent with there also being no low-frequency component in the fully fixed case.

\clearpage

Figure~\ref{fig:XTmed} shows the space--time behavior of the flow leading to the observed lift and motion for several of the specific cases.  Figures~\ref{fig:XTmed} (a) and (b) show the strong similarity between the unconstrained $\delta$ evolution (figure~\ref{fig:XTmed} a) versus contained motion at a similar $\delta=\delta_c$ (figure~\ref{fig:XTmed} b) for $\alpha = -5$.  Both show both the high- and low-frequency components.  The traveling wave instability manifests as a high-frequency, spatially correlated perturbation.  It is associated with a synchronized change in the shape of the vortex structures; although each vortex remains distinct, their amplitudes and aspect ratios both change at this higher frequency.  The low frequency arises from the slow advection of these structures at a speed modestly faster than the precession.  For $\alpha=-2$ and $\delta_c = 0.10$ in figure~\ref{fig:XTmed} (c), the structures themselves do not rapidly fluctuate, so no high-frequency component is evident, although the low-frequency component is pronounced.  Their slow advection is similar to the low-frequency component in the $\alpha=-5$ cases in  (a) and (b).  In all these cases, new small vorticies form ahead of the precessing circle, grow as they advect slightly faster than the circle, and eventually weaken and dissappear as they catch up to its trailing side. 

For the unconstrained $\alpha = -2.5$ case, the circle precesses close to the container wall, with only two prominent and steady counter-rotating lift-promoting vortices filling the container opposite it (figure~\ref{fig:XTmed} d).  No instability or unsteadiness is apparent in this case, though this structure ressembles those that do become unstable with a long period for either larger $|\alpha|$ or larger $\delta_c$.  For small $\delta_c$, increasing $|\alpha|$ shows the similar high-frequency instability seen in other cases, though for $\alpha = -4$ and $\delta_c = 0.03$ the fluctuating vorticies themselves still precess in lockstep with the circle (figure~\ref{fig:XTmed} e), so there is no low frequency in this case.   

\section{Summary \& Discussion}
\label{s:conclusions}

For weak activity, mobility of the object decreases the stability of the suspension relative to the corresponding fixed geometry, resulting in a lower threshold for sustained flow.  These cases also all transport the object toward a nearby wall, with active shearing stress in the space between it and the wall maintained by the shear strain rate due to its counter-rolling motion along the wall.  Attraction toward the wall is due to the concomitant tensile normal component produced in the gap.  As the object approaches, viscous resistance decreases its rotational mobility, which for sufficiently weak activity stabilizes the entire system, parking the object near the wall.  The parked distance is larger for weaker activity.  Though only a circle-in-circle configuration was considered here, parking is anticipated to be a feature of many such systems:  if suspension instability depends on the mobility of the object, it will be suppressed if the object is transported to a region where its mobility is low.  It is noteworthy that this directed and self-terminating transport arises without more complex chemoattractants or similar mechanisms that might organize motion in biological systems.  

In contrast, when the activity is strong, the mobility of the object is not essential for the instability of the suspension.  In this case, the object is advected chaotically, although not following any simple random-walk-like statistical distribution, and it never reaches the container walls.  The lift force preventing contact results derives from the greater activity in the active bulk fluid region versus the relatively suppressed activity in the fluid that is temporarily confined between the object and the wall.  This mechanism is observed to be robust:  the object remained well off the wall even in long simulations for a range of activity strengths.  Similarly, objects initialized near the wall were quickly lifted away from it (not shown), and objects held fixed near the wall experience a lift force away from it.  

For modest activity strengths, lift force fluctuations were associated with  wave-like instabilities, as observed in similar fixed geometries, and with precession along the wall.  For greater activity, these create net lift but do not overcome the gap driven attraction toward the wall for weaker activity.   In essence, when activity is concentrated close to the wall under the object, enhanced there by the high strain rate in that region, it draws the object closer to the wall.  In contrast, modestly stronger $|\alpha|$ supports more bulk region active flow, independently of the object, and draws it further from the wall.  Net lift decreases and changes sign with increase distance from the wall, which leads in cases to nearly constant stand-off distances, for which motion along the wall is indefinitely persistent in a single direction.  For still weaker but sufficient activity to maintain flow in absense of object mobility, the net result is a persistant wallward force, only countered by the usual wall-normal lubrication resistance and which can be anticipated to eventually lead to contact.

Collectively, these behaviors suggest scenarios where modulations of the overall activity of the suspension could lead to sequences of the events that could achieve useful transport objectives.   For example, a brief period of high activity (e.g., $\alpha \lesssim -10$) will pick up an object from a wall.   Following that, modest activity (e.g., $-5 < \alpha < -2$) could then transport it some distance in a deterministic way, following the contour of a container wall, as for the $\alpha = -5$ case.  Two scenarios could then park the object.  In the simpler case, $|\alpha|$ would diminish to stabilize the suspension so all motion stops.  More interestingly, the object could also autonomically park at a point where it encounters a region of confinement that sufficiently diminishes its hydrodynamic mobility.

There are many caveats to keep in mind with so simple of a model.  In our particular case, we recognize that a small enough spacing between the object will violate the coarse graining approach of the suspension model used.  For example steric interaction might cause additional repulsive forces, or alignment might be forced directly by the geometry.   Similarly three-dimensional effects will surely alter the details of any such flow, though the mechanisms discussed here are anticipated to be robust to dimensionality.  Such investigations and generalization are left to further investigations.  It will be interesting if any of these mechanisms can be identified in experiments.

\bigskip
The author is ever grateful for thoughtful input from Professor R.~H.~Ewoldt, who made comments on a draft of this paper.

\clearpage

\newpage

\bibliographystyle{jfm}
\bibliography{activesus}

\end{document}